\documentclass[aps,pra,twocolumn,longbibliography,superscriptaddress,floatfix]{revtex4-1}
\usepackage{amsfonts}
\usepackage{amsmath}
\usepackage{mathtools}
\usepackage{graphicx}
\usepackage{epsfig}
\usepackage{dcolumn}
\usepackage{bm}
\usepackage{amsmath}
\usepackage{graphicx}
\usepackage{ulem}
\usepackage{epstopdf}
\usepackage{subfigure}
\usepackage{color}
\usepackage{amsthm}
\usepackage{newlfont}
\usepackage{graphicx}
\usepackage{epstopdf}
\usepackage{appendix}
\usepackage{url}

\usepackage{breakurl}
\usepackage[breaklinks=true]{hyperref}
\usepackage{textcomp}
\usepackage{appendix}
\usepackage{multirow}	
\usepackage{color}
\usepackage{amssymb}
\usepackage{epsfig}
\usepackage{bm}
\usepackage[american]{babel}
\usepackage{braket}
\usepackage{soul}
\usepackage{float}
\hypersetup{colorlinks=true,linkcolor=blue,citecolor=blue,filecolor=blue,urlcolor=blue,pdfstartview=FitH}

\newsavebox{\graphicsbox}

\DeclareGraphicsExtensions{.pdf,.png,.jpg}
 
\newcommand{\curl}{\nabla \times}
\newcommand{\dive}{\nabla \cdot}

\newcommand{\be}{\begin{equation}}
\newcommand{\ee}{\end{equation}}
\newcommand{\ba}{\begin{eqnarray}}
\newcommand{\ea}{\end{eqnarray}}

\newcommand{\vect}[1]{\mathbf{#1}}

\newcommand{\bas}{\begin{eqnarray*}}
\newcommand{\eas}{\end{eqnarray*}}

\begin{document}

\title{Decay of two-dimensional quantum turbulence in binary Bose-Einstein condensates}
\author{Thudiyangal Mithun}
 \email{mthudiyangal@umass.edu}
 \affiliation{Department of Mathematics and Statistics, University of Massachusetts, Amherst MA 01003-4515, USA}   
\author{Kenichi Kasamatsu}
 \email{kenichi@phys.kindai.ac.jp}
 \affiliation{Department of Physics, Kindai University, Higashi-Osaka, Osaka 577-8502, Japan}
 \author{Bishwajyoti Dey}
  \email{bdey@physics.unipune.ac.in}
 \affiliation{Department of Physics, SP Pune University - Pune 411007, India}
 \author{Panayotis G. Kevrekidis}
   \email{kevrekid@math.umass.edu}
 \affiliation{Department of Mathematics and Statistics, University of Massachusetts, Amherst MA 01003-4515, USA}
\affiliation{Mathematical Institute, University of Oxford, Oxford, UK}

\begin{abstract}
{We study two-dimensional quantum turbulence in miscible binary Bose-Einstein condensates in either a harmonic trap or a steep-wall trap through the numerical simulations of the Gross-Pitaevskii equations. 
The turbulence is generated through a Gaussian stirring potential. 
When the condensates have unequal intra-component coupling strengths or asymmetric trap frequencies, the turbulent condensates undergo a dramatic decay dynamics  
to an interlaced array of vortex-antidark structures, a quasi-equilibrium state, of like-signed vortices with an extended size of the vortex core. The time of formation of this state is shortened when the parameter asymmetry of the intra-component couplings or the trap frequencies are enhanced.
The corresponding spectrum of the incompressible kinetic energy
exhibits two
noteworthy features: 
(i) a $k^{-3}$ power-law around the range of the wave number determined by the spin healing length (the size of the extended vortex-core) and (ii) a flat region around the range of the wave number determined by the density healing length.
The latter is associated with the small scale phase fluctuation relegated outside the Thomas-Fermi radius and is more prominent as the strength of intercomponent interaction approaches the strength of intra-component interaction. 
We also study the impact of the inter-component interaction to the cluster formation of like-signed vortices in an elliptical steep-wall trap, finding that the inter-component coupling gives rise to the decay of the clustered configuration. 
}
\end{abstract}

\maketitle
 %
\section{Introduction}
Turbulence is a complex dynamical behavior of a chaotic dynamical system, which connects the two distinct physical properties, namely, order and chaos~\cite{holmes}. In a two-dimensional (2D) fluid, there are two notable predictions in the turbulence theory: i) The existence of a negative temperature regime and the associated formation of clusters of point vortices predicted by Onsager \cite{onsager1949statistical}, ii) The existence of inverse energy cascade, an energy flow towards the largest spatial length, predicted by Kraichnan \cite{kraichnan1967inertial,kraichnan1975statistical}. These two predictions are focal to the understanding of turbulence in 2D fluids. 

The precise control over the parameters such as the trapping frequencies and atomic interactions renders Bose-Einstein condensates (BECs)  one of the widely used nonlinear systems to study the turbulent dynamics in quantum fluids, where the turbulence
is referred to as quantum turbulence \cite{Eyink_2006,Tsubota_2008,tsubota2013quantized,Barenghi4647,Parker_2017,Allen_2014,White_2014:a,Tsatsos_2016}. 
In 2D quantum fluids, a topological excitation is a vortex with a quantized circulation around the vortex core with a finite size. A remarkable feature of the 2D quantum turbulence is the existence of Kolmogorov's $k^{-5/3}$ law in the incompressible kinetic energy spectrum, which has a similarity to the energy cascade in classical fluids \cite{kraichnan1980two,navon2016emergence,navon2019synthetic}, where $k$ is the wave number. Furthermore, the spectrum shows a $k^{-3}$ dependence for length
scales smaller than the vortex core size determined by the healing length \cite{Bradley_2012}. While the initial stage of the turbulent dynamics is driven by the annihilation of the oppositely circulating vortices, the final stage goes to the negative temperature state caused by the ``evaporative heating" of the vortex system, where the annihilation of oppositely circulated vortices ceases \cite{Neely_2013,Billam_2014} and exhibits a $k^3$ scaling in the range of the small wave number \cite{Reeves_2014a}. In the negative temperature state, and in the presence of trap conditions that
allow for this (e.g. steep-wall traps allow for this, while parabolic ones
suppress it~\cite{Groszek_2016a}), the like-signed vortices accumulate to form giant vortex clusters (also known as Onsager vortex clusters). These clusters stay on the two opposite sides of a bounded condensate.  
Recently, by initiating the turbulent dynamics of the vortices,  two landmark experiments reported in Refs.~\cite{Gauthier1264,Johnstone1267}
have shown for the first time the existence of the negative temperature state and the Onsager vortex cluster. It has been proposed that the cluster formation of single species vortices is also possible in the dilute atomic gases \cite{Valani_2018}, and 
relevant considerations have been extended also to the finite temperature condensates \cite{10.21468/SciPostPhys.8.3.039}.

The multi-component BEC setting, either of the same atomic species \cite{Myatt_1997,Hall_1998,Maddaloni_2000,Satoshi_2010,Egorov_2013} or of the different atomic species \cite{Modugno_2002,Papp_2008,Thalhammer_2008,McCarron_2011}, enriches
significantly the phenomenology of vortices due to the presence of two competing energy scales of intra- and inter-component interactions \cite{doi:10.1142/S0217979205029602}. A highly notable feature is that the core of the one vortex can fill with the density of the other component, resulting in the formation of interlaced vortex patterns or vortex-bright structures \cite{Mueller_2002,Kasamatsu_2003,Law_2010}.
In the miscible multi-component case where the components co-exist
(rather than phase-separate), it is more relevant to refer to these
states as vortex-antidark solitons~\cite{ionut}.
Such vortices have larger core size and nontrivial vortex-vortex interaction \cite{Eto_2011,Kasamatsushortrange_2016} as compared with those in the single-component BEC. Hence, it is natural to inquire whether the turbulent dynamics in a binary condensate may exhibit unprecedented features. Furthermore, the two-component system gives the freedom of investigating the turbulence under both symmetric and asymmetric setup of parameters involved, where the asymmetry
can represent the cases of unequal intra-component strength \cite{Egorov_2013, cornell1998having} and asymmetric trap frequency \cite{Gauthier1264}. 

In this paper, we study the vortex turbulence in a 2D two-component BEC. Depending on the strength of the intra- ($g_{11}$ and $g_{22}$) and inter-component interactions ($g_{12}$), the system resides either in a miscible regime ($\sqrt{g_{11} g_{22}} > g_{12}$) or in an immiscible one ($\sqrt{g_{11} g_{22}} < g_{12}$) \cite{Myatt_1997,Hall_1998,Maddaloni_2000,Modugno_2002,Satoshi_2010,Egorov_2013}. Recently,  studies of turbulent dynamics in a binary condensate have been reported in \cite{MarkusUniversal2013,kobyakov2014turbulence,Han_2018,Han_2019}. Han and Tsubota found that different spatial distributions of vortices in each component arose from the initially phase imprinted vortices; the Onsager cluster formation takes place for the case of small inter-component coupling strength (compared to the intra-component one), while for a large inter-component strength the system exhibits a phase separated state where the components (and hence their vortices) may sit at two opposite poles~\cite{tick}
\footnote{it should be noted though that radial phase separation is possible as well, see
e.g.~\cite{ionut2} for a recent discussion and azimuthal variations thereof.}.
Importantly, the turbulent dynamics in two-component BECs may highly deviate from this phenomenology for the following reasons: i) an initial state used in the simulations of Refs.~\cite{Han_2018,Han_2019}, where vortices and anti-vortices are distributed evenly and randomly over the condensate, is difficult to obtain in  experiments, ii) The asymmetry in the parameters is likely to manifest itself in the experimental
dynamics~\cite{Egorov_2013}. 

In this work, we present  turbulent dynamics in two-component BECs
induced via a stirring scheme, that is commonly used in experiments
\cite{Neely_2013,Gauthier1264,Johnstone1267,Kwon_2014,Kwon_2015,seo2017observation,Thompson_2013}.
{ It is worth noting that~\cite{Thompson_2013} discussed
  experimental evidence for the power-law in driven BECs, while more
  recently turbulent Na-K bosonic mixtures have been studied in~\cite{Castilho_2019}.}
Here, we investigate the relevant phenomenology in miscible two-component BECs with asymmetric parameter settings. Since it is known that the trap geometry plays a significant role in the vortex cluster formation \cite{Groszek_2016a,Gauthier1264}, we implement the dynamics in a harmonic trap and also in a steep-wall trap \cite{Kwon_2014,Stagg_2015,Groszek_2016a}. 
We find that the initial turbulence generated via a stirring potential decays to the interlaced vortex-antidark structures mentioned above which, in turn,
bear a large size of the vortex core. 
{ This interlaced structure can be regarded as a quasi-equilibrium state, 
{since the time development of inter- and intra-component energy relaxes}
and the density profile at a given moment is similar to the interlaced vortex lattice \cite{Kasamatsu_2003}.
The corresponding incompressible spectrum develops a $k^{-3}$
power-law for the wave numbers determined by the inverse of the spin healing length, $\xi_s$ and a flat region for the range of the wave number determined by the density healing length, $\xi$.
{It is unlike} the well known $k^{-5/3}$- and $k^{-3}$-power laws for IR and UV regimes, respectively, of the two-dimensional Gross-Pitaevskii turbulence \cite{White_2014:a,Tsatsos_2016,Reeves_2014a}. 
The former $k^{-3}$ power-law seen around $k_s =2\pi/\xi_s$
is associated with the vortex core properties \cite{Bradley_2012};
while, the latter flat region is caused by the bottleneck effect of the incompressible kinetic energy flow, where the small scale vortex fluctuations accumulate around the condensate periphery.} 
In the case of the steep-wall trap, where formation of the Onsager cluster characterized by the large dipole moment of the vortex charges is expected in a single-component BEC \cite{Gauthier1264,Johnstone1267}, the presence of the inter-component coupling also causes the decay of vortices, preventing the persistence of the cluster configuration. 

The paper is organized as follows. After introducing the formulation of the problem in Sec.~\ref{sc:model}, 
we first study the turbulent dynamics of miscible two-component BECs in a harmonic potential in Sec.~\ref{harmonicturb}. 
In Sec.~\ref{SteepWall}, we consider the turbulence in a steep-wall trap, discussing the cluster formation of vortices and anti-vortices. 
Section~\ref{sec:conclu} is devoted to the conclusion. 

\section{Theoretical model of binary BECs}\label{sc:model}
We begin with the effective 2D Gross-Pitaevskii (GP) energy functional $E[\Psi_1,\Psi_2]=\int \mathcal{E}_\text{2D}(\bm{r})d^2 r$ expressed in
terms of the condensate wave functions $\Psi_j$ for the $j$-th component ($j=1,2$), where the energy density is
\begin{align}\label{eq:GPEF}
\mathcal{E}_\text{2D}(\bm{r}) = \sum_{j=1}^2 \left[ \frac{\hbar^2}{2m_j}|\nabla\Psi_j|^2 +V_j\left(\bm{r} \right) |\Psi_j|^2 + \frac{g_{jj}}{2} |\Psi_j|^4 \right] \nonumber \\ 
+ g_{12} |\Psi_1|^2 |\Psi_2|^2.
\end{align}
Here, the wave functions obey the normalization $\int d^2 r |\Psi_j|^2 = N_j$ with the particle number $N_j$ in the 2D system. 
The parameter $m_j$ represents the atomic mass of the $j$-th component. 
{The 2D interaction strengths $g_{jk}$ are related with a 3D coupling constant $g_{jk}^{\text{3D}}$ as $g_{jk} = g^{\text{3D}}_{jk} \int |\psi(z)|^4 dz / \int |\psi(z)|^2 dz$ with the longitudinal component of the wave function being $\psi(z)$. Here, $g^{\text{3D}}_{jj}= 4\pi\hbar^2 a_j/m_j$ is the intracomponent interaction strength and $g^{\text{3D}}_{12}=2\pi\hbar^2 a_{12}(m_1+m_2)/(m_1 m_2)$ the intercomponent one with the corresponding $s$-wave scattering lengths $a_{j}$ and $a_{12}$.}
Throughout the paper we consider the case of equal particle numbers $N_1=N_2 \equiv N$
and equal masses $m_1=m_2=m$; for completeness,
we also consider briefly the case $N_1 \neq N_2$ in
Appendix \ref{gq_ge_g2}. The mass equality suggests our focus on
a scenario of two hyperfine states of the same gas, in particular
$^{87}$Rb as discussed below~\cite{siambook}.

The one-body potential $V_j$ consists of  two parts denoted as $V_{Tj}(\bm{r})$ and $V_s(\bm{r})$;
\ba\label{eq:trap}
V_{Tj}(\bm{r}) =\frac{1}{2} m \omega_r^2 R_0^2 \bigg(\frac{\sqrt{(1+\epsilon_{xj}) x^2+(1+\epsilon_{yj}) y^2}}{R_0}\bigg)^{\alpha}, \label{trappot}
\ea
and 
\begin{equation}
V_s(x,y,t)=V_0 \exp \left[ -\frac{ (x-x_0(t))^2+(y-y_0(t))^2}{\sigma_0^2} \right],
 \label{eq:stir_pot}
\end{equation}
where $\omega_r$ is the radial harmonic frequency, $\epsilon_{xj}$ and $\epsilon_{yj}$ represent the trap anisotropy along the $x$- and $y$-directions, respectively, and $R_0$ is the typical size of the potential. 
For $\alpha=2$, Eq.~\eqref{eq:trap} represents a harmonic-oscillator potential, while for a large $\alpha$ it can be considered as a steep-wall potential. 
{The additional potential $V_s(\bm{r})$ of Eq.~\eqref{eq:stir_pot} represents a Gaussian stirring obstacle having a strength $V_0$ and a width $\sigma_0$. 
This can be created by a blue-detuned laser beam directed axially along the trap \cite{PhysRevLett.83.2502,PhysRevLett.85.2228}}.

From Eq. \eqref{eq:GPEF} we get the time-dependent GP equations (GPE)
\be
\begin{split}
i \hbar \frac{\partial \Psi_j}{\partial t} &= \bigg[-\frac{\hbar^2 \nabla^2}{2m} +V_j(\vect{r}) +g_{jj} |\Psi_j|^2 + g_{12} |\Psi_{3-j}|^2\bigg] \Psi_j.
\label{eq:2GP}
\end{split}
\ee
In the following, we denote the physical quantities in units of the radial harmonic oscillator, i.e., the length, time, energy are scaled by $a_0$, $1/\omega_r$, $\hbar \omega_r$, respectively, where $a_0 = \sqrt{\hbar/(m \omega_r)}$ is the radial harmonic oscillator length. The wave function is scaled as $a_0^{-1}\sqrt{N}$, which leads to $\int d^2 r |\Psi_j|^2 = 1$ and the dimensionless coupling constants $\tilde{g}_{jk}=g_{jk} N m/\hbar^2$. Then, the Thomas-Fermi radius is $R_\text{TF} = \sqrt{2 \mu/(m \omega_r^2)} = \sqrt{2 \tilde{\mu}} a_{0}$ and the healing length $\xi =\hbar/\sqrt{2m \mu}  = a_0/\sqrt{2 \tilde{\mu}}$ with $\mu = \tilde{\mu} \hbar \omega_r$.  
In Eq.~\eqref{trappot}, we take the size of the trap potential as $R_0 = R_\text{TF}$ for convenience. 

\section{Vortex turbulence in a harmonic trap}\label{harmonicturb}
 {Our motivating example is that of a mixture of 2D BECs
   of $^{87}$Rb atoms in the different hyperfine spin states, e.g.,
   $|F=1,m_F=-1\rangle$ and $|F= 2, m_F= 1\rangle$}.
 In a harmonic trap ($\alpha=2$) with the frequency $\omega_r=2\pi\times 15$ Hz and the aspect ratio $\lambda=\omega_z/\omega_r=10$. Choosing the $s$-wave scattering length $a_1 = 100a_\text{B}$ \cite{Egorov_2013} ($a_\text{B}$  is the Bohr radius), and $N^\text{3D}\approx6.5\times 10^4$, we get the parameter values as $a_0 \approx 2.7\mu$m and $\tilde{g}_{11}=g_{11} N m / \hbar^2 = 4\pi N^\text{3D} (a_1 /a_0) \sqrt{\lambda/(2 \pi)} \approx 2000$, {where $N = N^\text{3D} \sqrt{\lambda}/a_0 = 7.4 \times 10^{4} / \mu m$} \cite{Groszek_2016a}. 
The inter-component coupling strength is chosen as $0 < g_{12} < \sqrt{g_{11} g_{22}}$, being repulsive and in the miscible regime~\cite{siambook}.  

In order to generate the vortices, we use a stirring technique
{ with the use of repulsive Gaussian potential of Eq.~\eqref{eq:stir_pot}}  \cite{Weiler_2008,Reeves_2012,Neely_2013,Kwon_2015,zhu2018surface,Groszek_2018}. It has a strength of $V_0= 1.2 \mu \approx 42.28 \hbar \omega_r$ and a width $\sigma_0 =  0.1 R_\text{TF}$. In Eq.~\eqref{eq:stir_pot}, $x_0(t) = r_{0} \cos(v t/r_{0})=r_{0} \cos( 2 \pi t/T)$ and $y_{0}(t)=r_{0} \sin( 2 \pi t/T)$, where $T$ is the period and $v$ is the velocity of the obstacle \cite{Reeves_2012,Groszek_2018}. 
Since it is found that for a harmonically trapped condensate the maximum excitation depends on the position of the obstacle, we fix $r_{0} \approx 0.4R_{0}$, corresponding to the location where the energy required to form a vortex dipole is minimal \cite{Reeves_2012,Zhou_2004}. We further fix $v= 0.6c_s$, where $c_s=\sqrt{\mu/m}=\sqrt{\tilde{\mu}}a_0 \omega_r$ is the velocity of the sound wave (Bogoliubov speed of sound).   
 
The numerical simulations are performed as follows. We first get the initial stationary solution through the imaginary time propagation of the GPE \eqref{eq:2GP} in the presence of the static obstacle of Eq.~\eqref{eq:stir_pot}. Next, the condensate is evolved via real-time simulations, being stirred by the potential of Eq.~\eqref{eq:stir_pot} for two periods, where the obstacle strength is ramped down to zero in the second period (see Appendix \ref{stirr_ini}). Just after that, we
set the time $t=0$, corresponding to the end of the preparation stage and the beginning of our evolution observations. 
We use a split-step fast-Fourier scheme for the numerical simulation \cite{Mithun_2019}. 
{In the simulation, we consider the simulation domain $[-L/2:L/2] \times [-L/2:L/2]$ with $M \times M$ grid points. }
{We take $M=1024$ and $L=40$, unless otherwise mentioned, and the
  time step $\Delta t$ in such a way that the width of the spatial
  grids $\Delta x = L/M < \xi/a_0$ and the time step satisfy $\Delta t
  < (\Delta x)^2/2$. The selection of the spatial and temporal
  discretizations and method have been made so as to ensure a relative
  norm $|N(t)-N(0)|/N(0)$ and a energy $|E(t)-E(0)|/E(0)$ less than
  $10^{-2}$, over the temporal horizon of our numerical simulations.}

The stirring potential can generate vortices via two mechanisms. One is the vortex--anti-vortex pair nucleation which occurs at the low density region induced by the repulsive Gaussian potential. Although the considered impenetrable obstacle with $V_0/\mu > 1$ is able to emit a single vortex into the condensate, even when the co-produced partner (anti-vortex) is well inside the obstacle-induced zero-density region \cite{Sasaki_2010,Reeves_2012,Reeves_2015}, a vortex and an anti-vortex are always emitted simultaneously from the obstacle in our setting. The other
mechanism is the vortex entrance from outside of the condensate boundary due to the random distribution of phase in the low-density periphery, {where the energy cost for vortex formation is minimal.}
Nevertheless, we confirmed in our simulation that the second scenario is less probable, as shown in Appendix \ref{stirr_ini} \cite{animation}.

Now, we analyze both the vortex dynamics and the energy spectra. To calculate the spectra we take the average over 4 different initial conditions and these initial conditions are obtained by changing $\sigma_0$ and $V_0$ by small amounts. 
 
\subsection{Vortex dynamics 
in turbulent binary BECs}
\begin{figure*}[!htbp]  
 \includegraphics[scale=1.0,width=0.95\textwidth]{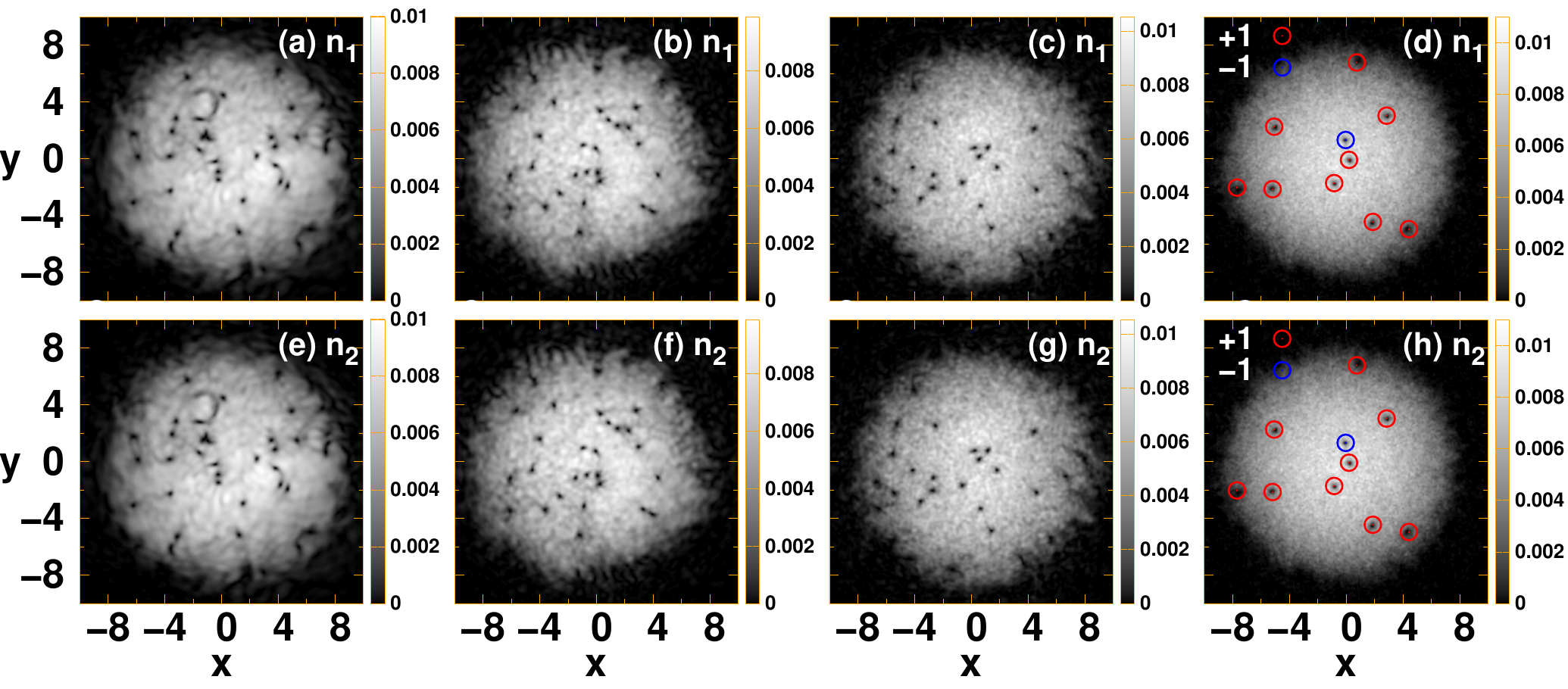} \\
 \includegraphics[scale=1.0,width=0.95\textwidth]{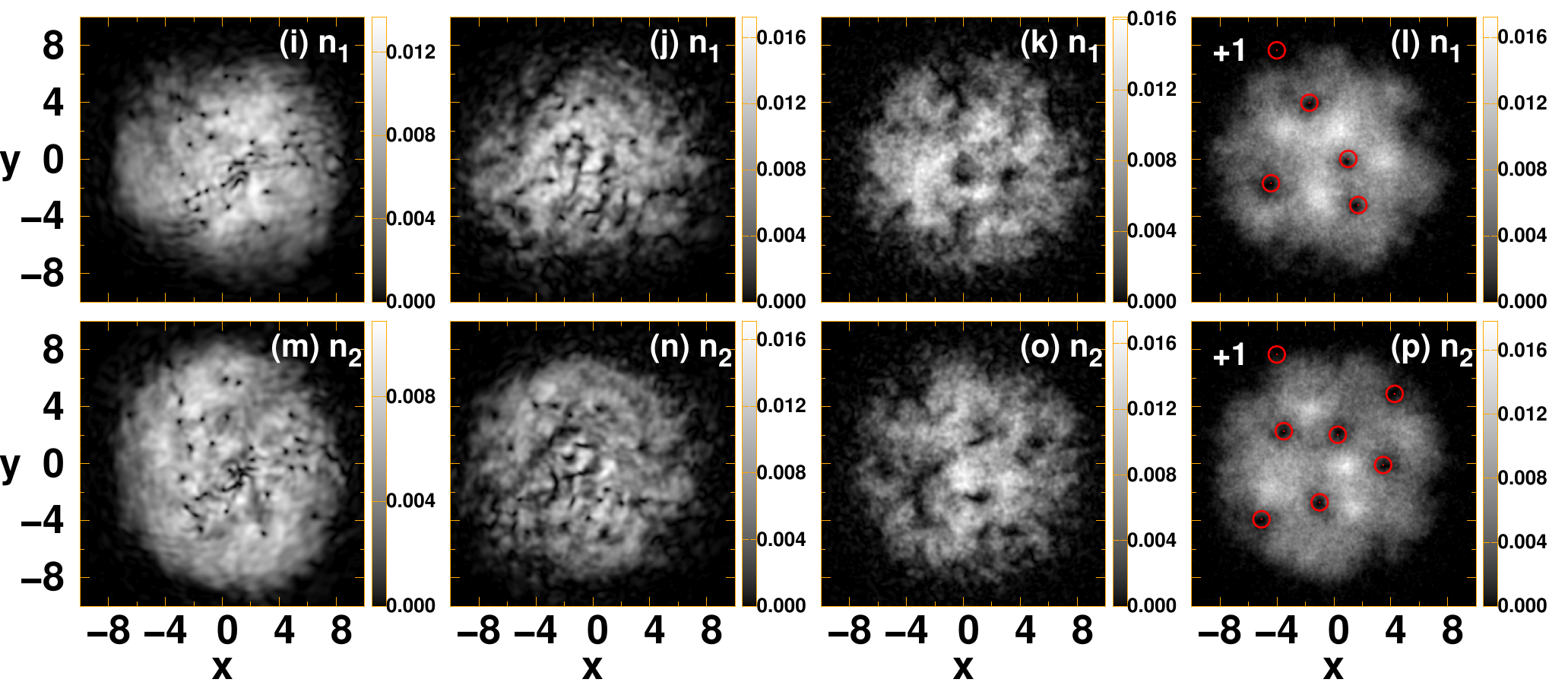} 
 \caption{\label{fig1b}{\footnotesize {} The first and second rows of the panels show the evolution of the density of the first component (1st row) and the second one (2nd row) for the isotropically trapping case of $\epsilon_{y1}=0$  at $t=0$ (a,e), $t=10$ (b,f), $t=50$ (c,g) and $t=400$ (d,h). The third and fourth rows of the panels show the evolution of the density of the first component (top) and the second one (bottom) for $\epsilon_{y1}=0.025$  at $t=0$ (i,m), $t=10$ (j,n), $t=50$ (k,o) and $t=400$ (l,p). The parameters are, $\tilde{g}=2000$, $g_{12}=0.95g$, $M=1024$ and $L=40$.}}.
   \end{figure*}

   As a parametric example for our numerical demonstration, we set $g_{12} = 0.95 g_{11} $ and  $g_{22} = g_{11} \equiv g $; recall that in such systems the ability to
   tune scattering lengths via Feshbach resonances exists and has been used
   to move, e.g., from immiscible to miscible regimes~\cite{Papp_2008}. For this set of $g_{ij}$'s, we get $\tilde{\mu} =35.23$, $\xi \approx 0.119 a_0$, $R_\text{TF} \approx 8.39 a_0$, and $c_s\approx 5.93 a_0 \omega_r$. By stirring the obstacle potential, the vortices and anti-vortices are emitted from it and eventually form a turbulent state. 
For this set of parameters, however, we noticed that the turbulent dynamics and the energy spectra are similar to that of a single-component case \cite{Groszek_2016a, Reeves_2012}. This is due to the fact that the two components behave in the same manner under the symmetric choice of the parameters [see Fig.~\ref{fig1b}(k-r)] and, as a result, the vortices in both  components are always co-located. 
The incompressible kinetic energy spectra exhibit the $k^{-5/3}$ power law in the infrared (IR) region $k\xi < 1$ and $k^{-3}$ power law in the ultraviolet (UV) region $k\xi > 1$; see, e.g. Refs.~\cite{Bradley_2012,Neely_2013,Tsatsos_2016}. 
   %
 \begin{figure}[!htbp]  
 \includegraphics[scale=1.0,width=0.49\textwidth]{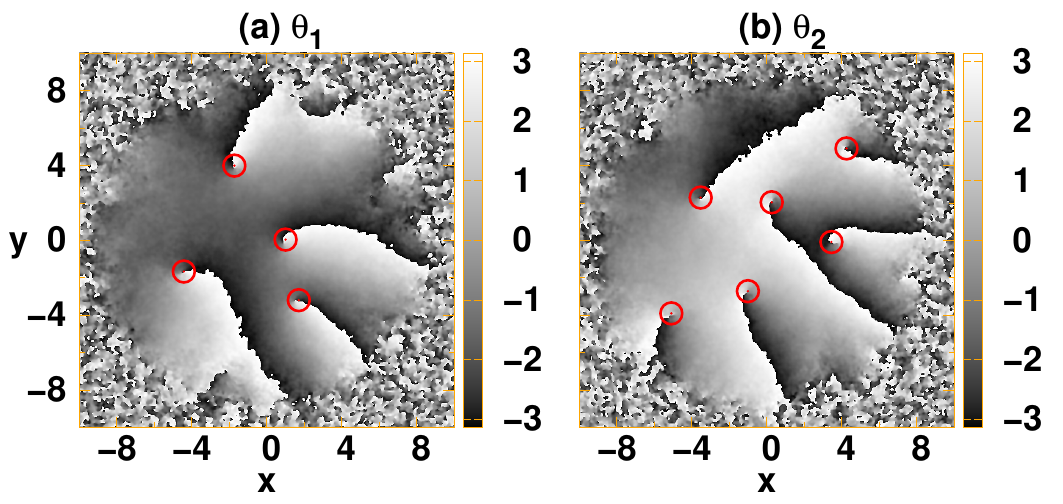}  
 \caption{\label{fig1bb}{\footnotesize {}The phase profile, (a) and (b), corresponds to the (l) and (p) in Fig.~\ref{fig1b}, respectively.}}
   \end{figure}

In reality, there are ingredients that can break the parameter symmetry between the two components. In order to break this symmetry, we introduce a small anisotropy in the trapping potential of Eq.~\eqref{trappot} within the first component as $\epsilon_{y1}=0.025$ while the other $\epsilon_{x(y)j}=0$. 
%
%
An introduction of the parameter anisotropy dramatically changes the dynamics as seen in Fig.~\ref{fig1b}(i)-(p); contrary to the symmetric case (a)-(h). Here the snapshots of the density of the first and second components are shown in the upper (i-l) and the middle (m-p) panels, respectively.  
We see that an initial turbulent structure undergoes gradual change into a so-called interlaced vortex-antidark structures \cite{ionut}, where the density of one component sits in the core of a vortex in the other component \cite{anderson_VB,Mueller_2002,Schweikhard_2004,Kasamatsu_2003,cornell1998having}
within our miscible configuration.
 {In this quasi-equilibrium state of vortex-antidark
   solitary waves, vortices continue to rearrange their positions as
   time evolves. On the other hand, such vortex states are formed by
   minimizing the inter-component interaction energy (shown in
   following paragraph) similar to the formation of a well ordered
   interlaced vortex lattice state \cite{Kasamatsu_2003}. Hence, we
   hereafter refer to this state as an interlaced vortex lattice state
   although the obtained states are not genuinely crystallized. Moreover, the vortices in this state are singly quantized ones} with the counterclockwise winding, as seen in Fig.~\ref{fig1bb}(a) and (b).
The size of the vortex cores in the interlaced lattice state is determined by the spin healing length 
\begin{equation}
 \xi_s =\xi \sqrt{ \frac{g+g_{12}}{g-g_{12}} }
 \label{eq:spin_heal}
\end{equation}
instead of the mass healing length $\xi = a_0/\sqrt{2\tilde{\mu}}$ \cite{Eto_2011}. Thus, the vortices have an extended core due to the spin healing length
when $g_{12}$ is nearly equal to $g$. 

 \begin{figure}[!htbp]  
  \includegraphics[width=0.49\textwidth]{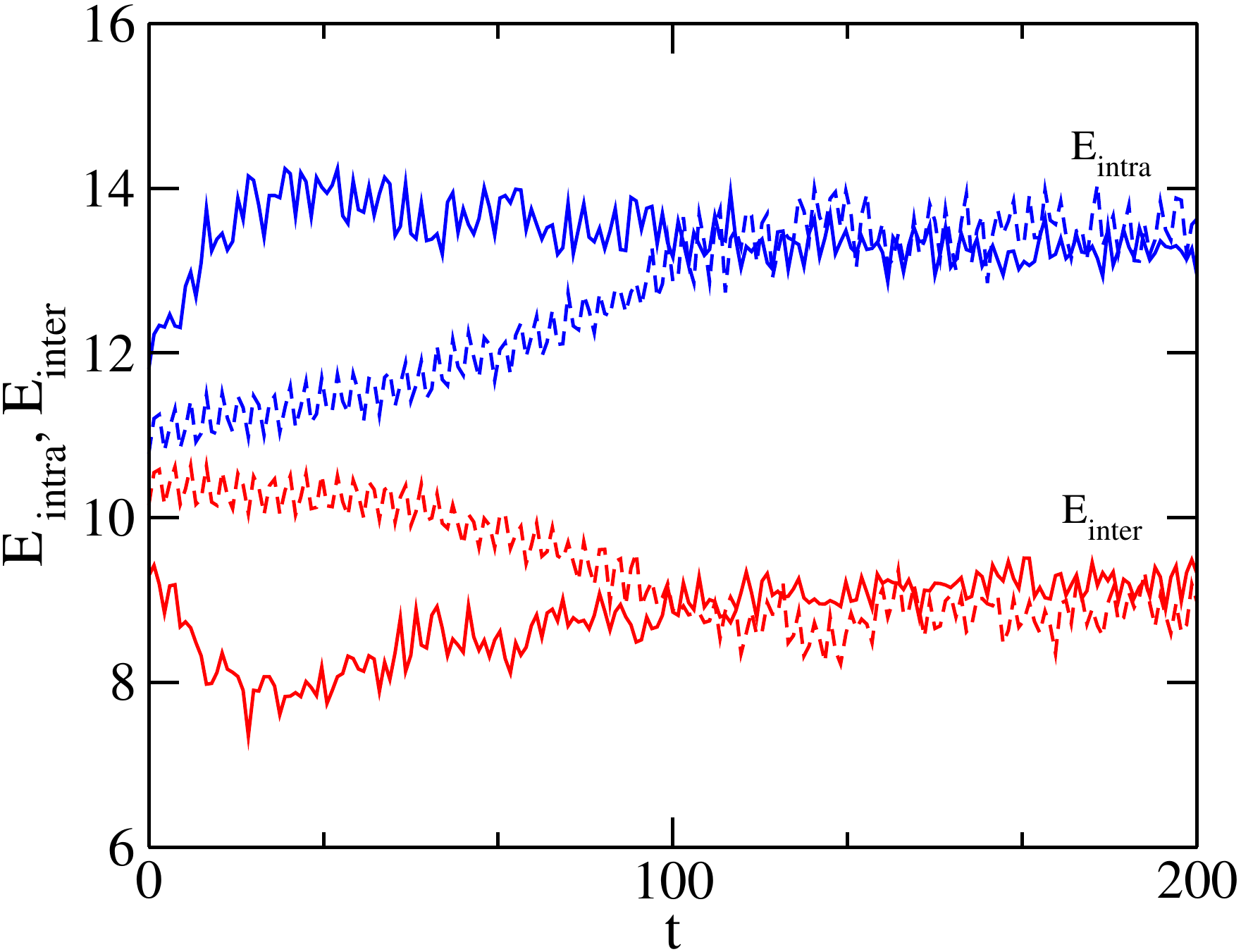}
  \caption{\label{EinterEintra}{\footnotesize  The evolution of the intra- $E_\text{intra}(t)$ and inter-component energies $E_\text{inter}(t)$ 
for $\epsilon_{y1}=0.025$ by the solid curves (corresponding to the Fig.~\ref{fig1b}) and $\epsilon_{y1}=0.005$ by the dashed curves. The other parameters are the same with those in Fig.~\ref{fig1b}.}}
   \end{figure}
We also find that the formation of the interlaced vortex lattice state
is depending on the values of $g_{12}$. To see this, we first calculate the inter and intra-component energy. For an interlaced vortex structure, the inter-component interaction energy is minimized~\cite{Kasamatsu_2003}. Figure~\ref{EinterEintra} shows the evolution of the intra- $E_\text{intra}(t)= \sum_{j=1}^2 (g_{jj}/2) \int d\mathbf{r} |\Psi_j|^4  
$ and inter-component energies $E_\text{inter}(t)=g_{12} \int d\mathbf{r} |\Psi_1|^2 |\Psi_2|^2$. It displays that initially the inter-component energy decreases with time; the intra-component interaction energy concurrently increases. 
This process is associated with the effective phase separation due to the relative displacement of the vortex positions of each component. Subsequently, the inter-component energy increases and saturates close to its value at $t=0$. {We noticed that this energy exchange process that leads to the phase separation is occurring only at higher $g_{12}$ as shown in the Appendix \ref{diff_ene}. It indicates that the  interlaced vortex lattice state
is favorable only at larger values of $g_{12}$ and the increase in $E_\text{intra}$ at the earlier times
  reflects the large local density variation during the phase separation process.}
To address the formation of interlaced vortex lattice state in more detail, we calculate the energy spectra of the compressible and incompressible kinetic energies 
as shown in the next subsection. 
It is noticed that even for smaller values of the anisotropy ($\epsilon_{y1} \sim 0.005$) the results remain similar, yet the time required to form such interlaced lattice varies.
{Indeed, as shown in Fig.~\ref{EinterEintra}, the relaxation time of the energies toward the quasi-equilibrium becomes longer as the trap anisotropy $\epsilon_{y1}$ becomes smaller, and presumably goes to infinity in the limit of $\epsilon_{y1} = 0$.}

 \begin{figure}[!htbp]  
  \includegraphics[width=0.49\textwidth]{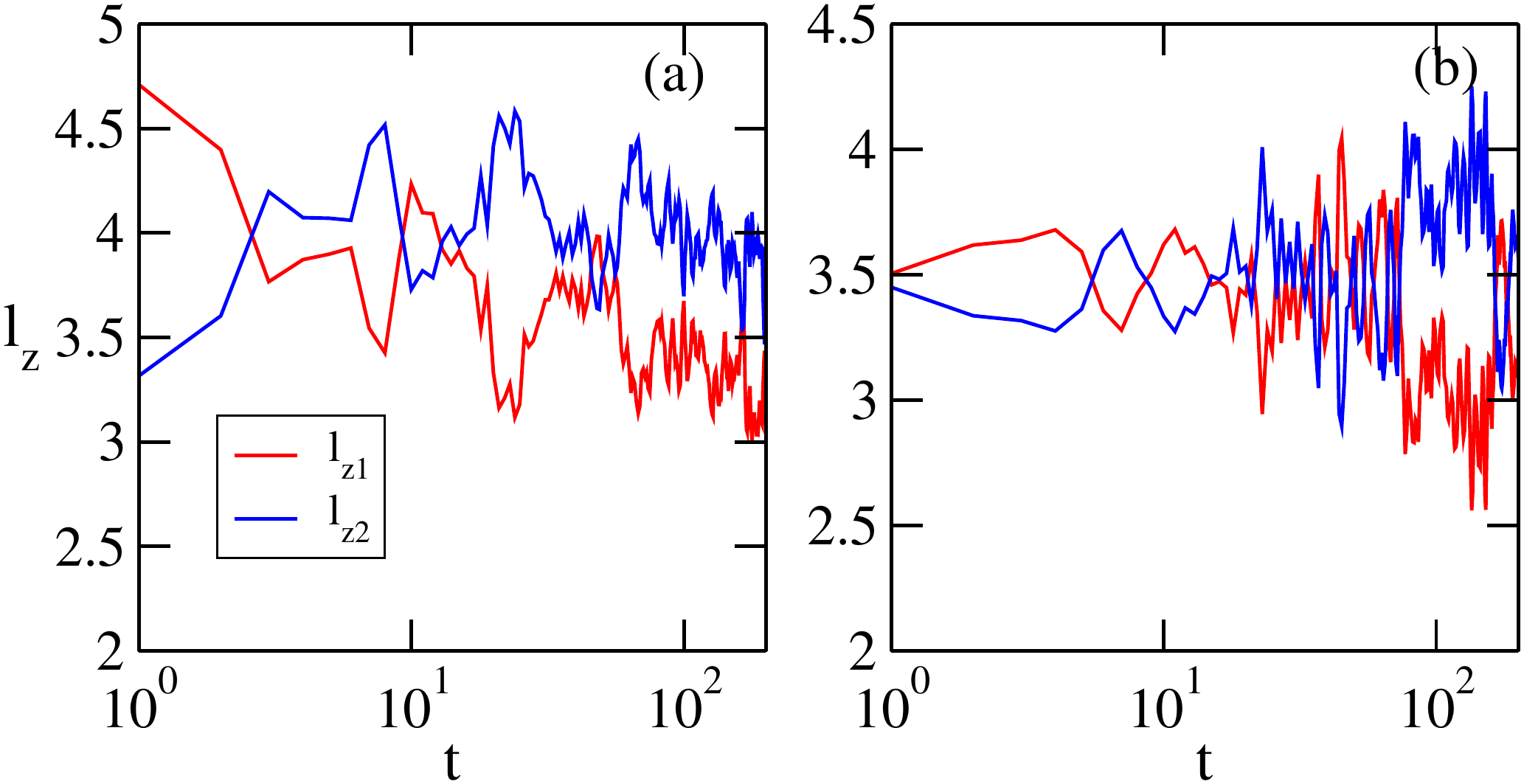}
  \caption{\label{angular}{\footnotesize  The evolution of the angular momentum per particle, (a) in the presence of the trap anisotropy $\epsilon_{y1}=0.025$ (corresponding to  Fig.~\ref{fig1b}), and (b) with a small difference between the intra-component coupling strengths as $g_{11} = 0.975 g_{22}$, but  without the trap anisotropy $\epsilon_{xj}=\epsilon_{yj}=0$. The parameters are, $g_{12}=0.95g_{11}$ and $M=1024$.}}
   \end{figure}
   In order to get a further insight into the turbulent dynamics, we calculate the angular momentum per particle
\begin{eqnarray}
   l_{zi}=-i\hbar\int \int dx dy \Psi_i^{\ast}\big(x \partial_y - y \partial_x \big) \Psi_i
  \label{angl_mom}
\end{eqnarray}
of the $i$-th component. 
Figure~\ref{angular}(a) shows the time evolution of $l_{zi}$ for the components
$i=1,2$ corresponding to Fig.~\ref{fig1b}. The angular momenta of both  components are monotonically increased during the stirring process and the time of the vanishing stirring potential determines their value at $t=0$. 
Although the number of vortices and that of anti-vortices are almost equal at $t=0$, 
the nonuniform distribution of vortices and anti-vortices results in finite positive angular momentum. This is stemming from the counterclockwise rotation of the obstacle during stirring. In this case, the (counterclockwise) vortices are distributed on average in the inner region, where the condensate density $|\Psi_i|^2$ is high, while the (clockwise) anti-vortices are in outer region. In a similar vein, we observed that the vortex distribution obtained from a clockwise  moving obstacle has a finite negative angular momentum. The initial difference in the angular momenta between the components is due to the trap anisotropy which breaks the rotational symmetry. The two components can exchange their angular momenta due to the presence of the inter-component mean-field coupling. At the same time, the magnitude of the total angular momentum decays slowly as time evolves. This is because the time derivative of the angular momentum is non-vanishing when there are asymmetries in the trapping potential or the nonlinear mean-field energy densities \cite{Mukherjee_2020}. 
Both contributions are found to play a role. The nonlinear one reflects the (radial) symmetry breaking induced by the stirring, while the one associated with the confinement reflects the possible deviation from radial symmetry of the trapping potential.
During the dynamical process, from turbulence to the quasi-equilibrium lattice configuration, the system either ejects the anti-vortices through the periphery of the condensate due to the initial stirring, or annihilation of vortex anti-vortex pairs occurs, emitting small-amplitude
(phonon) wave packets within the condensates. The energy dissipation via the vortex-phonon interaction takes place and it eventually leads to the quasi-equilibrium configuration of vortices, although the total energy is conserved during the time development of the GPE. Similar relaxation dynamics can be seen in the single-component BEC in a rotating potential \cite{Lobo_2004,Parker_2005}. 
When the stirring period is increased such as 3 or 4 periods, the initial net angular momentum at $t=0$ is also increased, 
so that the quasi-equilibrium configuration possesses more vortices than those in Fig.~\ref{fig1b}.

As another example of parameter asymmetry among the components, we study the turbulent dynamics and the angular momentum evolution for the asymmetric intra-component coupling strength $g_{11}=0.975 g_{22}$ \cite{Schweikhard_2004,Matthews_1999} by setting $\epsilon_{xj} =\epsilon_{yj}= 0$;
see again Eqs.~(10)-(11) in~\cite{Mukherjee_2020}. 
The simulation result shows that the dynamics is similar to Fig.~\ref{fig1b}, where the initial turbulent state undergoes a dynamical transition into the interlaced vortex lattice configuration (see Appendix \ref{gq_ge_g2}).  
The evolution of the angular momenta in Fig.~\ref{angular}(b) shows that the exchange process of the angular momentum eventually causes an imbalance of the angular momentum in the quasi-equilibrium state, where the number of the remaining vortices in the second component is more than that of the first component. This is in
line with the dynamical robustness of the vortices in the second component when
it bears a larger intra-component strength $g_{22} > g_{11}$~\cite{Garcia_2000,Garcia_2000:a}.

Interestingly, such a dynamically turbulent stage and the subsequent formation of large core vortices have been observed in the JILA experiment of a two-component condensate \cite{Schweikhard_2004}, where the asymmetry among the components exists due to the population difference and the different intra-component strengths. In the experiment, a fraction of the first component with a vortex lattice, which was initially prepared, was coherently transferred to the second component. Then, an interlaced vortex lattice emerged dynamically through a transient turbulent state. The transition time from the turbulence to the interlaced lattice was about a few seconds, which is in reasonable agreement with our numerical results, where the lattice structure appears after $t \sim 100$, i.e., $t \sim 1$ sec in the physical units. 

Finite size effects are also crucial for the   interlaced vortex lattice formation.
To address the finite size effect, we perform a numerical experiment in a homogeneous system without a trap by considering a periodic boundary condition, and
by keeping the parameters $g_{11} = 0.975 g_{22}$ and $g_{12} = 0.95 g_{11}$. 
Due to the periodic boundary condition, the only mechanism of energy dissipation is vortex anti-vortex annihilation and, 
as a result, equal numbers of vortices and anti-vortices are expected to be
maintained during the time evolution. 
The result indicates that the vortices almost completely disappear through the pair annihilation in the final quasi-steady state (see Appendix \ref{ang_homo}). 
Thus, the external trap plays an important role in the formation of the interlaced vortex lattice structure. 

 %
\subsection{Kinetic Energy Spectra}
\begin{figure}[!htbp]  
 \includegraphics[scale=1.0,width=0.45\textwidth]{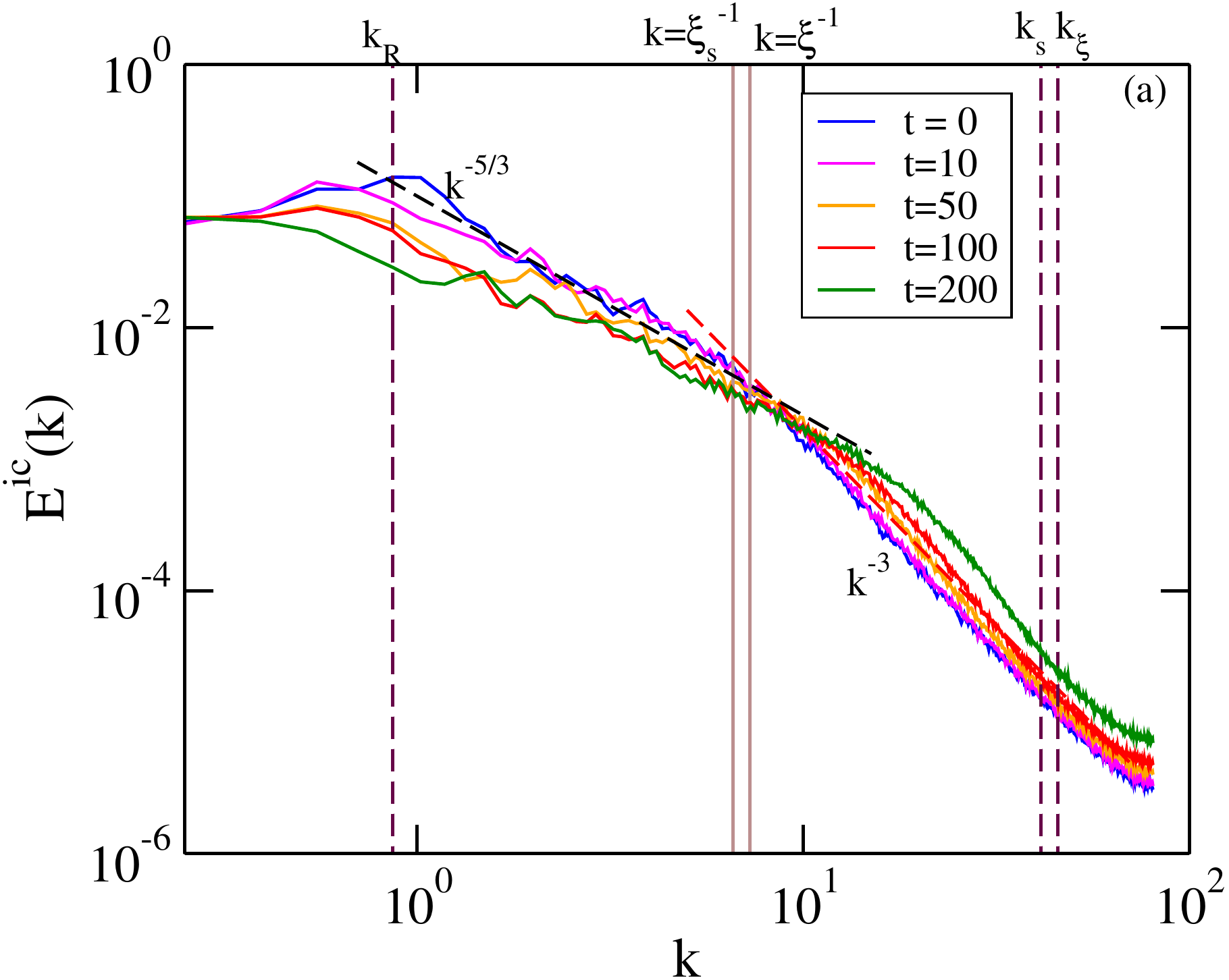}
  \includegraphics[scale=1.0,width=0.45\textwidth]{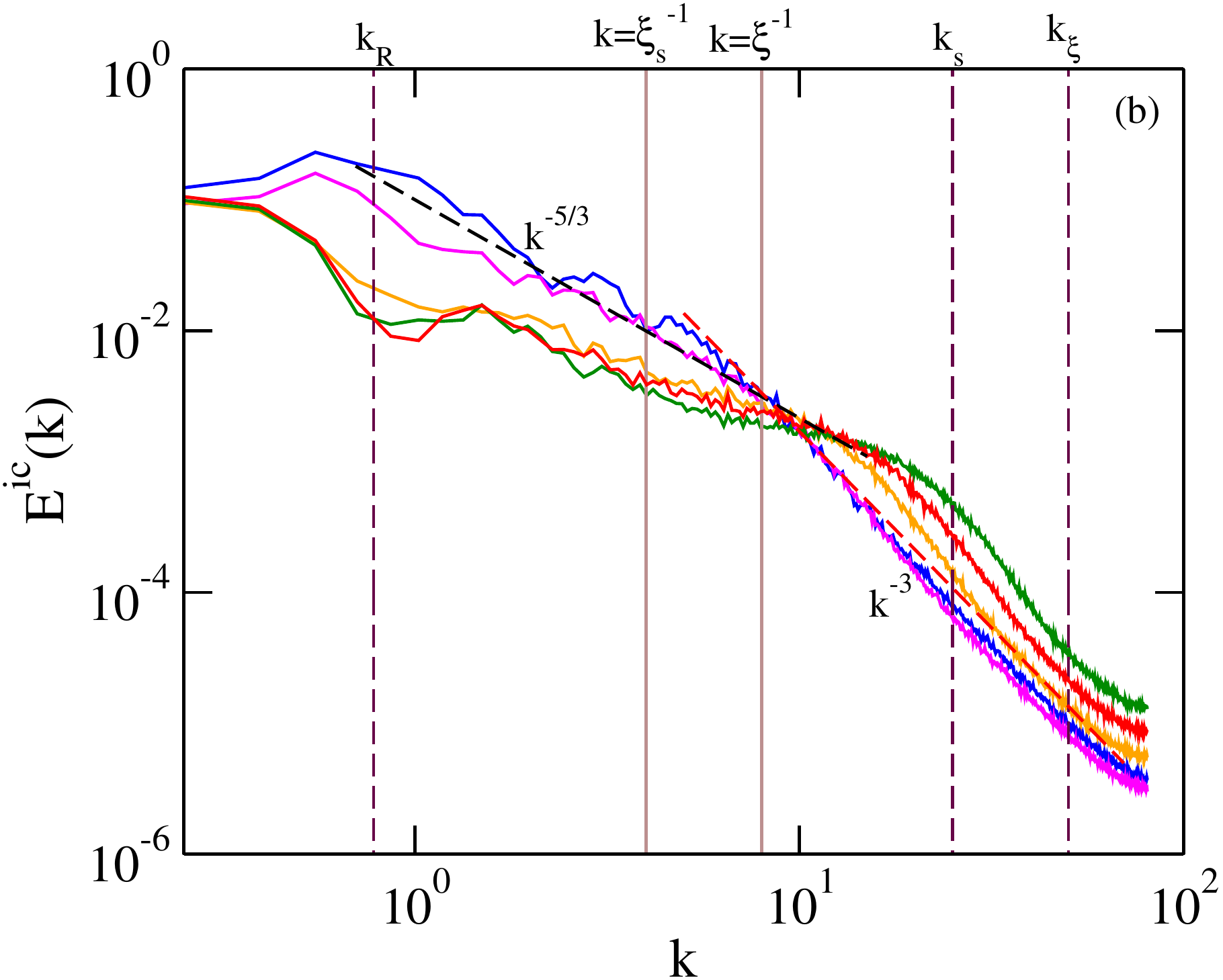}
   \includegraphics[scale=1.0,width=0.45\textwidth]{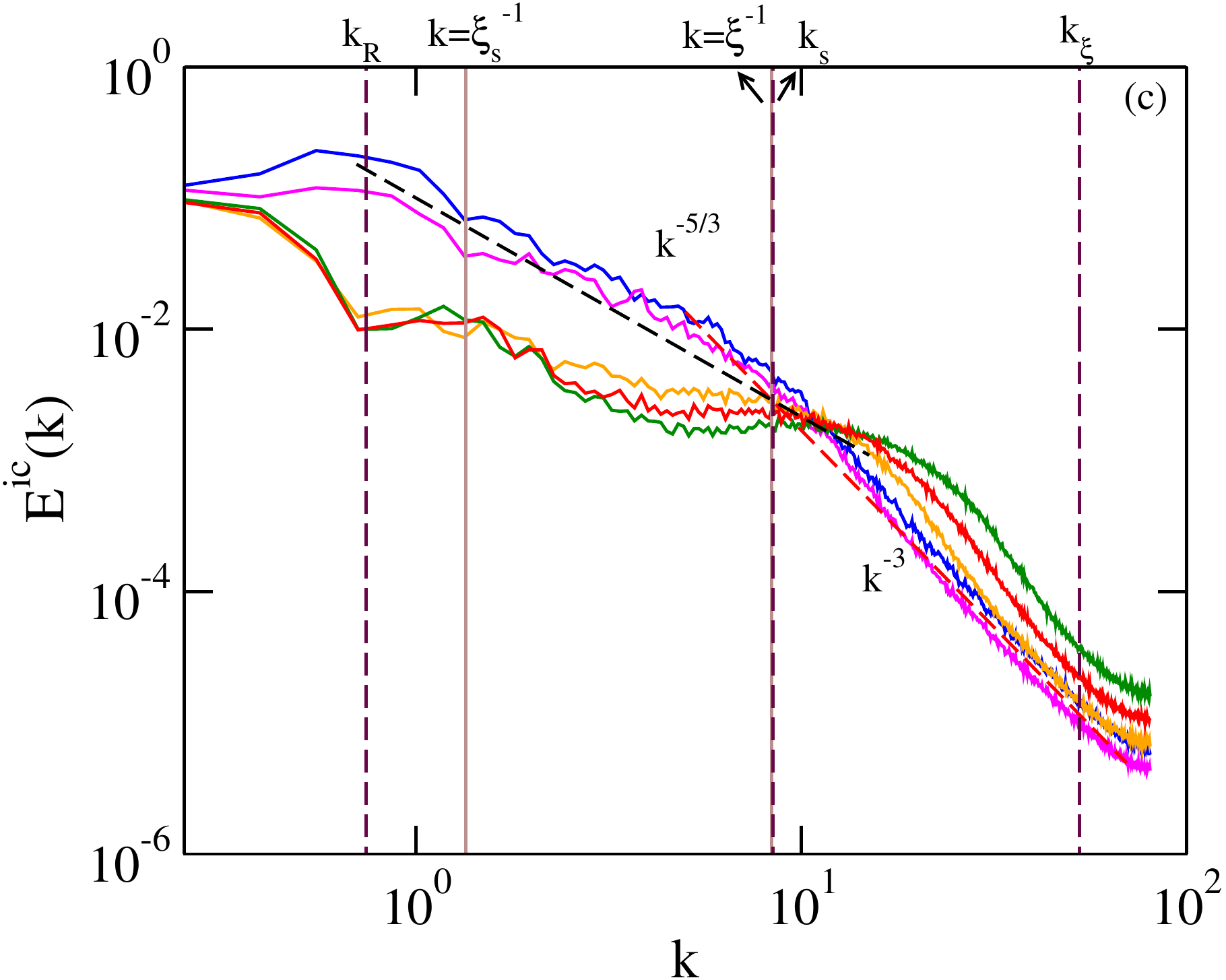}
  \caption{\label{fig1ba}{\footnotesize  
      The incompressible kinetic energy spectrum of the first component in a harmonic trap ($\alpha=2$) with a small anisotropy $\epsilon_{y1}=0.025$ at several different times for (a) $g_{12} = 0.1 g_{11}$, (b) $g_{12} = 0.6 g_{11}$, and (c) $g_{12} = 0.95 g_{11}$. The
    spectrum   of the second component shows a similar trend. The black and red dashed lines serve as guide to the eye  for the $k^{-5/3}$ and $k^{-3}$ power laws, respectively. The vertical maroon dashed lines (from left to right) represent $k_R$, $k_s (= 2 \pi/\xi_s)$ and $k_{\xi}$; the vertical brown solid lines (from left to right) represent $k =\xi_s^{-1}$ and $k = \xi^{-1}$. Here an average over 4 different initial conditions is considered.}}.
   \end{figure}

In order to study the characteristics of the emergent
quantum turbulence (as a result of our preparation procedure), 
we calculate the incompressible and compressible kinetic energy spectrum, $E^\text{ic}(k)$ and $E^\text{c}(k)$  \cite{Numasato_2010,tsubota2013quantized,Horng_2009}, for the case of $\epsilon_{y1}=0.025$, $g_{11} = g_{22}$, and the various values of $g_{12}$ (see Appendix \ref{spectra_calc}).
The incompressible fluid part of the condensate represents the divergence free component of the condensate velocity. The spectral behavior in the UV
(large $k$) region represents the contribution from the vortex core, while that in the IR (small $k$) region indicates the  largest scales involved
(of the order of the condensate size). Figure~\ref{fig1ba}(a) shows $E^\text{ic}(k)$ of the $\Psi_1$-component at several different times for a weak inter-component coupling $g_{12} = 0.1 g_{11}$. The spectrum at each time exhibits a behavior similar to a 2D single-component BEC \cite{Simula_2014}. 
In the UV regime at $k > \xi^{-1}$ determined by the mass healing length, the spectrum exhibits the power-law $k^{-3}$ and this scaling continues up to $k_{\xi}\sim 2\pi/\xi$, which is determined by the core profile of a single vortex \cite{Bradley_2012}. 
In the regime of $k_R < k  < \xi^{-1}$, the spectrum clearly exhibits the Kolmogorov power-law $\sim k^{-5/3}$, a characteristic of the inverse energy cascade, where $k_R= 2 \pi/R_\text{TF}$.  In this regime, a vortex--anti-vortex annihilation process strongly affects the spectral behavior due to the sound wave emission \cite{Parker_2005,Billam_2015,Simula_2014}. 

\begin{figure}[!htbp]  
  \includegraphics[width=0.49\textwidth]{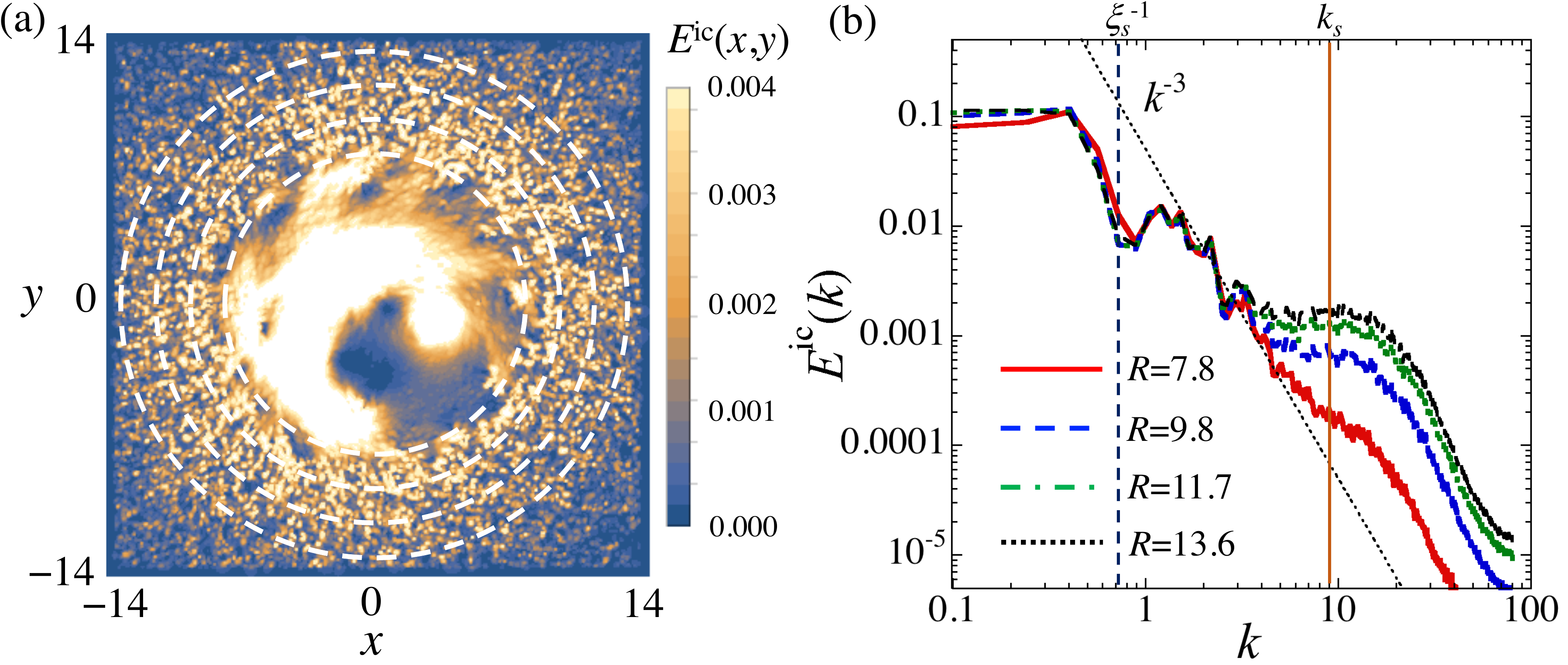}
  \caption{\label{energyspecdist}{\footnotesize (a) The 2D spatial distribution of the density of the incompressible kinetic energy $E^\text{ic}(\bm{r}) = n(\bm{r}) |\bm{u}^\text{ic} (\bm{r})|^2 / 2$ at $t=200$ for $g_{12} = 0.95 g_{11}$ and $\epsilon_y = 0.025$, corresponding to the result in Fig.~\ref{fig1ba}(c). 
Circles represents the region to calculate the incompressible energy spectrum $E^\text{ic}(k)$ in (b), where the radius is $R=$7.8, 9.8, 11.7, 13.6 from the inner to outer circle. 
The panel (b) shows $E^\text{ic}(k)$ calculated from $E^\text{ic}(k)$ within the region of the different $R$. 
As a guide, $k^{-3}$, $k=\xi_s^{-1}$ and $k=k_s$ lines are drawn. 
  }}
   \end{figure}

{ It can be seen that in Fig.~\ref{fig1ba}(a) the magnitude of $E^\text{ic}(k)$ is slightly decreased for $k_R < k < \xi^{-1}$ while it is increased for $\xi^{-1} < k < k_{\xi}$. 
This feature is more visible at higher $g_{12}$ as shown in Figs.~\ref{fig1ba}(b) and (c), where the spectrum exhibits a plateau around $k \sim \xi^{-1}$. 
To see what happens, we plot in Fig.~\ref{energyspecdist}(a) that the density of the incompressible kinetic energy in the real space $E^\text{ic}(\bm{r}) = n(\bm{r}) |\bm{u}^\text{ic} (\bm{r})|^2 / 2$, 
corresponding to the energy spectrum at $t=200$ of Fig.~\ref{fig1ba}(c) ($g_{12} = 0.95g_{11}$). 
The energy density exhibits clear spatial separation of the large-scale structure at the central region and the small-scale one in the periphery. 
This small-scale structure is the origin of the plateau of $E^\text{ic}(k)$ for the high-wave number. 
The plateau can be also seen in the turbulence in 3D condensates \cite{sasa2011energy}, known as the bottleneck effect
{ and these small scale fluctuations can be suppressed by using phenomenological dissipation \cite{Reeves_2012,Kobayashi_2005}}.
When we truncate $E^\text{ic}(\bm{r})$ outside of a certain radius and calculate the energy spectrum from it, the magnitude of $E^\text{ic}(k)$ at the high wave number is suppressed, as shown in Fig.~\ref{energyspecdist}(b). 
If we wipe out the small-scale fluctuation in the periphery, the
spectrum shows the $k^{-3}$ power law for
$\xi_s^{-1} < k < k_s$, associated with the extended vortex core in the
quasi-equilibrium state, the core size being determined by the
spin-healing length of Eq.~\eqref{eq:spin_heal}. 
The absence of the $k^{-5/3}$ power law at the later evolution stage is consistent with our observation that the turbulence decays into the quasi-equilibrium state.    }
  
 \begin{figure}[!htbp]  
 \includegraphics[scale=1.0,width=0.45\textwidth]{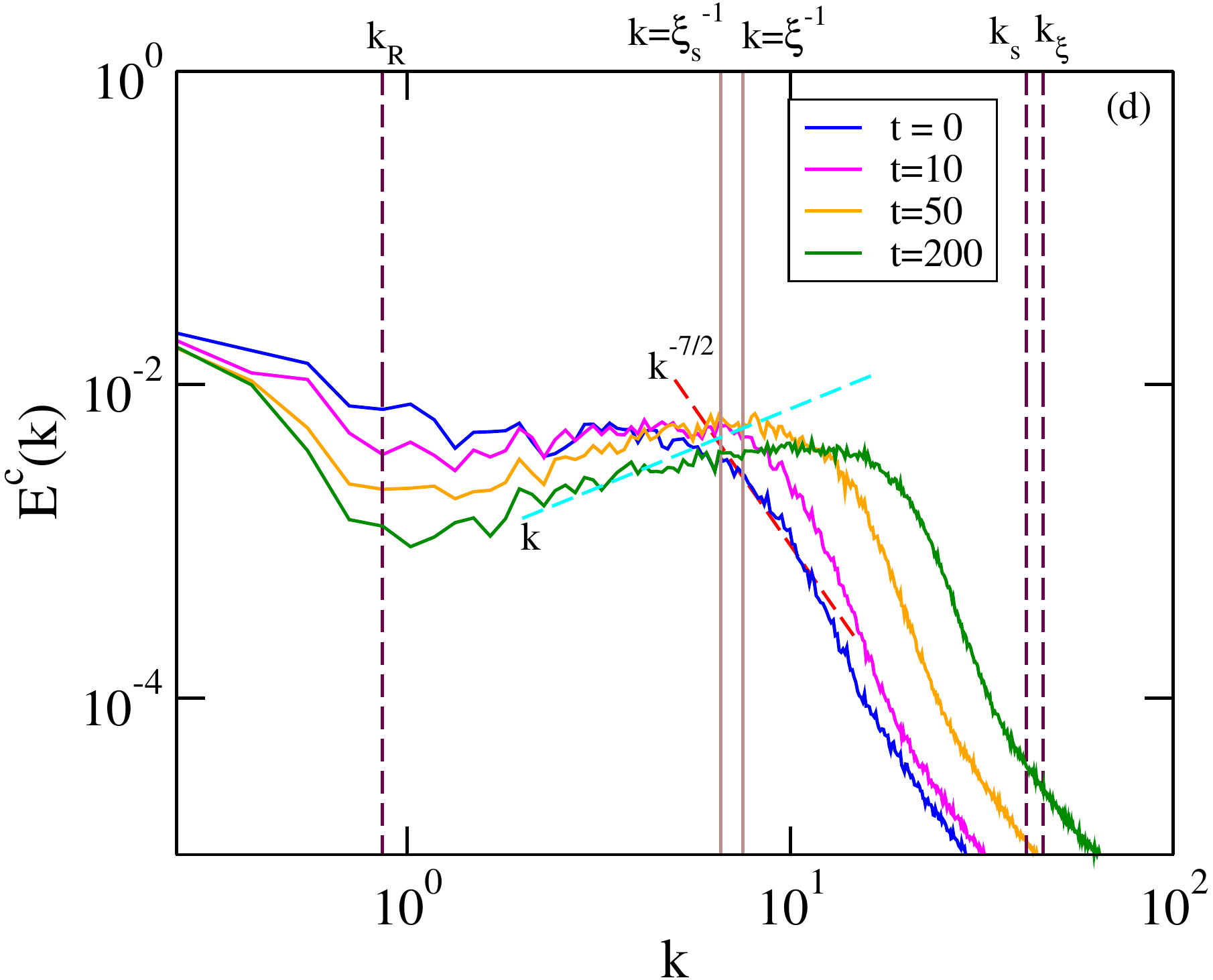}
 \includegraphics[scale=1.0,width=0.45\textwidth]{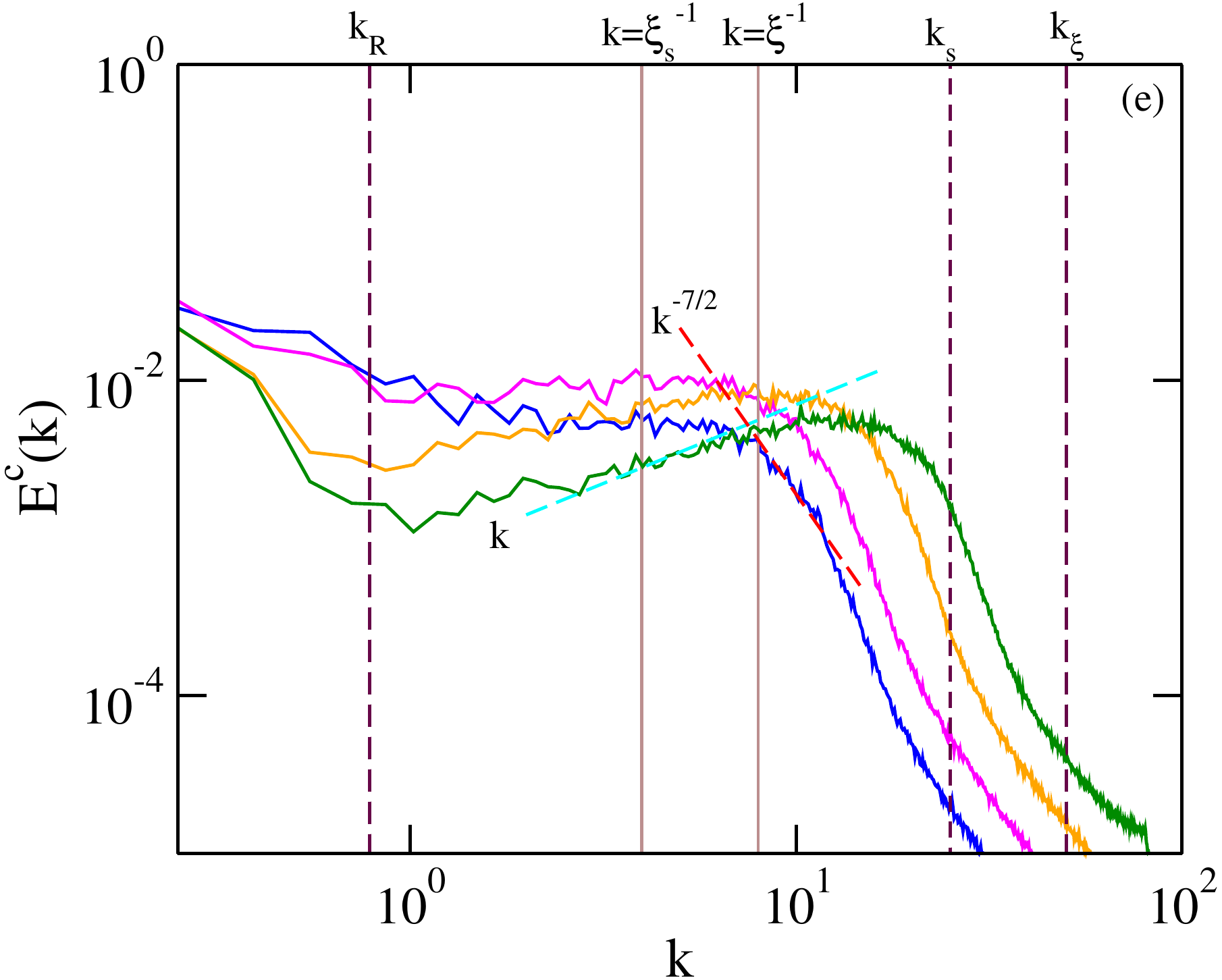}
 \includegraphics[scale=1.0,width=0.45\textwidth]{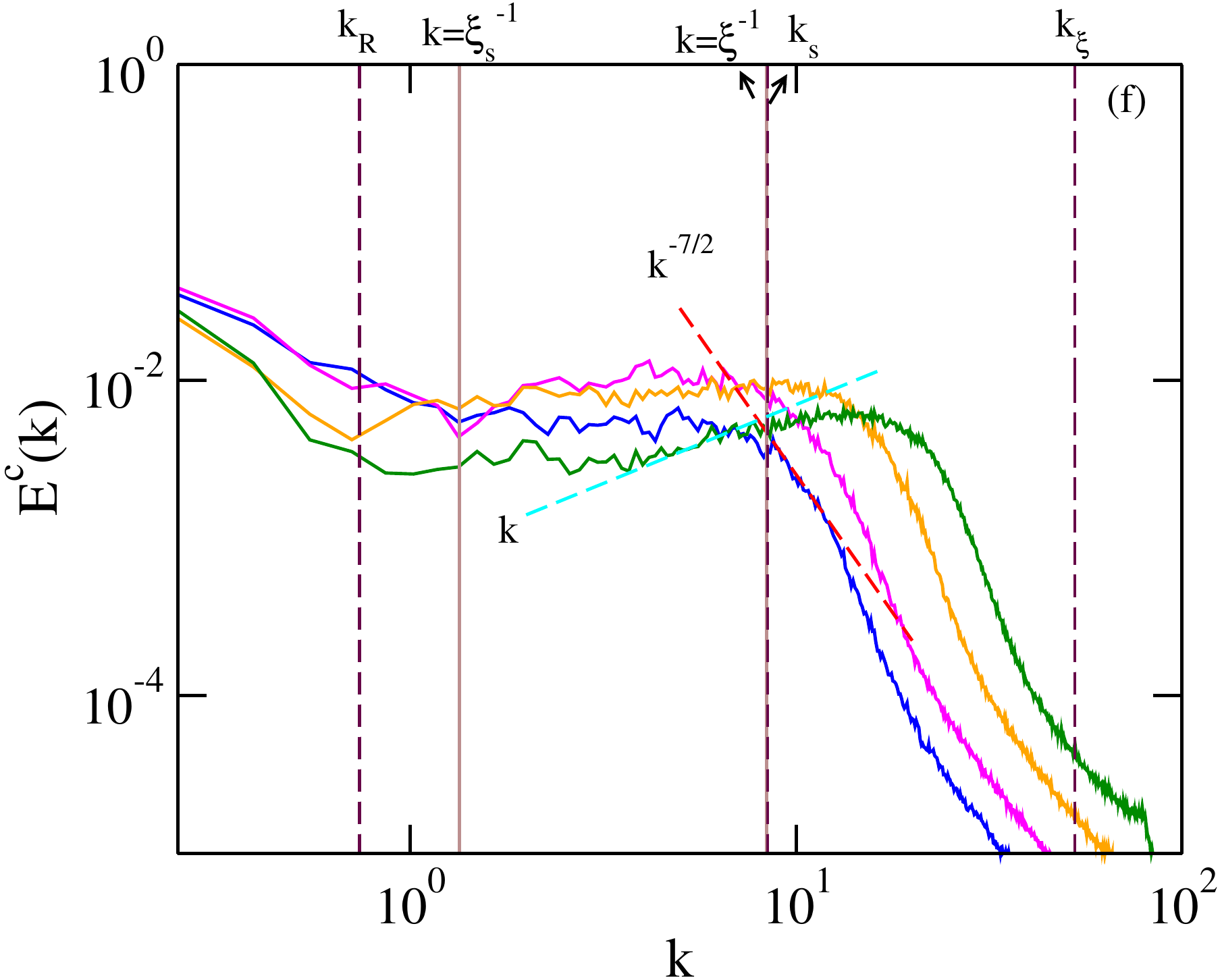}
  \caption{\label{fig6bb}{\footnotesize  
      The compressible kinetic energy spectrum of the first component in a harmonic trap ($\alpha=2$) with a small anisotropy $\epsilon_{y1}=0.025$ at several different times for (a) $g_{12} = 0.1 g_{11}$, (b) $g_{12} = 0.6 g_{11}$, and (c) $g_{12} = 0.95 g_{11}$. The
      spectrum of the second component shows a similar trend. The red and cyan dashed lines serve to guide the eye for the $k^{-7/2}$ and $k$ power laws, respectively. The vertical maroon dashed lines (from left to right) represent $k_R$, $k_s$ and $k_{\xi}$; (from left to right)  the dash-dotted lines represent $k =\xi_s^{-1}$ and $k = \xi^{-1}$. Here averages over 4 different initial conditions are considered.}}.
   \end{figure}
{ Next, we turn to the compressible energy spectrum, for which typical results for the same parameters with Fig.~\ref{fig1ba} are shown in Fig.~\ref{fig6bb}. 
Here, the early-time stage of the spectrum exhibits the  $k^{-7/2}$-power law for the UV region, which is consistent with the turbulence in a single-component BEC for {a clustering regime} \cite{Reeves_2012}. As time evolves the spectrum develops a $k$-power law in the IR region corresponds to the equilibration of the sound waves \cite{Numasato_2010}. Though, the $k^{-3/2}$ power-law is reported for a single-component case in the IR region \cite{Reeves_2012,Numasato_2010} for a small region around $k=\xi^{-1}$, the spectra
of the two-component system does not show a clear evidence for that, especially at higher inter-component strengths $\sqrt{g_{12}g_{21}}$. But, it is to be noted that the $k^{-3/2}$ reported in the Ref. \cite{Reeves_2012} for a single component case in the clustered regime corresponds to the limit $\sqrt{g_{12}g_{21}} \rightarrow 0$. We see the development of such a power law for the wave numbers around $k=\xi^{-1}$ in this limit, but not highlighted in Fig.~\ref{fig6bb} as it is not prominent. }


 \begin{figure}[!htbp]  
 \includegraphics[scale=1.0,width=0.48\textwidth]{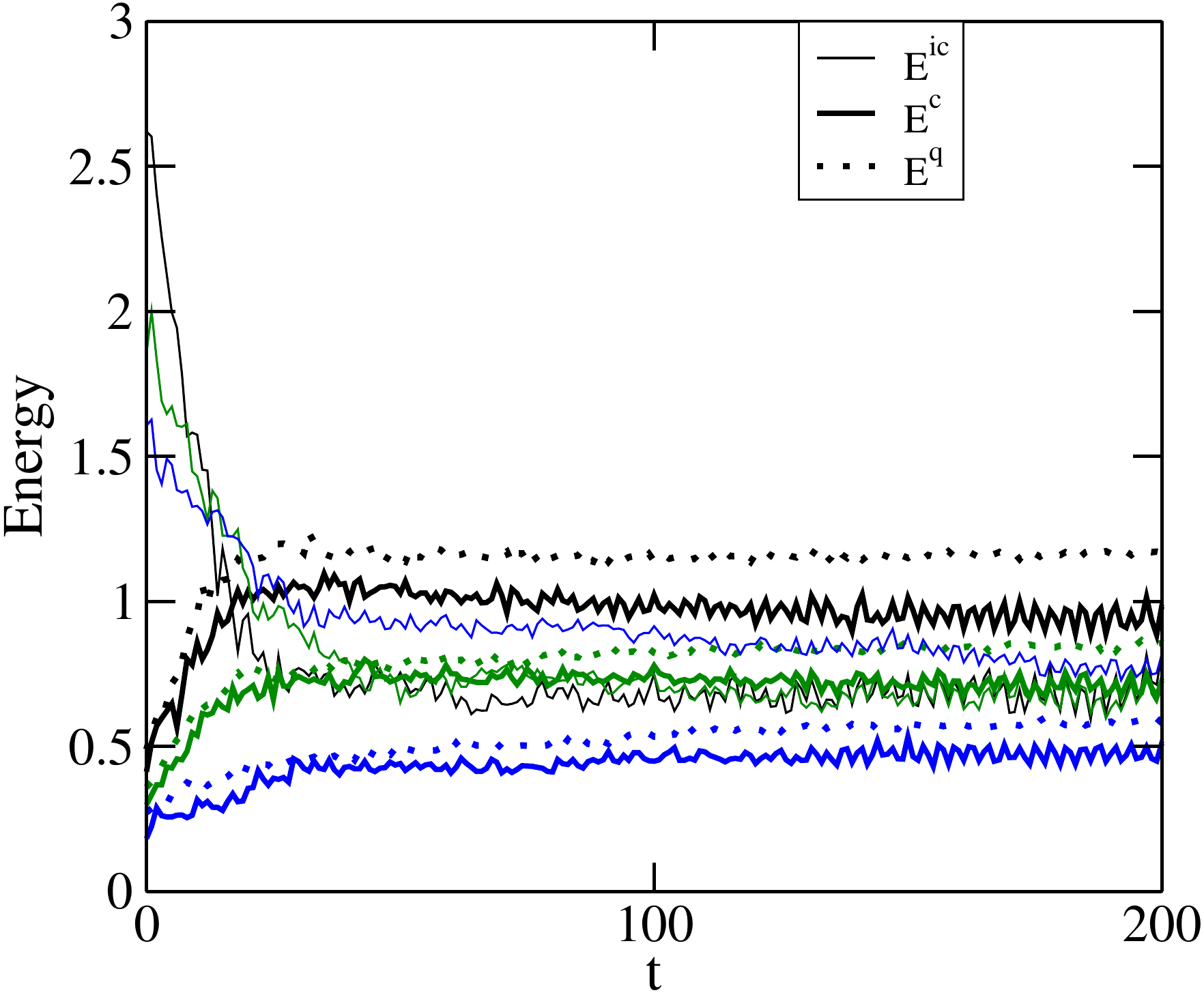} 
  \caption{\label{fig:energy_vort_g12}{\footnotesize  
 The evolution of the compressible $E^\text{c}$ (solid thin lines), the incompressible $E^\text{ic}$ (solid thick lines), and the quantum pressure $E^{q}$ (dotted lines) energies. Here, the black, green, and blue curves indicate the results for $g_{12}=0.95g$, $g_{12}=0.6g$, and $g_{12}=0.1g$, respectively. 
 }}.
  \end{figure}

  {Finally, we show in Fig.~\ref{fig:energy_vort_g12} the development of the {compressible, incompressible and quantum pressure energies, $E^\text{c}$, $E^\text{ic}$ and $E^q$} respectively, with respect to time. Just after $t=0$, the incompressible energy is decaying, while the energy of the quantum pressure, $E_q$ (see Appendix \ref{spectra_calc} for the relevant definition) and compressible energy are increased. This fast process of the energy exchange at the initial stage indicates the higher rate of vortex-anti-vortex annihilation process. At later times, both the $E^\text{ic}$ and $E^\text{c}$ nearly saturate with their behavior being essentially independent of $g_{12}$.
  {Further, the steep increase in $E^q$ at the initial times for higher $g_{12}$ is consistent with the increase in $E_\text{intra}$ discussed in the previous section due to the large density variation.} Since the compressible energy dominates the kinetic energy of the system for higher $g_{12}$, the incompressible energy, responsible for the vortex motion, can be relaxed  by the bigger bath of the sound waves.} 

\section{Vortex Cluster Formation}\label{SteepWall}
It is well-known that the systems having a bounded energy spectrum with more than one conserved quantity exhibit a negative temperature regime \cite{Mosk:2005}.
The existence of the negative temperature restricts the thermalization of an isolated system. A well-known example for this case is a bounded 2D fluid with a large number of point vortices as indicated by Onsager \cite{onsager1949statistical}. In the negative temperature regime, the like-signed vortices condense to form a giant vortex cluster.  One of the main contributions in the further development of Onsager's theory on the existence of the negative absolute temperatures and the associated vortex cluster formation is from Kraichnan \cite{kraichnan1967inertial,kraichnan1975statistical}, who conjectured that clusters of like-signed vortices originate from the incompressible kinetic energy cascade of a 2D system. {Hints of} signatures of such clustered states of like-signed vortices were reported in Ref. \cite{Neely_2013}, conducted in a 2D trapped dilute atomic gases.
Although many theoretical investigations had connected this cluster formation with the negative temperature, experimental evidence showcasing the connection between the negative temperature and the vortex cluster was absent until the recent discovery of such states in the two remarkable experiments reported in \cite{Gauthier1264,Johnstone1267}.

{ It has been shown that the formation of clustered vortices occurs depending on the initial vortex configuration \cite{Billam_2014} or via an evaporative heating mechanism that removes the low-energy vortex dipoles from the condensates through  vortex pair annihilation \cite{Simula_2014,Groszek_2016a}. }
Here, we study the cluster formation of the two-component BECs, especially the impact of the inter-component coupling $g_{12}$. 
One of the main factors that affect the cluster formation is the vortex-sound coupling. An efficient way to reduce such coupling is to consider a non-circular geometric trap with a non-circular obstacle \cite{White_2014,Stagg_2015,Gauthier1264,Pakter_2018}. Since it is found that a harmonic trap suppresses the cluster formation \cite{Groszek_2016a}, we consider an elliptical steep-wall trap with $\epsilon_{xj}=0.3$, $\epsilon_{yj}=-0.3$, and $\alpha=50$.
The vortex nucleation is caused by the non-circular shaped Gaussian obstacle, which has the form 
\begin{equation}
V_s(x,y,t)=V_0 \exp \left[ -\frac{ d_s^2(x-x_0(t))^2+y^2}{\sigma^2}\right],
 \label{eq:stir_pot:m2}
\end{equation}
with $d_s=3$ and $x_0(t) = 0.6R_\text{TF} \sin(2\pi t/T)$, where $R_\text{TF} \approx 8.34a_0$. We sweep the condensate with an obstacle of strength $V_0=15\mu$ for a half of the period $T$ with velocity $v=0.4v_s$. 
Here we ramp down the obstacle to zero during the range from $t=T/4$ to $t=T/2$.

Of the numerous measures of this clustered states listed in the references \cite{Yatsuyanagi_2005,White_2012,Neely_2013,Reeves_2013,Billam_2014,Simula_2014,Groszek_2016a,Skaugen_2016}, we use the vortex dipole moment to detect such states \cite{Simula_2014}. The dipole moment is defined as
\begin{eqnarray}
  d=|\bm{d}|=\sum_i q_i r_i,
  \label{dipole}
\end{eqnarray}
  where $q_i=\pm h/m$ and $r_i$ is the position of the vortex and detected by measuring the Jacobian field \cite{Mazenko_2001,Angheluta_2012,Skaugen_2016}. {Here, the vortex positions of the wave function $\Psi$ are mapping to density of vortices $\rho_v(\bm{r},t)$ as 
\ba\label{eq:vort_det}
\rho_v(\bm{r},t) =\delta(\Psi)\mathcal{D}(\bm{r},t), 
\ea
where the Jacobian determinant $\mathcal{D}$ is
\ba\label{eq:vort_det2}
\mathcal{D}(\bm{r},t)=
\begin{vmatrix}
\partial_x \text{Re} \Psi &  \partial_y \text{Re}\Psi  \\ 
\partial_x \text{Im}\Psi  & \partial_y \text{Im}\Psi 
\end{vmatrix} 
= \text{Im}(\partial_x  \Psi^{\ast}  \partial_y  \Psi).
\ea
The position of vortices can be determined from nonzero values of the Jacobian field, while  the rotational direction can be determined from its sign.}
Here, $+q_i$ indicates the charge of a vortex and $-q_i$ represents the charge of an anti-vortex. 
%
 \begin{figure*}[!htbp]  
 \includegraphics[scale=1.0,width=0.95\textwidth]{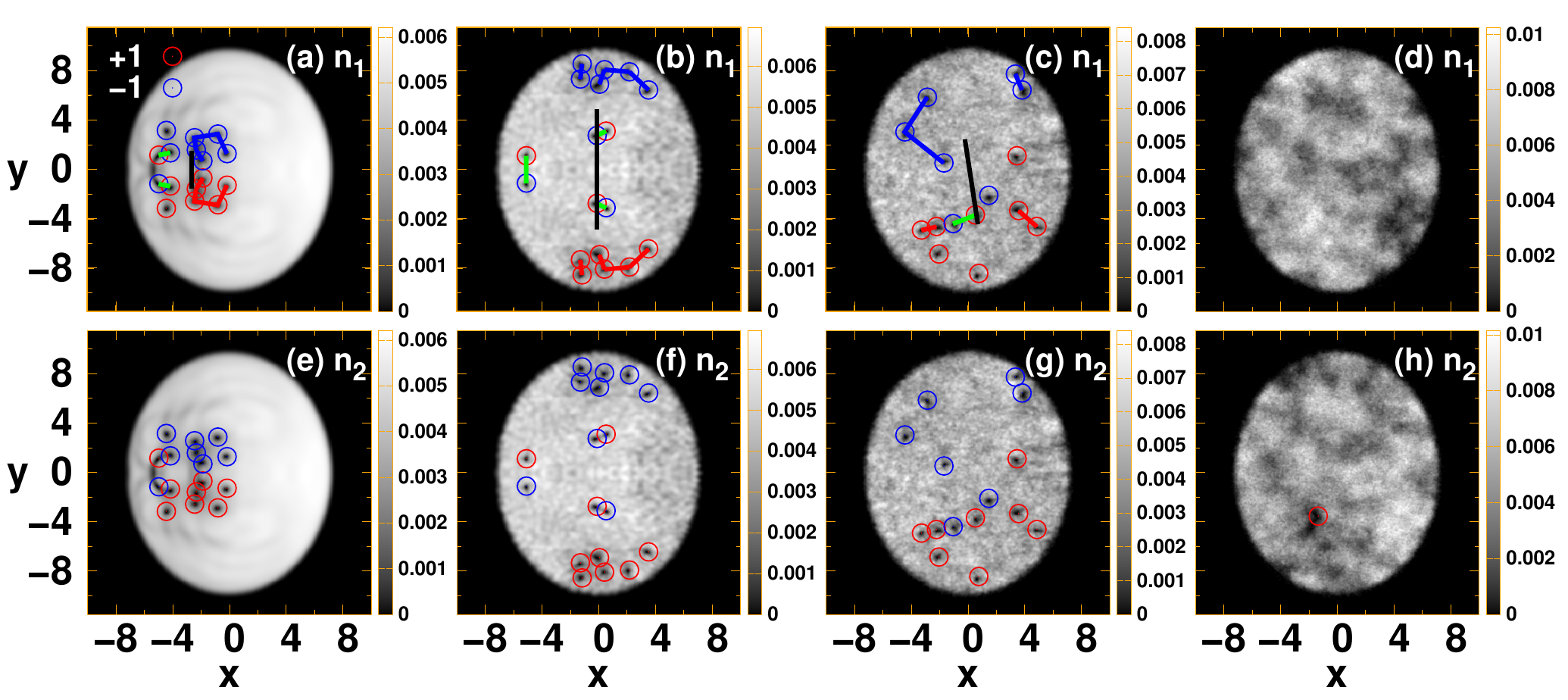}
  \includegraphics[scale=1.0,width=0.95\textwidth]{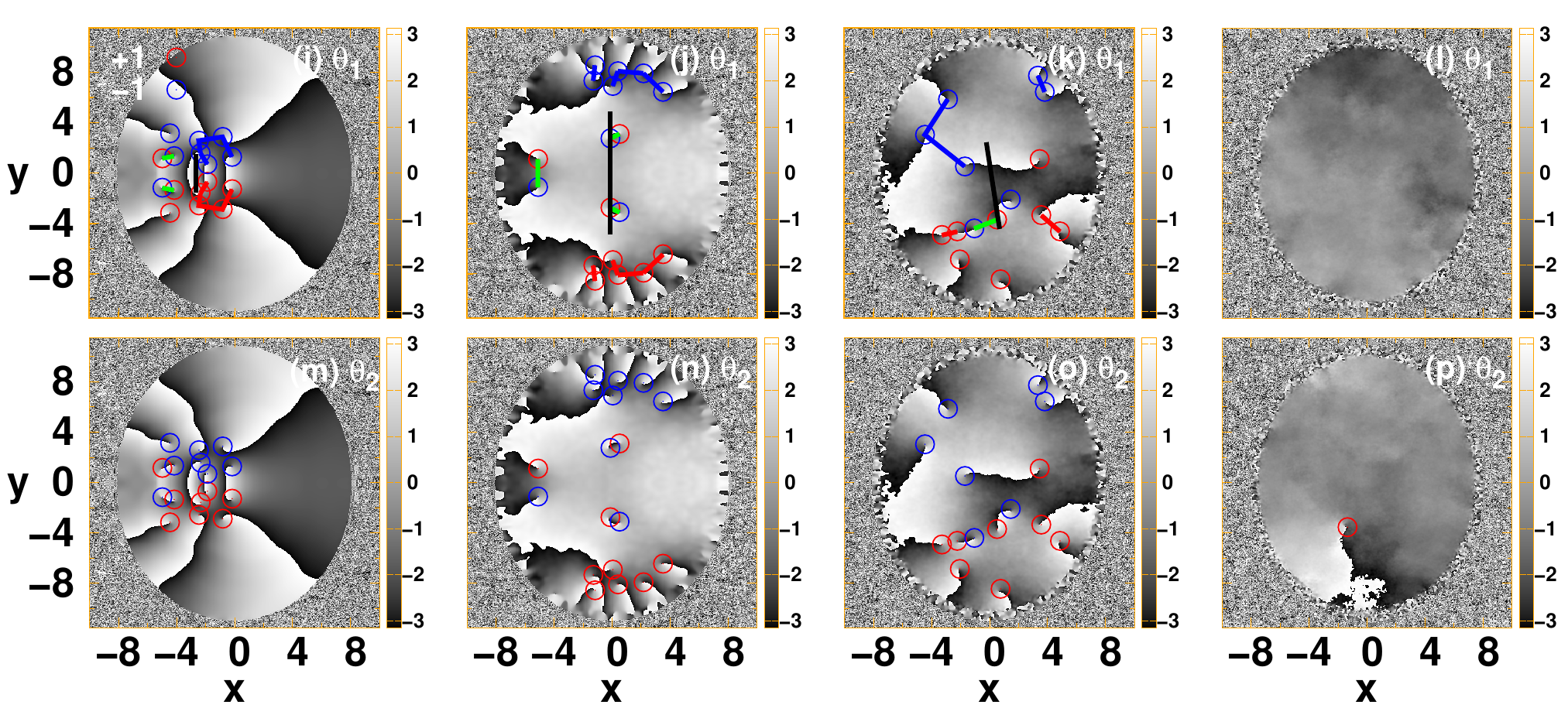}
  \caption{\label{fig_on3}{\footnotesize  
  Vortex dynamics of a binary BEC in an elliptic steep-wall trap with $\epsilon_{xj} = 0.3$,  $\epsilon_{yj} = -0.3$ and $\alpha=50$. 
      The snapshots of the density of the both the components at a,e) $t=0$, b,f) $t=10$, and c,g) $t=50$ and d,h) $t=500$. The red, blue, green and black lines represent the vortex cluster, anti-vortex cluster, dipoles (i.e., lines connecting the vortices in
      a dipole) and the dipole moment (see the definition in Eq.~(\ref{dipole})), respectively. The corresponding phase profiles are shown in the lower
      two rows of panels. {The parameters are $g_{11}=0.975 g_{22}$, $g_{12}=0.95 g_{11}$, $\tilde{g}_{22}=2000$, $M=512$ and $L=30$.} }}.
   \end{figure*}
%
 \begin{figure}[!htbp]  
 \includegraphics[scale=1.0,width=0.4\textwidth]{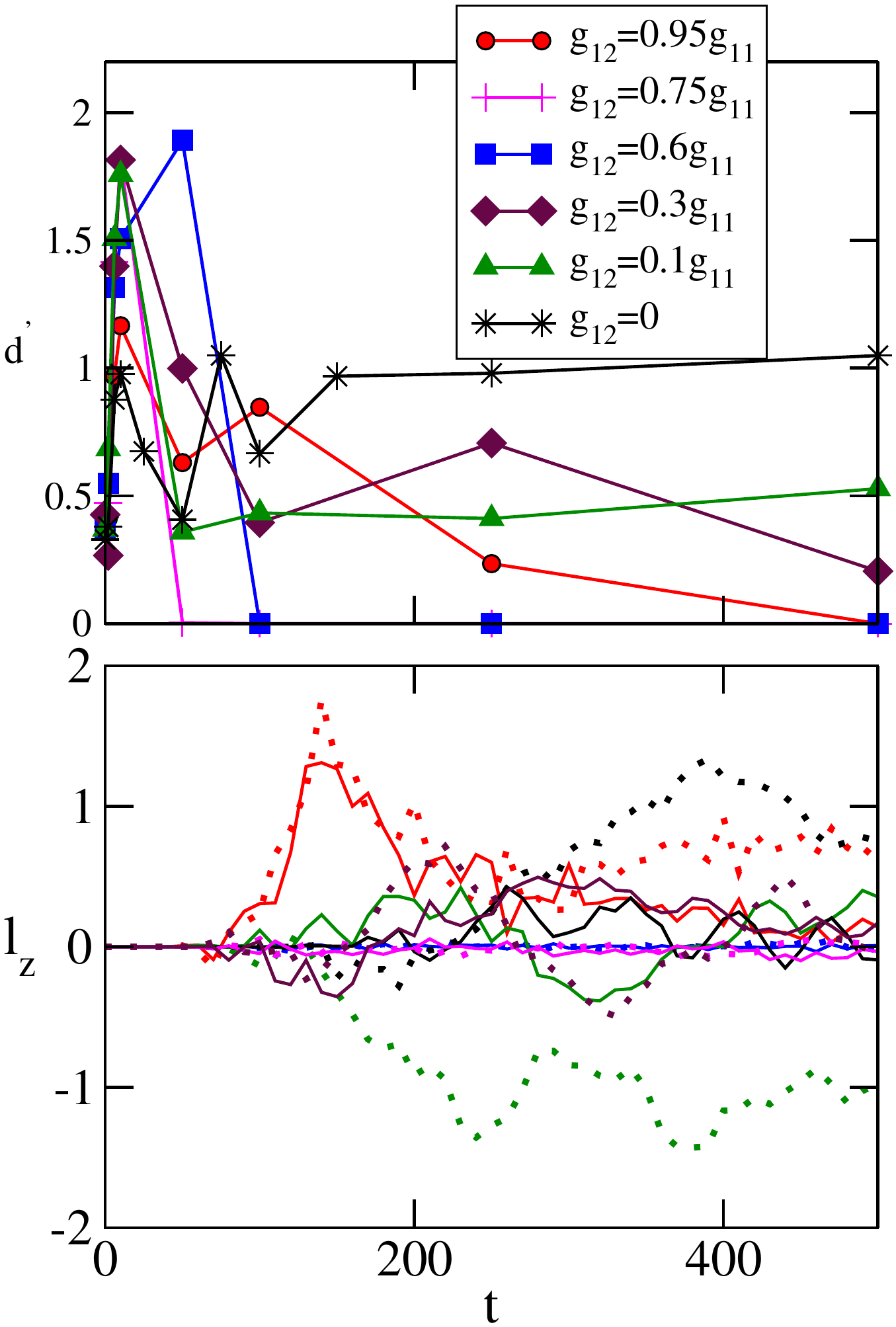}
  \caption{\label{angular3}{\footnotesize  The evolution of the dipole moment $d'$  of the first component and the angular momentum per particle for several values of $g_{12}$. In the lower panel, the solid lines represent the $l_{z1}$, while the dotted lines represent $l_{z2}$. The other parameters are $g_{11}=0.975 g_{22}$, $\tilde{g}_{22}=2000$, $M=512$ and $L=30$.}}.
   \end{figure}

   Since we have already seen the formation of large-core vortices in the harmonic-trap  resulting from the initial stirring for an anisotropic condensate in the previous section, here we investigate the turbulent dynamics for  $g_{11}=0.975 g_{22}$ in an elliptical steep-wall trap. Figure~\ref{fig_on3} shows the vortex turbulent dynamics at different times for the miscible case with $g_{12}=0.95 g_{11}$. {The upper panel (a-d) represents the density of the first component, while the bottom panel (e-h) represents that of the second component. The corresponding phase profiles are shown in (i-l) and (m-p), respectively}. Though cluster formation is apparent in the initial stage of the dynamics through a large dipole-moment (a) $d' \sim 0.37$, (b) $d' \sim 1.16$, (c) $d' \sim 0.63$, in the final stage it again leads to a quasi-equilibrium vortex-antidark structure with $d' = 0$ that persists throughout our simulations. Here  $d'=2d/(N_v R_0)$, where $N_v$ is the sum of vortices and anti-vortices. 
Due to the nearly zero angular momentum at $t=0$, shown in Fig.~\ref{angular3}(b), the number of vortices is also nearly zero.  
On the other hand, for $g_{11}= g_{22}$ we see the vortex clusters even at larger times (see Appendix \ref{g1_g2}). 
 \begin{figure}[!htbp]  
 \includegraphics[scale=1.0,width=0.5\textwidth]{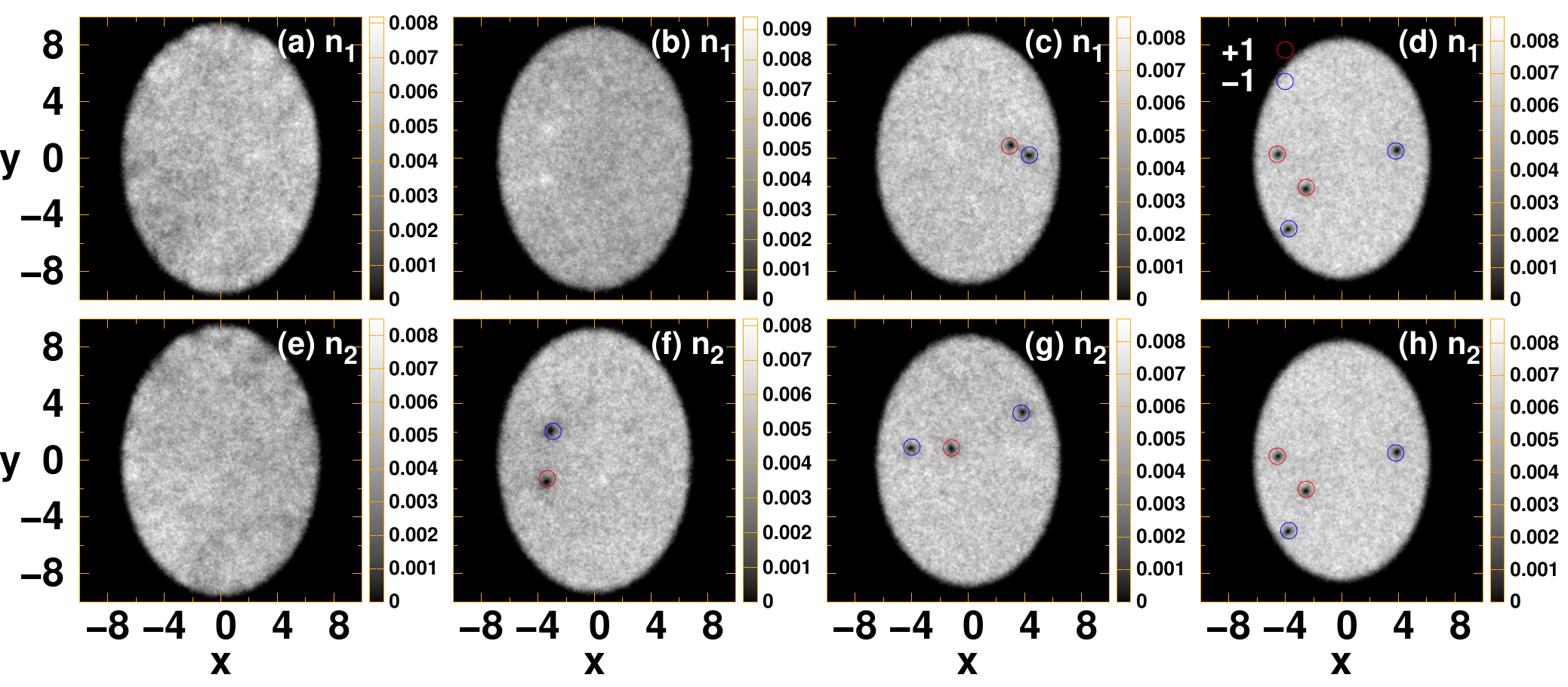}
  \caption{\label{fig_on3a}{\footnotesize  
      The snapshots of the density of the both the first (top panel) and second components (bottom panel) at $t=500$ for a,e) $g_{12}=0.75g_{11}$, b,f) $g_{12}=0.6g_{11}$, and c,g) $g_{12}=0.3g_{11}$ and d,h) $g_{12}=0.1g_{11}$.  The other parameters are $g_{11}=0.975 g_{22}$, $\tilde{g}_{22}=2000$, $M=512$ and $L=30$.}}.
   \end{figure}

In Fig.~\ref{angular3}, we show the evolution of the dipole moment $d'$ and the time dependence of the angular momentum per particle for different values of $g_{12}$. 
For higher values of $g_{12}$ the dipole moment goes to zero, corresponding to the quasi-equilibrium state without clusters as shown in Fig.~\ref{fig_on3}. On the other hand, for the lower values of $g_{12}$ the dipole moment remains finite even at larger times. The transition to the quasi equilibrium state for higher $g_{12}$ can be further understood from the angular momentum. Though initially both  components have the same angular momentum, the rate of angular momentum transfer among the components for larger $g_{12}$ is higher. {Since $g_{22}>g_{11}$ the final angular momentum (vortices) prefers to remain in the $\Psi_2$-component, which is consistent with the argument of the dynamical stability of the
  corresponding states~\cite{Garcia_2000,Garcia_2000:a}. This may lead to the
  long time persistence of isolated vortex-antidark structures.}

The snapshots of the density of the both the first (top panel) and second components (bottom panel) at $t=500$ shown in Fig.~\ref{fig_on3a} further elucidate the transition. {The disappearance of vortices at higher $g_{12}$ is a crucial factor  preventing the cluster formation.}

\section{Conclusions and Future Challenges}\label{sec:conclu}
We have investigated the two-dimensional quantum turbulence of miscible binary BECs, modeled by the GP equation. We considered both the symmetric and asymmetric setup of the system parameters where the asymmetry is introduced through the difference of the trap frequencies or that of the intra-component interaction strength. We followed an analogous stirring mechanism to the one that has been
previously used in the experiment of a single-component BEC to initiate the turbulent dynamics \cite{Neely_2013,Stagg_2015}. 

The initially generated vortices that resulted from the stirring are located at the same position in both components for the symmetric situation throughout its dynamical evolution and exhibit a similar type of energy spectra as that of a single-component condensate \cite{Bradley_2012}. 
In the asymmetric situation deviating slightly from the isotropic regime, however, as time increases, we see the increased core size of vortices with the same unit charge and the formation
of vortex-antidark solitonic lattices with the components mutually filling each other (i.e., where one has a dip associated with a vortex, the other has a bump).
Before forming this vortex-antidark state  
the system passes through a turbulent stage in which transferring of angular momentum among the components occurs.
This process occurs at the cost of inter-component energy. Interestingly,
a related dynamical turbulent stage may be directly connected with the
observations of the JILA experiment of a binary condensate \cite{Schweikhard_2004}, where the asymmetry among the components was due to the population difference and the distinct intra-component interaction strengths \cite{Egorov_2013, cornell1998having}. 
Furthermore, the spectra at the initial stage of turbulence dynamics feature similar power-laws, $k^{-5/3}$ power-law for the small wave number regime ($k \xi <1$) and $k^{-3}$ for the large wave number ($k \xi >1)$ regime, as in  the symmetric case. 
{We found that the decay of the turbulent state at the later quasi-equilibrium stage is caused by the spatial separation of the incompressible energy density, where the small scale components are accumulated at the periphery of the trapped condensate. 
Then the spectra are characterized by the $k^{-3}$ power law in the $k$-range associated with the spin-healing length and the plateau in the
range of the wave number determined by the density healing length
due to the bottleneck effect. 
This feature is enhanced for larger intercomponent coupling strength $g_{12}$. 
We also found that the decay behavior of the turbulence significantly suppresses the evolution toward the vortex cluster formation in the case of the steep-wall trap. }

The measurement of the $s$-wave scattering lengths for a binary condensate of $^{87}$Rb shows an asymmetry in the intra-component interaction strengths \cite{Egorov_2013, cornell1998having}. Moreover, 
the ability of designing not only anisotropic potentials, but, in principle,
arbitrary confining conditions is within reach in BEC experiments~\cite{boshier}.
Hence, the dynamics discussed here for the asymmetric case
should be directly accessible experimentally. Our results also point to the fact that the inter-component interaction strengths  shift the infinite temperature line, beyond which we expect the negative temperature. Similar results are reported in~\cite{Han_2018, Han_2019}; this is a direction worth exploring further. In yet another vein,
recent work has started exploring further solitary wave structures involving
more than two components~\cite{bersano,chai2019magnetic}. Appreciating
the possible scenarios in such a generalized setting involving also the
spin degree of freedom and associated magnetic excitations may be
of interest in its own right. {Additionally, in a multi-component system, there exist two phonon branches, density (in-phase) wave and spin (out-of-phase) wave (See the equation 2 in \cite{kim2020observation}). For an asymmetric set up in the limit $g_{12} \rightarrow g$, the energy of the spin-wave mode is lowered
  and can thus be excited much easily. Hence, it is interesting to see
  the contribution of the density-wave and the spin-wave components to
  the compressible energy. We are currently working on that and
  relevant results will be presented elsewhere. {Finally,
    the anisotropy between the components can be introduced via mass
    of the components too, and a  preliminary study in this direction
    is reported in Ref.~\cite{Castilho_2019}, where Na-K bosonic
    mixtures is considered. Also, it is to be noted that to incorporate
    the quantum effects such as quantum correlations and associated
    fragmentation and the finite temperature effects,  a
    beyond-mean-field
    model has to be considered. While here we have restricted our
    considerations to large atom numbers and near-zero temperatures,
    so that the mean-field setting provides a valid approximation, it
    is relevant to extend earlier works, such as
    e.g. Refs. \cite{Norrie_2006,
      tsatsos2014vortex,weiner2017phantom,katsimiga2017dark} in these
    interesting directions for the multi-component system.}
%
\section{acknowledgments}
This material is based upon work supported by the US National Science
Foundation under Grants No. PHY-1602994 and DMS-1809074
(PGK). PGK also acknowledges support from the Leverhulme Trust via a
Visiting Fellowship and thanks the Mathematical Institute of the University
of Oxford for its hospitality during part of this work.
The work of K.K. is partly supported by KAKENHI from the Japan Society for the Promotion of Science (JSPS) Grant-in- Aid for Scientific Research (KAKENHI Grant No. 18K03472). BD acknowledges the Science and Engineering Research Board, Government of India for funding, Grant No. EMR/2016/002627.
%
\appendix

\section{Vortex nucleation during the stirring procedure} \label{stirr_ini}
 \begin{figure}[!htbp]  
 \includegraphics[scale=1.0,width=0.5\textwidth]{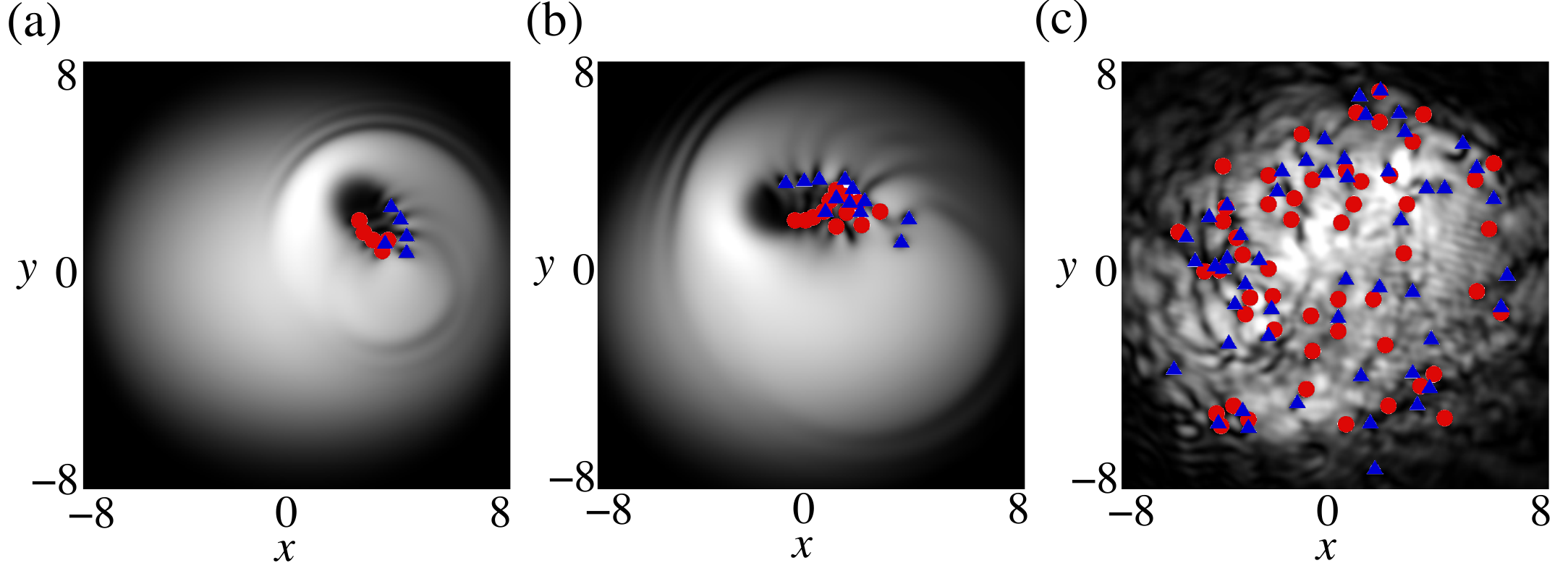}
  \caption{\label{fig_pre_stirr}{\footnotesize  
Snapshots of the density of the first component just after the obstacle starts to move (a) and (b), and that for $t=0$ (c) after completing the 2nd period of the stirring. The positions of vortices and anti-vortices are plotted by (red) circles and (light blue) triangles, respectively. Here, the parameter values correspond to those in Fig.~\ref{fig1b}. 
}}
   \end{figure}
Figure~\ref{fig_pre_stirr} shows the snapshots of the density of the first component during the stirring process in Sec.~\ref{harmonicturb}, which is before our $t=0$. 
The obstacle induces counterclockwise rotation centered at a radius $r_0 = 0.4 R_\text{TF}$ beyond the critical velocities for vortex nucleation. 
The snapshots show that vortices are nucleated at the zero density region at the obstacle, in the form of vortex--anti-vortex pairs. Note that, as shown in Fig.~\ref{fig_pre_stirr}(a) and (b), the vortices with counterclockwise circulation are emitted into the inner region of the condensate, while the anti-vortices with clockwise circulation are into the opposite outer side. 
This imbalance of the vortex and anti-vortex distribution is responsible for the nonzero angular momentum at $t=0$.

\section{Turbulent dynamics for several different parameters}
{ In this section, we show some numerical results not presented in the main text. }

\subsection{Turbulent dynamics for $g_{11} \ne g_{22}$  or $N_1 \ne N_2$}\label{gq_ge_g2}
 \begin{figure}[!htbp]  
 \includegraphics[scale=1.0,width=0.5\textwidth]{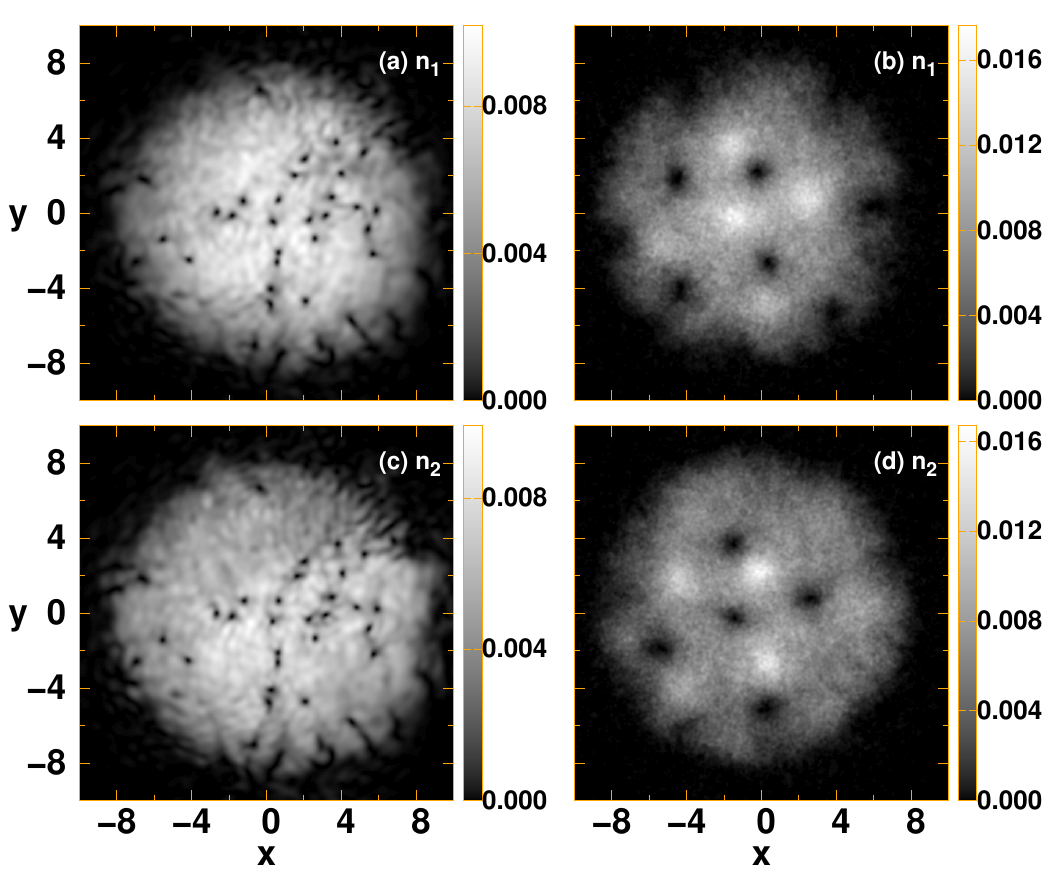} 
  \caption{\label{fig1c1}{\footnotesize  
  The density of the first (top panels) and second (bottom panels) components  at (a,c) $t=0$ and (b,d) $t=600$. {The parameters are $g_{11}=0.975 g_{22}$,
  $g_{12}=0.95 g_{11}$, $\tilde{g}_{22}=2000$, $N_1 = N_2$, $M=1024$, $L=40$ and
  $\epsilon_{xj}=\epsilon_{yj}=0.0$.}}}.
   \end{figure}
Here, we show the turbulent dynamics and the subsequent quasi-equilibrium states for the cases of $g_{11} \ne g_{22}$ or $N_1 \neq N_2$ without the trap asymmetry ($\epsilon_{y1}=0$). The stirring procedure is the same as before. Figure~\ref{fig1c1} shows the density of the first and second components at (a,c) $t=0$ and (b,d) $t=600$ for $g_{11}=0.975 g_{22}$ and $g_{12}=0.95 g_{11}$, $\tilde{g}_{22}=2000$, while $N_1 = N_2$. 
The dynamics exhibits a behavior similar to Fig.~\ref{fig1b} and leads to the 
formation of interlaced lattice state
of vortices as in Fig.~\ref{fig1c1}(b) and (d). 
We also show the turbulent dynamics when the particle number is slightly different $N_{1} \ne N_{2}$; Fig.~\ref{fig1c11} shows the density of the first and second components at (a,c) $t=0$ and (b,d) $t=400$. Here, the dimensionless $g_{ij}$'s assume the values indicated in the caption. When the population difference is small, we have observed similar dynamics as in the previous case. 
 \begin{figure}[!htbp]  
 \includegraphics[scale=1.0,width=0.5\textwidth]{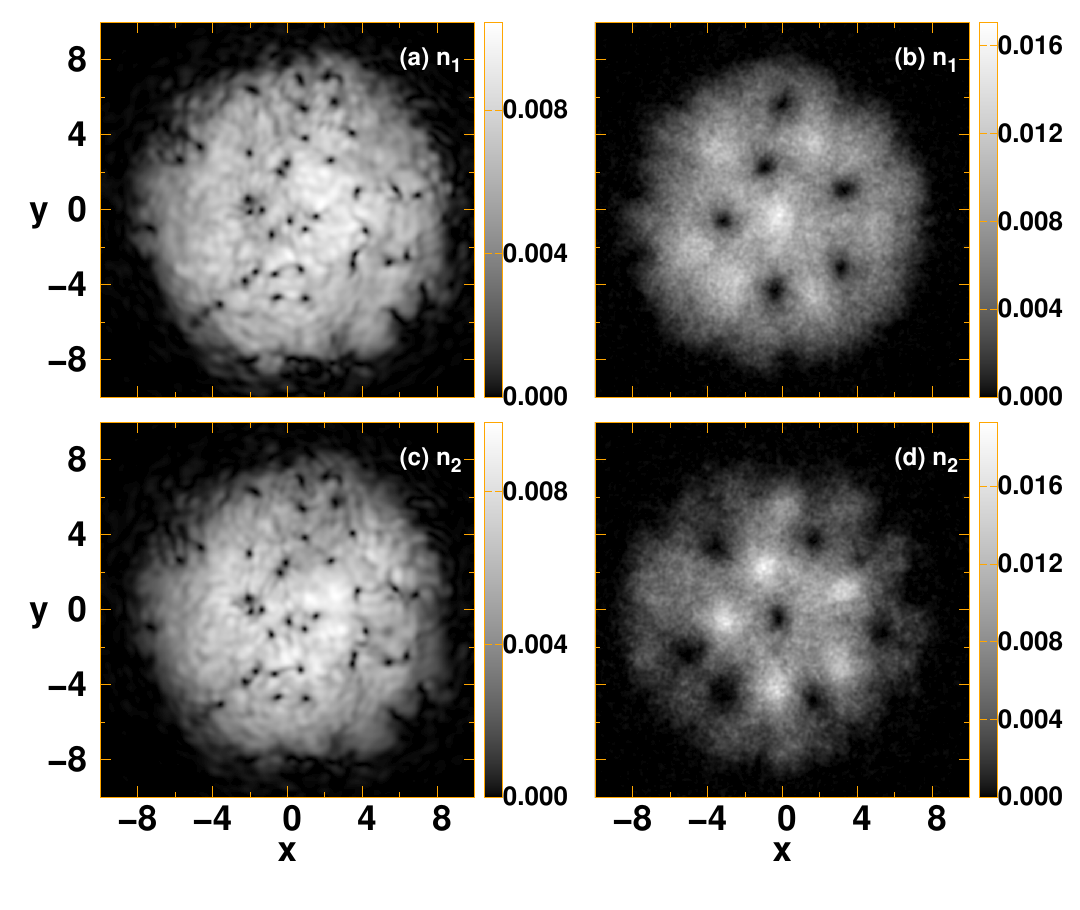} 
  \caption{\label{fig1c11}{\footnotesize  
  The density of the first (top panels) and second (bottom panels) components at (a,c) $t=0$ and (b,d) $t=400$. { The parameters are $g_{11} = 2200$, $g_{12} = 1710$, $g_{21} = 2090$, $g_{22} = 1800$, $M=1024$, $N_1=1.1$, $N_2=0.9$, $M=1024$, $L=40$ and $\epsilon_{xj}=\epsilon_{yj}=0.0$.}}}.
   \end{figure}

 \subsection{Turbulent dynamics and angular momentum evolution for the homogenous condensates}\label{ang_homo}
 In this Appendix we demonstrate the turbulent dynamics for the homogeneous system, where the stirring procedure is made in a way similar to the trapped system. 
We evolve the initial wave function $\psi_i=\sqrt{\mu/(g_{11}+g_{12})}$, with $\mu = 34.78$ by setting $V_j=0$ in Eq.~\eqref{eq:2GP}. 
We implement  periodic boundary condition. 
Since there is no low density region, as seen in the outside of the Thomas-Fermi radius in the trapped system, the vortex--anti-vortex annihilation is an only mechanism of the decay of the vortex excitation, and as a result, equal numbers of vortices and those of anti-vortices are expected in the final state. 

  \begin{figure*}[!htbp]  
 \includegraphics[scale=1.0,width=0.95\textwidth]{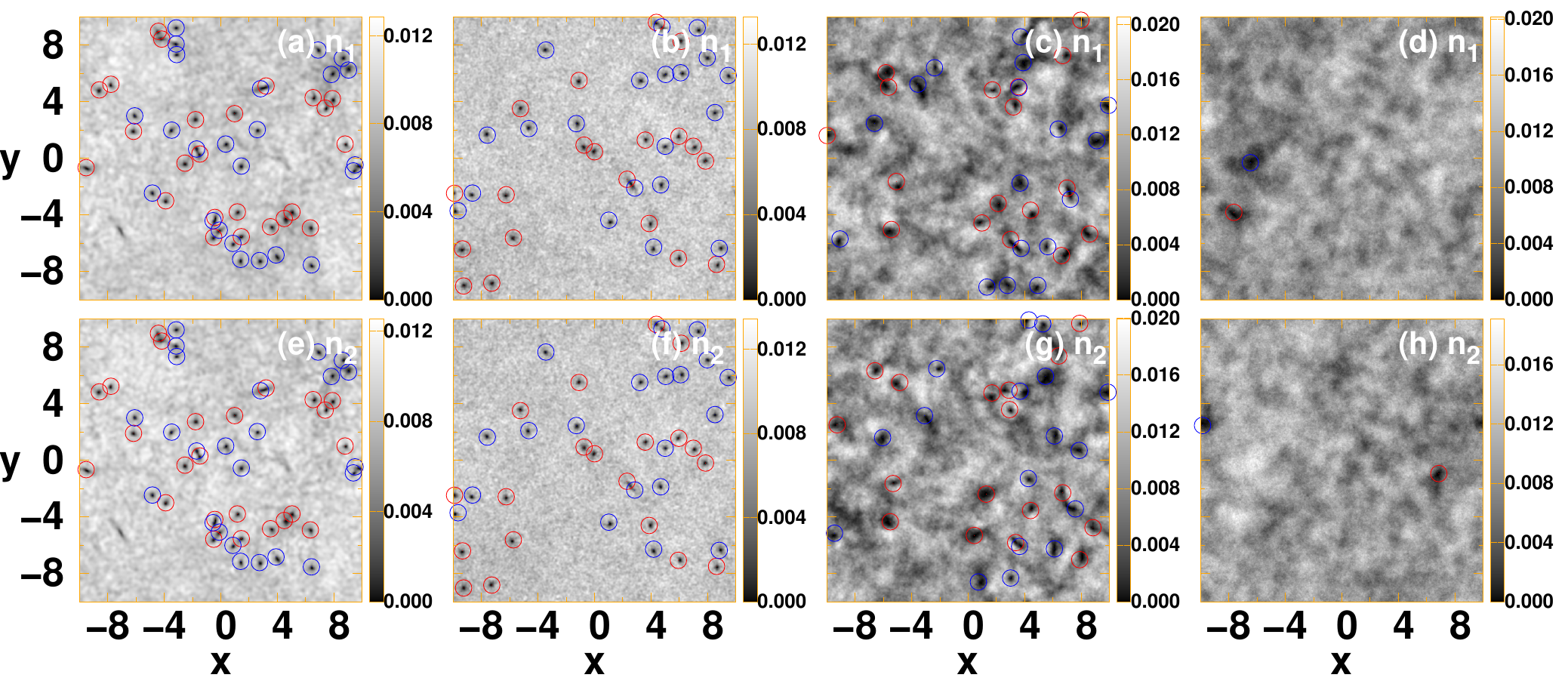} 
  \includegraphics[scale=1.0,width=0.95\textwidth]{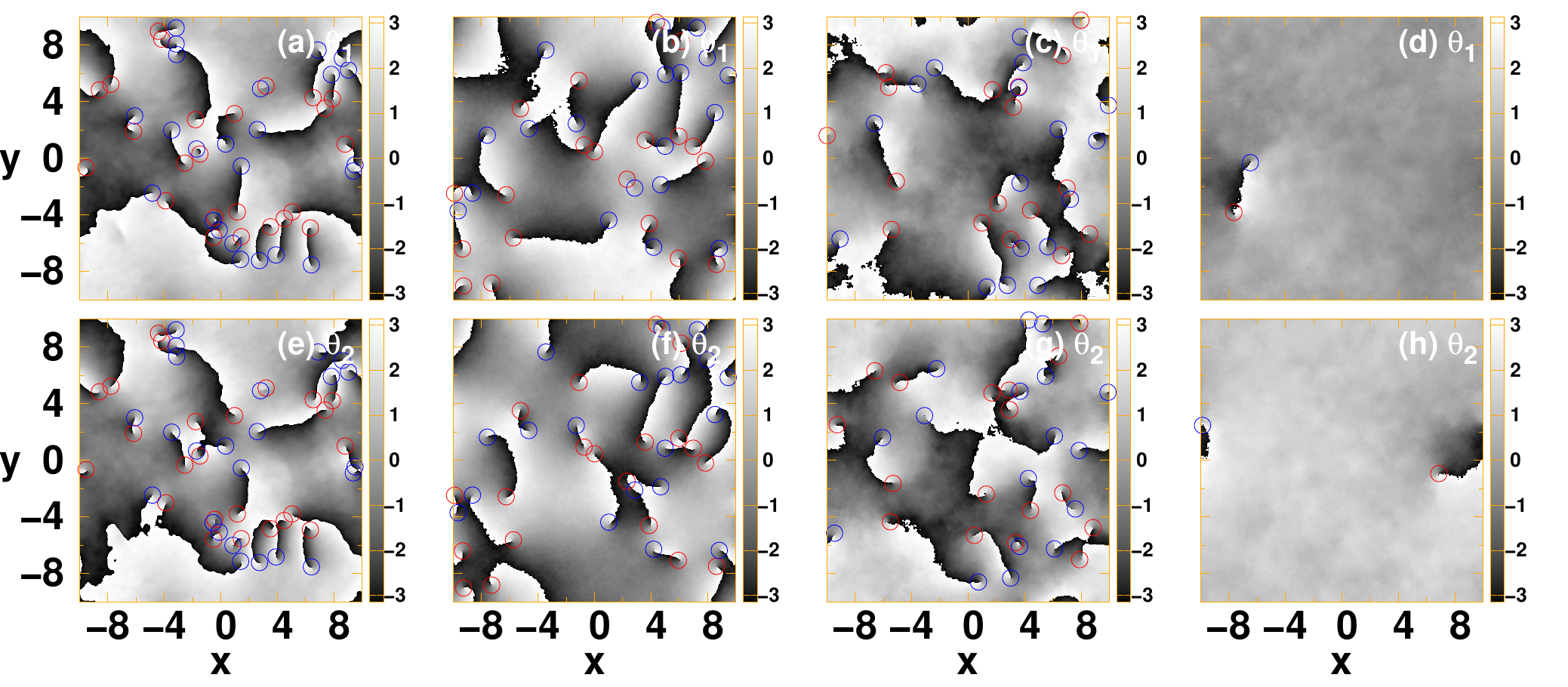} 
  \caption{\label{fig1c3}{\footnotesize  
Snapshots of the time development for the homogeneous binary condensates after the stirring. 
The first and the second rows show the density of the first component and that of the second component, respectively, at (a,e) $t=0$, (b,f) $t=50$, (c,g) $t=100$ and (d,h) $t=600$ for $M = 512$ grid points. 
The third and fourth rows show the corresponding phase profiles. { The parameters are $g_{11}=0.975 g_{22}$, $g_{12}=0.95 g_{11}$ and $\epsilon_{xj}=\epsilon_{yj}=0.0$.}}}
\end{figure*}
Figures~\ref{fig1c3} and \ref{angular1c3} show the vortex dynamics and the corresponding angular momentum transfer, respectively. 
As seen in the trapped system, the snapshots of the density exhibit the transient dynamics from the turbulent state to the dark-antidark structure. 
Subsequently, the scale of the density variation is determined by the spin healing length given by the formula \cite{Eto_2011}
\begin{equation}
 \xi_s^2 = \frac{1}{2}\left(\frac{g_{22}}{\mu_1g_{22}-\mu_2g_{12}}+\frac{g_{12}}{\mu_2g_{11}-\mu_1g_{12}}\right).
 \label{eq:spin_heala}
\end{equation}
However, the phase profiles show that the vortices do not survive in the long time
dynamics, due to the fact that equal numbers of vortices and antivortices undergo  pair-annihilations. 
This behavior can be understood from the evolution of the angular momentum. 
There is a finite angular momentum at $t=0$, caused by the introduction of the stirring potential that breaks the rotational symmetry of the system. 
After the long-time evolution, the angular momentum eventually goes to zero, 
although a small oscillation can be seen for the first component, which is caused by the survived vortex and anti-vortex seen in the phase profile of Fig.~\ref{fig1c3}(d).
 \begin{figure}[!htbp]  
 \includegraphics[scale=1.0,width=0.48\textwidth]{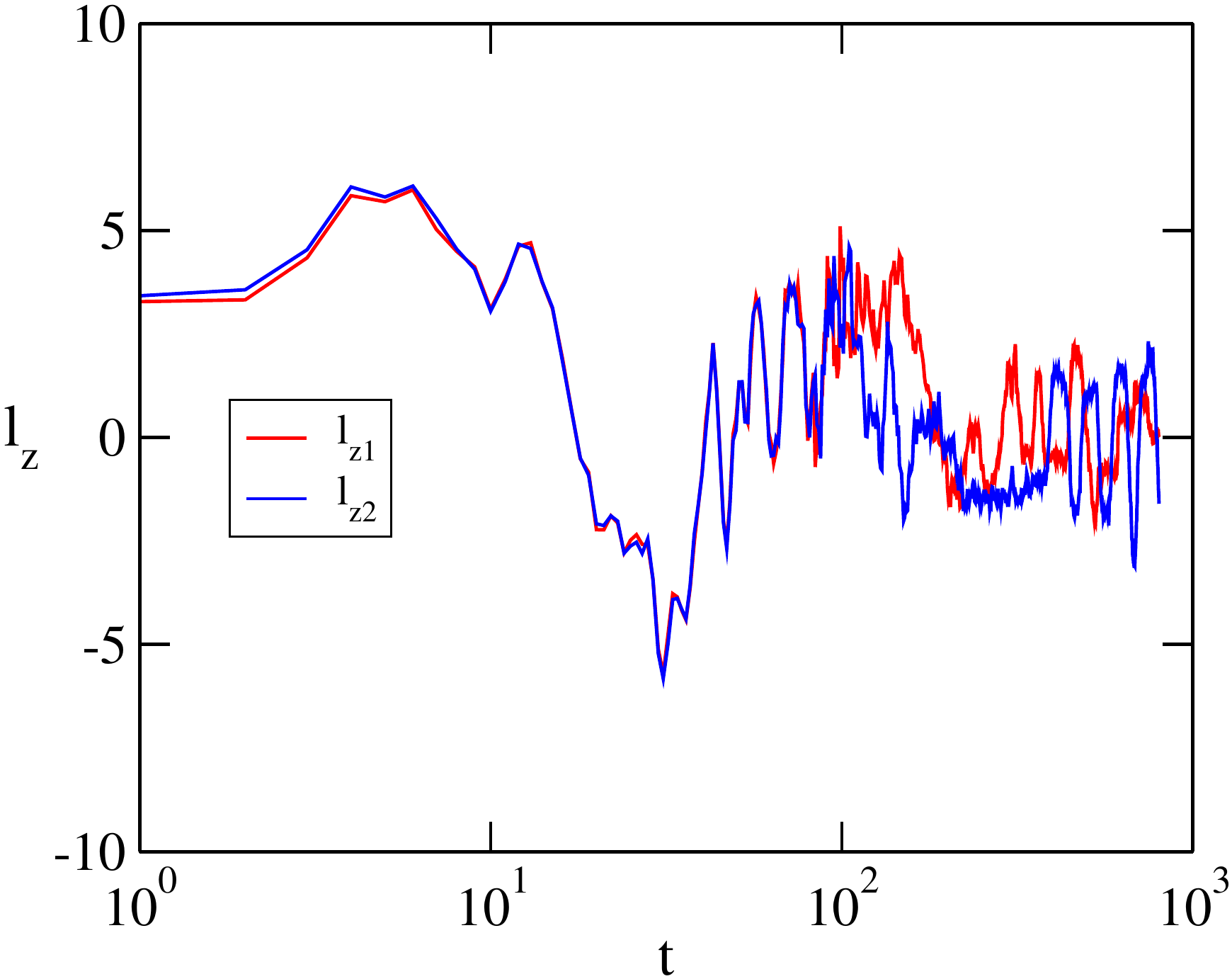}
  \caption{\label{angular1c3}{\footnotesize  The angular momentum per particle corresponding to the Fig.~\ref{fig1c3}. }}.
   \end{figure}
    %
%
 \subsection{Vortex turbulent dynamics for several values of $g_{12}$}\label{diff_ene}
       \begin{figure}[!htbp]  
 \includegraphics[scale=1.0,width=0.48\textwidth]{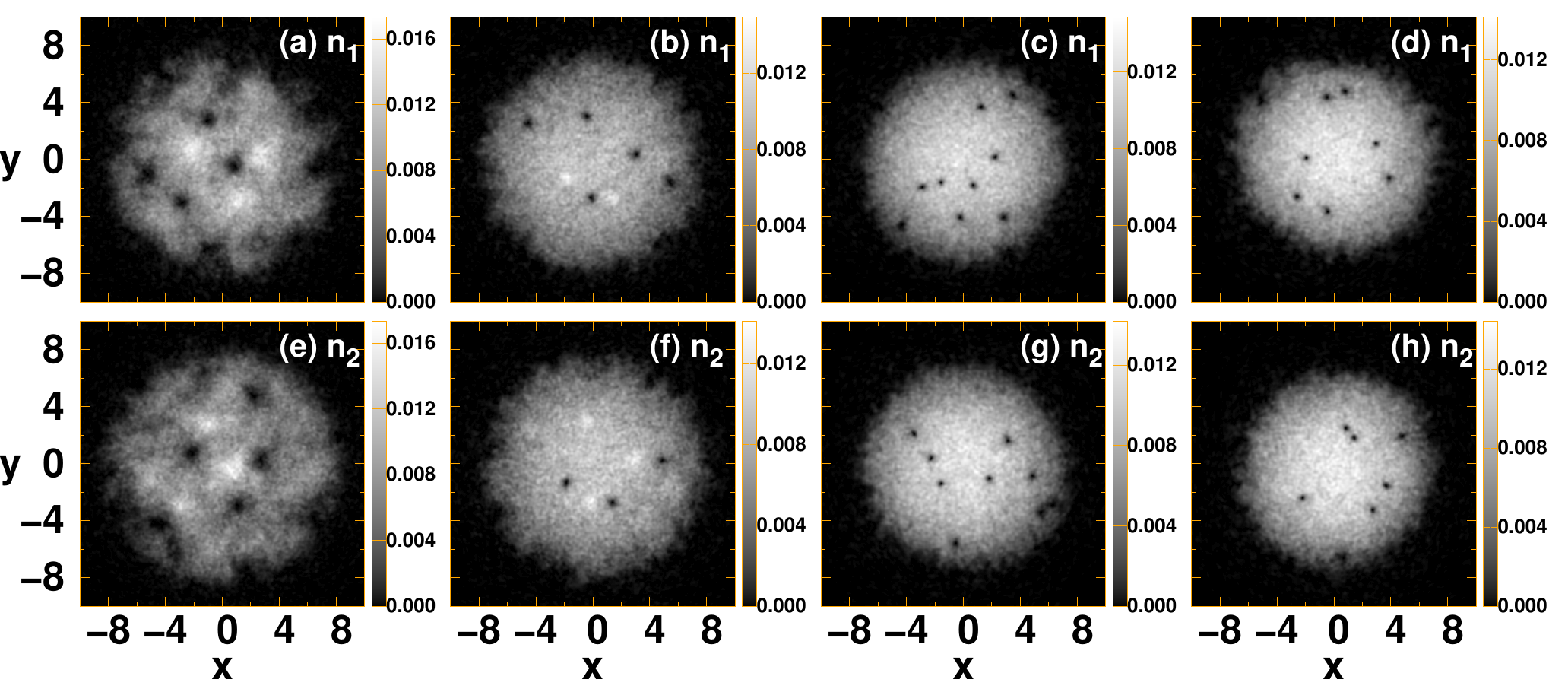}
  \caption{\label{density_g12}{\footnotesize  The density of the first and second components for (a and d) $g_{12}=0.95g$, (b and e) $g_{12}=0.6g$, (c and g) $g_{12}=0.1g$ and (d and h) $g_{12}=0$ at {$t=200$ for $M = 1024$ grid points.}}}.
   \end{figure}
Here we discuss the $g_{12}$-dependence of the turbulent dynamics. We set $g_{11} = g_{22}$ and $\epsilon_{y1} = 0.025$ and the vortices are generated by the stirring potential in the same way as before. 
Figure~\ref{density_g12} shows the condensate density at $t=200$ for different strength of $g_{12}$. It shows the clear interlaced lattice state
of the like-signed vortices for higher $g_{12}$. With decreasing $g_{12}$, the vortex-antidark
lattice structure disappears and the vortex structure resembles  that in a single-component condensate. Also, the vortices feature chaotic motions which cannot be interpreted as an interlaced lattice state.

  \section{Numerical Calculation of Energy Spectra} \label{spectra_calc}

To calculate the energy spectra \cite{Eyink_2006,RevModPhys.71.S383,kraichnan1980two,Numasato_2010}, we do the  decomposition as follows. 
The kinetic energy term $| \nabla \Psi|^2 / 2 $ in the Hamiltonian Eq.~\eqref{eq:GPEF} can be written as 
\ba\label{eq:vort_sp1aa}
\frac{1}{2}|\nabla\Psi|^2= \frac{1}{2} \left( n|\bm{u}|^2+ \left| \nabla \sqrt{n} \right|^2 \right), 
\ea
where the Madelung transformation $\Psi=\sqrt{n} e^{i \phi}$ yields the condensate density $n=|\Psi|^2$ and the superfluid velocity $\bm{u}=\nabla \phi$. 
We do not consider the index $j$ to represent the components. 
Here the first and second terms represent the density of the kinetic energy ($E_\text{ke}$) and the quantum pressure ($E^{q}$), respectively, where 
the energies are given by 
\ba\label{eq:vort_sp1bb}
E_\text{ke}=\frac{1}{2}\int n|\bm{u}|^2 d^2 r,~~~E^q=\frac{1}{2}\int |\nabla \sqrt{n}|^2 d^2 r. 
\ea
The velocity vector $\bm{u}$ now can be written as a sum over a solenoidal part (incompressible) $\bm{u}^\text{ic} $ and an irrotational (compressible) part $\bm{u}^\text{c}$ as
\ba\label{eq:vort_sp1}
\bm{u}=\bm{u}^\text{ic}+\bm{u}^\text{c}, 
\ea
such that $\dive{\bm{u}^\text{ic}} = 0$ and $\curl{\bm{u}^\text{c}} = 0$.  We next define the scalar potential $\Phi$ and the vector potential $\bm{A}$ of the velocity field which satisfy the relations   
\ba\label{eq:vort_sp1a}
\sqrt{n} {\bm{u}^\text{ic}} = \curl{\bm{A}}, \quad \sqrt{n}  {\bm{u}^\text{c}} = {\nabla} \Phi
\ea
respectively.  Taking the divergence of the equation for the scalar potential we obtain 
\ba\label{eq:vort_sp1b}
\nabla^2{\Phi} = \dive{ \left( \sqrt{n} {\bm{u}^\text{c} } \right) } = \dive{\left(  \sqrt{n} {\bm{u}} \right)} . 
\ea 
From this Poisson equation we numerically determine the scalar potential $\Phi$ \cite{Horng_2009}.
On applying the Fourier transform to the Eq.~\eqref{eq:vort_sp1b} we get
\ba\label{eq:vort_sp1c}
{\tilde{\Phi}} =  -\frac{\mathcal{F}[\dive{\sqrt{n} {\bm{u}}}]}{k_x^2+k_y^2}. 
\ea
After taking the inverse Fourier transform of $\tilde{\Phi}$, we get $\sqrt{n} {\bm{u}^c}$ from Eq.~\eqref{eq:vort_sp1a}. Further we can find ${\sqrt{n} \bm{u}^\text{ic}}$ from Eq.~\eqref{eq:vort_sp1}. 
   
 The compressible and incompressible kinetic energies are then
 \ba\label{eq:vort_sp2}
E^\text{ic,c} = \frac{1}{2}\int d^2 r |\sqrt{n} \bm{u}^\text{ic,c}(\bm{r})|^2, 
\ea
In the $k$-space, the total incompressible and compressible kinetic energy $E_{kin}^\text{ic,i}$ 
is represented by
 \ba\label{eq:vort_sp3}
E^\text{ic,c} = \frac{L^2}{2}\sum_{j=x,y}\int d^2 k |\mathcal{F}_j(\bm{k})^\text{ic,c}|^2, 
\ea
where $\mathcal{F}_j(\bm{k})$ is the Fourier transform of $\sqrt{n} u_j$ of the $j$-th component of $\bm{u}= (u_x, u_y)$.
We can modify Eq.~\eqref{eq:vort_sp3} as 
 \ba\label{eq:vort_sp4}
E^\text{ic,c}(k)= \frac{k}{2}\sum_{j=x,y}\int_0^{2\pi} d\phi_k |\mathcal{F}_j(\bm{k})^\text{ic,c}|^2, 
\ea
where we consider the polar coordinates and $k=\sqrt{k_x^2+k_y^2}$.
We numerically integrate over the $k$-shell (summing over the grid points) to find $E^\text{ic,c}(k)$. 
Now to get the respective kinetic energy, we integrate $E^\text{ic,c}(k)$ with respect to $k$.
\section{Symmetric case with $g_{11}=g_{22}$ in steep-wall trap}\label{g1_g2}
 \begin{figure}[!htbp]  
 \includegraphics[scale=1.0,width=0.5\textwidth]{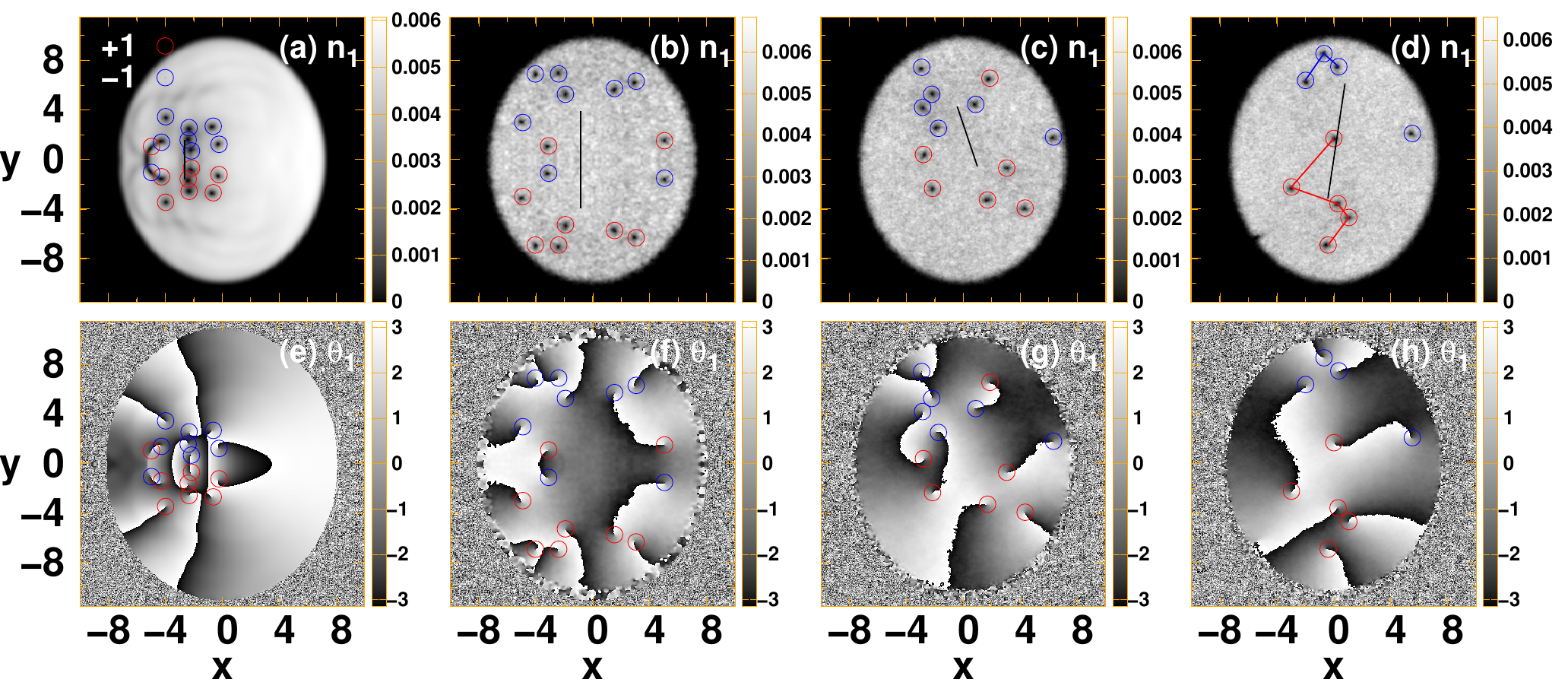}
  \caption{\label{fig_on2}{\footnotesize  
The density of the first component at (a) $t=0$, (b) $t=50$, (c) $t=250$ and (d) $t=500$. The red, blue and black lines represent the vortex cluster, anti-vortex cluster and the dipole moment, respectively. Here, the exponent of the trap potential is $\alpha=50$. The parameters are $g_{22}=g_{11}$, $g_{12}=0.95g_{11}$, $\tilde{g}_{22}=2000$, $L=30$, and $M=512$.}}.
   \end{figure}
   Figure~\ref{fig_on2} shows the vortex turbulent dynamics at different times for the symmetric choice of the intra-component couplings $g_{22}=g_{11}=2000\hbar^2/m$ and the miscible regime $g_{12}=0.95g_{11}$. Here, both the components behave in the same manner, the dynamics mimics those of the single-component BEC. The measured dipole moment (a) $d' \sim 0.31$, (b) $d' \sim 0.79$, (c) $d' \sim 0.50$, (d) $d' \sim 0.90$ shows that even for the smaller time, the magnitude of the dipole moment is much greater than zero. Here  $d'=2d/(N_v R_0)$, where $N_v$ is the sum of the vortices and anti-vortices. The formation of vortex cluster can be clearly
   discerned in the figure.
 %
 \bibliographystyle{apsrev4}
\let\itshape\upshape
\normalem
\bibliography{reference1}

\providecommand{\noopsort}[1]{}\providecommand{\singleletter}[1]{#1}%
\begin{thebibliography}{98}%
\makeatletter
\providecommand \@ifxundefined [1]{%
 \@ifx{#1\undefined}
}%
\providecommand \@ifnum [1]{%
 \ifnum #1\expandafter \@firstoftwo
 \else \expandafter \@secondoftwo
 \fi
}%
\providecommand \@ifx [1]{%
 \ifx #1\expandafter \@firstoftwo
 \else \expandafter \@secondoftwo
 \fi
}%
\providecommand \natexlab [1]{#1}%
\providecommand \enquote  [1]{``#1''}%
\providecommand \bibnamefont  [1]{#1}%
\providecommand \bibfnamefont [1]{#1}%
\providecommand \citenamefont [1]{#1}%
\providecommand \href@noop [0]{\@secondoftwo}%
\providecommand \href [0]{\begingroup \@sanitize@url \@href}%
\providecommand \@href[1]{\@@startlink{#1}\@@href}%
\providecommand \@@href[1]{\endgroup#1\@@endlink}%
\providecommand \@sanitize@url [0]{\catcode `\\12\catcode `\$12\catcode
  `\&12\catcode `\#12\catcode `\^12\catcode `\_12\catcode `\%12\relax}%
\providecommand \@@startlink[1]{}%
\providecommand \@@endlink[0]{}%
\providecommand \url  [0]{\begingroup\@sanitize@url \@url }%
\providecommand \@url [1]{\endgroup\@href {#1}{\urlprefix }}%
\providecommand \urlprefix  [0]{URL }%
\providecommand \Eprint [0]{\href }%
\providecommand \doibase [0]{http://dx.doi.org/}%
\providecommand \selectlanguage [0]{\@gobble}%
\providecommand \bibinfo  [0]{\@secondoftwo}%
\providecommand \bibfield  [0]{\@secondoftwo}%
\providecommand \translation [1]{[#1]}%
\providecommand \BibitemOpen [0]{}%
\providecommand \bibitemStop [0]{}%
\providecommand \bibitemNoStop [0]{.\EOS\space}%
\providecommand \EOS [0]{\spacefactor3000\relax}%
\providecommand \BibitemShut  [1]{\csname bibitem#1\endcsname}%
\let\auto@bib@innerbib\@empty
\bibitem [{\citenamefont {Holmes}\ \emph {et~al.}(2012)\citenamefont {Holmes},
  \citenamefont {Lumley}, \citenamefont {Berkooz},\ and\ \citenamefont
  {Rowley}}]{holmes}%
  \BibitemOpen
  \bibfield  {author} {\bibinfo {author} {\bibfnamefont {P.}~\bibnamefont
  {Holmes}}, \bibinfo {author} {\bibfnamefont {J.}~\bibnamefont {Lumley}},
  \bibinfo {author} {\bibfnamefont {G.}~\bibnamefont {Berkooz}}, \ and\
  \bibinfo {author} {\bibfnamefont {C.}~\bibnamefont {Rowley}},\ }\href@noop {}
  {\emph {\bibinfo {title} {Turbulence, coherent structures, dynamical systems
  and symmetry}}}\ (\bibinfo  {publisher} {Cambridge University Press},\
  \bibinfo {address} {Cambridge, UK},\ \bibinfo {year} {2012})\BibitemShut
  {NoStop}%
\bibitem [{\citenamefont {Onsager}(1949)}]{onsager1949statistical}%
  \BibitemOpen
  \bibfield  {author} {\bibinfo {author} {\bibfnamefont {L.}~\bibnamefont
  {Onsager}},\ }\bibfield  {title} {\enquote {\bibinfo {title} {Statistical
  hydrodynamics},}\ }\href@noop {} {\bibfield  {journal} {\bibinfo  {journal}
  {Il Nuovo Cimento (1943-1954)}\ }\textbf {\bibinfo {volume} {6}},\ \bibinfo
  {pages} {279} (\bibinfo {year} {1949})}\BibitemShut {NoStop}%
\bibitem [{\citenamefont {Kraichnan}(1967)}]{kraichnan1967inertial}%
  \BibitemOpen
  \bibfield  {author} {\bibinfo {author} {\bibfnamefont {R.~H.}\ \bibnamefont
  {Kraichnan}},\ }\bibfield  {title} {\enquote {\bibinfo {title} {Inertial
  ranges in two-dimensional turbulence},}\ }\href@noop {} {\bibfield  {journal}
  {\bibinfo  {journal} {The Physics of Fluids}\ }\textbf {\bibinfo {volume}
  {10}},\ \bibinfo {pages} {1417} (\bibinfo {year} {1967})}\BibitemShut
  {NoStop}%
\bibitem [{\citenamefont {Kraichnan}(1975)}]{kraichnan1975statistical}%
  \BibitemOpen
  \bibfield  {author} {\bibinfo {author} {\bibfnamefont {R.~H.}\ \bibnamefont
  {Kraichnan}},\ }\bibfield  {title} {\enquote {\bibinfo {title} {Statistical
  dynamics of two-dimensional flow},}\ }\href@noop {} {\bibfield  {journal}
  {\bibinfo  {journal} {Journal of Fluid Mechanics}\ }\textbf {\bibinfo
  {volume} {67}},\ \bibinfo {pages} {155} (\bibinfo {year} {1975})}\BibitemShut
  {NoStop}%
\bibitem [{\citenamefont {Eyink}\ and\ \citenamefont
  {Sreenivasan}(2006)}]{Eyink_2006}%
  \BibitemOpen
  \bibfield  {author} {\bibinfo {author} {\bibfnamefont {G.~L.}\ \bibnamefont
  {Eyink}}\ and\ \bibinfo {author} {\bibfnamefont {K.~R.}\ \bibnamefont
  {Sreenivasan}},\ }\bibfield  {title} {\enquote {\bibinfo {title} {Onsager and
  the theory of hydrodynamic turbulence},}\ }\href {\doibase
  10.1103/RevModPhys.78.87} {\bibfield  {journal} {\bibinfo  {journal} {Rev.
  Mod. Phys.}\ }\textbf {\bibinfo {volume} {78}},\ \bibinfo {pages} {87}
  (\bibinfo {year} {2006})}\BibitemShut {NoStop}%
\bibitem [{\citenamefont {Tsubota}\ and\ \citenamefont
  {Kobayashi}(2008)}]{Tsubota_2008}%
  \BibitemOpen
  \bibfield  {author} {\bibinfo {author} {\bibfnamefont {M.}~\bibnamefont
  {Tsubota}}\ and\ \bibinfo {author} {\bibfnamefont {M.}~\bibnamefont
  {Kobayashi}},\ }\bibfield  {title} {\enquote {\bibinfo {title} {Quantum
  turbulence in trapped atomic Bose-Einstein condensates},}\ }\href@noop {}
  {\bibfield  {journal} {\bibinfo  {journal} {Journal of Low Temperature
  Physics}\ }\textbf {\bibinfo {volume} {150}},\ \bibinfo {pages} {402}
  (\bibinfo {year} {2008})}\BibitemShut {NoStop}%
\bibitem [{\citenamefont {Tsubota}\ and\ \citenamefont
  {Kasamatsu}(2013)}]{tsubota2013quantized}%
  \BibitemOpen
  \bibfield  {author} {\bibinfo {author} {\bibfnamefont {M.}~\bibnamefont
  {Tsubota}}\ and\ \bibinfo {author} {\bibfnamefont {K.}~\bibnamefont
  {Kasamatsu}},\ }\enquote {\bibinfo {title} {Quantized Vortices and Quantum
  Turbulence},}\ in\ \href {\doibase 10.1007/978-3-642-37569-9_13} {\emph
  {\bibinfo {booktitle} {Physics of Quantum Fluids: New Trends and Hot Topics
  in Atomic and Polariton Condensates}}},\ \bibinfo {editor} {edited by\
  \bibinfo {editor} {\bibfnamefont {A.}~\bibnamefont {Bramati}}\ and\ \bibinfo
  {editor} {\bibfnamefont {M.}~\bibnamefont {Modugno}}}\ (\bibinfo  {publisher}
  {Springer Berlin Heidelberg},\ \bibinfo {address} {Berlin, Heidelberg},\
  \bibinfo {year} {2013})\ pp.\ \bibinfo {pages} {283--299}\BibitemShut
  {NoStop}%
\bibitem [{\citenamefont {Barenghi}\ \emph {et~al.}(2014)\citenamefont
  {Barenghi}, \citenamefont {Skrbek},\ and\ \citenamefont
  {Sreenivasan}}]{Barenghi4647}%
  \BibitemOpen
  \bibfield  {author} {\bibinfo {author} {\bibfnamefont {C.~F.}\ \bibnamefont
  {Barenghi}}, \bibinfo {author} {\bibfnamefont {L.}~\bibnamefont {Skrbek}}, \
  and\ \bibinfo {author} {\bibfnamefont {K.~R.}\ \bibnamefont {Sreenivasan}},\
  }\bibfield  {title} {\enquote {\bibinfo {title} {Introduction to quantum
  turbulence},}\ }\href {\doibase 10.1073/pnas.1400033111} {\bibfield
  {journal} {\bibinfo  {journal} {Proceedings of the National Academy of
  Sciences}\ }\textbf {\bibinfo {volume} {111}},\ \bibinfo {pages} {4647}
  (\bibinfo {year} {2014})}\BibitemShut {NoStop}%
\bibitem [{\citenamefont {Parker}\ \emph {et~al.}(2017)\citenamefont {Parker},
  \citenamefont {Allen}, \citenamefont {Barenghi},\ and\ \citenamefont
  {Proukakis}}]{Parker_2017}%
  \BibitemOpen
  \bibfield  {author} {\bibinfo {author} {\bibfnamefont {N.~G.}\ \bibnamefont
  {Parker}}, \bibinfo {author} {\bibfnamefont {A.~J.}\ \bibnamefont {Allen}},
  \bibinfo {author} {\bibfnamefont {C.~F.}\ \bibnamefont {Barenghi}}, \ and\
  \bibinfo {author} {\bibfnamefont {N.~P.}\ \bibnamefont {Proukakis}},\
  }\enquote {\bibinfo {title} {Quantum Turbulence in Atomic Bose-Einstein
  Condensates},}\ in\ \href {\doibase 10.1017/9781316084366.019} {\emph
  {\bibinfo {booktitle} {Universal Themes of Bose-Einstein Condensation}}},\
  \bibinfo {editor} {edited by\ \bibinfo {editor} {\bibfnamefont {N.~P.}\
  \bibnamefont {Proukakis}}, \bibinfo {editor} {\bibfnamefont {D.~W.}\
  \bibnamefont {Snoke}}, \ and\ \bibinfo {editor} {\bibfnamefont {P.~B.}\
  \bibnamefont {Littlewood}}}\ (\bibinfo  {publisher} {Cambridge University
  Press},\ \bibinfo {year} {2017})\ pp.\ \bibinfo {pages}
  {348--370}\BibitemShut {NoStop}%
\bibitem [{\citenamefont {Allen}\ \emph {et~al.}(2014)\citenamefont {Allen},
  \citenamefont {Parker}, \citenamefont {Proukakis},\ and\ \citenamefont
  {Barenghi}}]{Allen_2014}%
  \BibitemOpen
  \bibfield  {author} {\bibinfo {author} {\bibfnamefont {A.~J.}\ \bibnamefont
  {Allen}}, \bibinfo {author} {\bibfnamefont {N.~G.}\ \bibnamefont {Parker}},
  \bibinfo {author} {\bibfnamefont {N.~P.}\ \bibnamefont {Proukakis}}, \ and\
  \bibinfo {author} {\bibfnamefont {C.~F.}\ \bibnamefont {Barenghi}},\
  }\bibfield  {title} {\enquote {\bibinfo {title} {Quantum turbulence in atomic
  Bose-Einstein condensates},}\ }\href
  {http://stacks.iop.org/1742-6596/544/i=1/a=012023} {\bibfield  {journal}
  {\bibinfo  {journal} {Journal of Physics: Conference Series}\ }\textbf
  {\bibinfo {volume} {544}},\ \bibinfo {pages} {012023} (\bibinfo {year}
  {2014})}\BibitemShut {NoStop}%
\bibitem [{\citenamefont {White}\ \emph
  {et~al.}(2014{\natexlab{a}})\citenamefont {White}, \citenamefont {Anderson},\
  and\ \citenamefont {Bagnato}}]{White_2014:a}%
  \BibitemOpen
  \bibfield  {author} {\bibinfo {author} {\bibfnamefont {A.~C.}\ \bibnamefont
  {White}}, \bibinfo {author} {\bibfnamefont {B.~P.}\ \bibnamefont {Anderson}},
  \ and\ \bibinfo {author} {\bibfnamefont {V.~S.}\ \bibnamefont {Bagnato}},\
  }\bibfield  {title} {\enquote {\bibinfo {title} {Vortices and turbulence in
  trapped atomic condensates},}\ }\href {\doibase 10.1073/pnas.1312737110}
  {\bibfield  {journal} {\bibinfo  {journal} {Proceedings of the National
  Academy of Sciences}\ }\textbf {\bibinfo {volume} {111}},\ \bibinfo {pages}
  {4719} (\bibinfo {year} {2014}{\natexlab{a}})}\BibitemShut {NoStop}%
\bibitem [{\citenamefont {Tsatsos}\ \emph {et~al.}(2016)\citenamefont
  {Tsatsos}, \citenamefont {Tavares}, \citenamefont {Cidrim}, \citenamefont
  {Fritsch}, \citenamefont {Caracanhas}, \citenamefont {dos Santos},
  \citenamefont {Barenghi},\ and\ \citenamefont {Bagnato}}]{Tsatsos_2016}%
  \BibitemOpen
  \bibfield  {author} {\bibinfo {author} {\bibfnamefont {M.~C.}\ \bibnamefont
  {Tsatsos}}, \bibinfo {author} {\bibfnamefont {P.~E.}\ \bibnamefont
  {Tavares}}, \bibinfo {author} {\bibfnamefont {A.}~\bibnamefont {Cidrim}},
  \bibinfo {author} {\bibfnamefont {A.~R.}\ \bibnamefont {Fritsch}}, \bibinfo
  {author} {\bibfnamefont {M.~A.}\ \bibnamefont {Caracanhas}}, \bibinfo
  {author} {\bibfnamefont {F.~E.~A.}\ \bibnamefont {dos Santos}}, \bibinfo
  {author} {\bibfnamefont {C.~F.}\ \bibnamefont {Barenghi}}, \ and\ \bibinfo
  {author} {\bibfnamefont {V.~S.}\ \bibnamefont {Bagnato}},\ }\bibfield
  {title} {\enquote {\bibinfo {title} {Quantum turbulence in trapped atomic
  Bose--Einstein condensates},}\ }\href@noop {} {\bibfield  {journal} {\bibinfo
   {journal} {Physics Reports}\ }\textbf {\bibinfo {volume} {622}},\ \bibinfo
  {pages} {1} (\bibinfo {year} {2016})}\BibitemShut {NoStop}%
\bibitem [{\citenamefont {Kraichnan}\ and\ \citenamefont
  {Montgomery}(1980)}]{kraichnan1980two}%
  \BibitemOpen
  \bibfield  {author} {\bibinfo {author} {\bibfnamefont {R.~H.}\ \bibnamefont
  {Kraichnan}}\ and\ \bibinfo {author} {\bibfnamefont {D.}~\bibnamefont
  {Montgomery}},\ }\bibfield  {title} {\enquote {\bibinfo {title}
  {Two-dimensional turbulence},}\ }\href@noop {} {\bibfield  {journal}
  {\bibinfo  {journal} {Reports on Progress in Physics}\ }\textbf {\bibinfo
  {volume} {43}},\ \bibinfo {pages} {547} (\bibinfo {year} {1980})}\BibitemShut
  {NoStop}%
\bibitem [{\citenamefont {Navon}\ \emph {et~al.}(2016)\citenamefont {Navon},
  \citenamefont {Gaunt}, \citenamefont {Smith},\ and\ \citenamefont
  {Hadzibabic}}]{navon2016emergence}%
  \BibitemOpen
  \bibfield  {author} {\bibinfo {author} {\bibfnamefont {N.}~\bibnamefont
  {Navon}}, \bibinfo {author} {\bibfnamefont {A.~L.}\ \bibnamefont {Gaunt}},
  \bibinfo {author} {\bibfnamefont {R.~P.}\ \bibnamefont {Smith}}, \ and\
  \bibinfo {author} {\bibfnamefont {Z.}~\bibnamefont {Hadzibabic}},\ }\bibfield
   {title} {\enquote {\bibinfo {title} {Emergence of a turbulent cascade in a
  quantum gas},}\ }\href@noop {} {\bibfield  {journal} {\bibinfo  {journal}
  {Nature}\ }\textbf {\bibinfo {volume} {539}},\ \bibinfo {pages} {72}
  (\bibinfo {year} {2016})}\BibitemShut {NoStop}%
\bibitem [{\citenamefont {Navon}\ \emph {et~al.}(2019)\citenamefont {Navon},
  \citenamefont {Eigen}, \citenamefont {Zhang}, \citenamefont {Lopes},
  \citenamefont {Gaunt}, \citenamefont {Fujimoto}, \citenamefont {Tsubota},
  \citenamefont {Smith},\ and\ \citenamefont
  {Hadzibabic}}]{navon2019synthetic}%
  \BibitemOpen
  \bibfield  {author} {\bibinfo {author} {\bibfnamefont {N.}~\bibnamefont
  {Navon}}, \bibinfo {author} {\bibfnamefont {C.}~\bibnamefont {Eigen}},
  \bibinfo {author} {\bibfnamefont {J.}~\bibnamefont {Zhang}}, \bibinfo
  {author} {\bibfnamefont {R.}~\bibnamefont {Lopes}}, \bibinfo {author}
  {\bibfnamefont {A.~L.}\ \bibnamefont {Gaunt}}, \bibinfo {author}
  {\bibfnamefont {K.}~\bibnamefont {Fujimoto}}, \bibinfo {author}
  {\bibfnamefont {M.}~\bibnamefont {Tsubota}}, \bibinfo {author} {\bibfnamefont
  {R.~P.}\ \bibnamefont {Smith}}, \ and\ \bibinfo {author} {\bibfnamefont
  {Z.}~\bibnamefont {Hadzibabic}},\ }\bibfield  {title} {\enquote {\bibinfo
  {title} {Synthetic dissipation and cascade fluxes in a turbulent quantum
  gas},}\ }\href@noop {} {\bibfield  {journal} {\bibinfo  {journal} {Science}\
  }\textbf {\bibinfo {volume} {366}},\ \bibinfo {pages} {382} (\bibinfo {year}
  {2019})}\BibitemShut {NoStop}%
\bibitem [{\citenamefont {Bradley}\ and\ \citenamefont
  {Anderson}(2012)}]{Bradley_2012}%
  \BibitemOpen
  \bibfield  {author} {\bibinfo {author} {\bibfnamefont {A.~S.}\ \bibnamefont
  {Bradley}}\ and\ \bibinfo {author} {\bibfnamefont {B.~P.}\ \bibnamefont
  {Anderson}},\ }\bibfield  {title} {\enquote {\bibinfo {title} {Energy Spectra
  of Vortex Distributions in Two-Dimensional Quantum Turbulence},}\ }\href
  {\doibase 10.1103/PhysRevX.2.041001} {\bibfield  {journal} {\bibinfo
  {journal} {Phys. Rev. X}\ }\textbf {\bibinfo {volume} {2}},\ \bibinfo {pages}
  {041001} (\bibinfo {year} {2012})}\BibitemShut {NoStop}%
\bibitem [{\citenamefont {Neely}\ \emph {et~al.}(2013)\citenamefont {Neely},
  \citenamefont {Bradley}, \citenamefont {Samson}, \citenamefont {Rooney},
  \citenamefont {Wright}, \citenamefont {Law}, \citenamefont
  {Carretero-Gonz\'alez}, \citenamefont {Kevrekidis}, \citenamefont {Davis},\
  and\ \citenamefont {Anderson}}]{Neely_2013}%
  \BibitemOpen
  \bibfield  {author} {\bibinfo {author} {\bibfnamefont {T.~W.}\ \bibnamefont
  {Neely}}, \bibinfo {author} {\bibfnamefont {A.~S.}\ \bibnamefont {Bradley}},
  \bibinfo {author} {\bibfnamefont {E.~C.}\ \bibnamefont {Samson}}, \bibinfo
  {author} {\bibfnamefont {S.~J.}\ \bibnamefont {Rooney}}, \bibinfo {author}
  {\bibfnamefont {E.~M.}\ \bibnamefont {Wright}}, \bibinfo {author}
  {\bibfnamefont {K.~J.~H.}\ \bibnamefont {Law}}, \bibinfo {author}
  {\bibfnamefont {R.}~\bibnamefont {Carretero-Gonz\'alez}}, \bibinfo {author}
  {\bibfnamefont {P.~G.}\ \bibnamefont {Kevrekidis}}, \bibinfo {author}
  {\bibfnamefont {M.~J.}\ \bibnamefont {Davis}}, \ and\ \bibinfo {author}
  {\bibfnamefont {B.~P.}\ \bibnamefont {Anderson}},\ }\bibfield  {title}
  {\enquote {\bibinfo {title} {Characteristics of Two-Dimensional Quantum
  Turbulence in a Compressible Superfluid},}\ }\href {\doibase
  10.1103/PhysRevLett.111.235301} {\bibfield  {journal} {\bibinfo  {journal}
  {Phys. Rev. Lett.}\ }\textbf {\bibinfo {volume} {111}},\ \bibinfo {pages}
  {235301} (\bibinfo {year} {2013})}\BibitemShut {NoStop}%
\bibitem [{\citenamefont {Billam}\ \emph {et~al.}(2014)\citenamefont {Billam},
  \citenamefont {Reeves}, \citenamefont {Anderson},\ and\ \citenamefont
  {Bradley}}]{Billam_2014}%
  \BibitemOpen
  \bibfield  {author} {\bibinfo {author} {\bibfnamefont {T.~P.}\ \bibnamefont
  {Billam}}, \bibinfo {author} {\bibfnamefont {M.~T.}\ \bibnamefont {Reeves}},
  \bibinfo {author} {\bibfnamefont {B.~P.}\ \bibnamefont {Anderson}}, \ and\
  \bibinfo {author} {\bibfnamefont {A.~S.}\ \bibnamefont {Bradley}},\
  }\bibfield  {title} {\enquote {\bibinfo {title} {Onsager-Kraichnan
  Condensation in Decaying Two-Dimensional Quantum Turbulence},}\ }\href
  {\doibase 10.1103/PhysRevLett.112.145301} {\bibfield  {journal} {\bibinfo
  {journal} {Phys. Rev. Lett.}\ }\textbf {\bibinfo {volume} {112}},\ \bibinfo
  {pages} {145301} (\bibinfo {year} {2014})}\BibitemShut {NoStop}%
\bibitem [{\citenamefont {Reeves}\ \emph {et~al.}(2014)\citenamefont {Reeves},
  \citenamefont {Billam}, \citenamefont {Anderson},\ and\ \citenamefont
  {Bradley}}]{Reeves_2014a}%
  \BibitemOpen
  \bibfield  {author} {\bibinfo {author} {\bibfnamefont {M.~T.}\ \bibnamefont
  {Reeves}}, \bibinfo {author} {\bibfnamefont {T.~P.}\ \bibnamefont {Billam}},
  \bibinfo {author} {\bibfnamefont {B.~P.}\ \bibnamefont {Anderson}}, \ and\
  \bibinfo {author} {\bibfnamefont {A.~S.}\ \bibnamefont {Bradley}},\
  }\bibfield  {title} {\enquote {\bibinfo {title} {Signatures of coherent
  vortex structures in a disordered two-dimensional quantum fluid},}\ }\href
  {\doibase 10.1103/PhysRevA.89.053631} {\bibfield  {journal} {\bibinfo
  {journal} {Phys. Rev. A}\ }\textbf {\bibinfo {volume} {89}},\ \bibinfo
  {pages} {053631} (\bibinfo {year} {2014})}\BibitemShut {NoStop}%
\bibitem [{\citenamefont {Groszek}\ \emph {et~al.}(2016)\citenamefont
  {Groszek}, \citenamefont {Simula}, \citenamefont {Paganin},\ and\
  \citenamefont {Helmerson}}]{Groszek_2016a}%
  \BibitemOpen
  \bibfield  {author} {\bibinfo {author} {\bibfnamefont {A.~J.}\ \bibnamefont
  {Groszek}}, \bibinfo {author} {\bibfnamefont {T.~P.}\ \bibnamefont {Simula}},
  \bibinfo {author} {\bibfnamefont {D.~M.}\ \bibnamefont {Paganin}}, \ and\
  \bibinfo {author} {\bibfnamefont {K.}~\bibnamefont {Helmerson}},\ }\bibfield
  {title} {\enquote {\bibinfo {title} {Onsager vortex formation in
  Bose-Einstein condensates in two-dimensional power-law traps},}\ }\href
  {\doibase 10.1103/PhysRevA.93.043614} {\bibfield  {journal} {\bibinfo
  {journal} {Phys. Rev. A}\ }\textbf {\bibinfo {volume} {93}},\ \bibinfo
  {pages} {043614} (\bibinfo {year} {2016})}\BibitemShut {NoStop}%
\bibitem [{\citenamefont {Gauthier}\ \emph {et~al.}(2019)\citenamefont
  {Gauthier}, \citenamefont {Reeves}, \citenamefont {Yu}, \citenamefont
  {Bradley}, \citenamefont {Baker}, \citenamefont {Bell}, \citenamefont
  {Rubinsztein-Dunlop}, \citenamefont {Davis},\ and\ \citenamefont
  {Neely}}]{Gauthier1264}%
  \BibitemOpen
  \bibfield  {author} {\bibinfo {author} {\bibfnamefont {G.}~\bibnamefont
  {Gauthier}}, \bibinfo {author} {\bibfnamefont {M.~T.}\ \bibnamefont
  {Reeves}}, \bibinfo {author} {\bibfnamefont {X.}~\bibnamefont {Yu}}, \bibinfo
  {author} {\bibfnamefont {A.~S.}\ \bibnamefont {Bradley}}, \bibinfo {author}
  {\bibfnamefont {M.~A.}\ \bibnamefont {Baker}}, \bibinfo {author}
  {\bibfnamefont {T.~A.}\ \bibnamefont {Bell}}, \bibinfo {author}
  {\bibfnamefont {H.}~\bibnamefont {Rubinsztein-Dunlop}}, \bibinfo {author}
  {\bibfnamefont {M.~J.}\ \bibnamefont {Davis}}, \ and\ \bibinfo {author}
  {\bibfnamefont {T.~W.}\ \bibnamefont {Neely}},\ }\bibfield  {title} {\enquote
  {\bibinfo {title} {Giant vortex clusters in a two-dimensional quantum
  fluid},}\ }\href {\doibase 10.1126/science.aat5718} {\bibfield  {journal}
  {\bibinfo  {journal} {Science}\ }\textbf {\bibinfo {volume} {364}},\ \bibinfo
  {pages} {1264} (\bibinfo {year} {2019})}\BibitemShut {NoStop}%
\bibitem [{\citenamefont {Johnstone}\ \emph {et~al.}(2019)\citenamefont
  {Johnstone}, \citenamefont {Groszek}, \citenamefont {Starkey}, \citenamefont
  {Billington}, \citenamefont {Simula},\ and\ \citenamefont
  {Helmerson}}]{Johnstone1267}%
  \BibitemOpen
  \bibfield  {author} {\bibinfo {author} {\bibfnamefont {S.~P.}\ \bibnamefont
  {Johnstone}}, \bibinfo {author} {\bibfnamefont {A.~J.}\ \bibnamefont
  {Groszek}}, \bibinfo {author} {\bibfnamefont {P.~T.}\ \bibnamefont
  {Starkey}}, \bibinfo {author} {\bibfnamefont {C.~J.}\ \bibnamefont
  {Billington}}, \bibinfo {author} {\bibfnamefont {T.~P.}\ \bibnamefont
  {Simula}}, \ and\ \bibinfo {author} {\bibfnamefont {K.}~\bibnamefont
  {Helmerson}},\ }\bibfield  {title} {\enquote {\bibinfo {title} {Evolution of
  large-scale flow from turbulence in a two-dimensional superfluid},}\ }\href
  {\doibase 10.1126/science.aat5793} {\bibfield  {journal} {\bibinfo  {journal}
  {Science}\ }\textbf {\bibinfo {volume} {364}},\ \bibinfo {pages} {1267}
  (\bibinfo {year} {2019})}\BibitemShut {NoStop}%
\bibitem [{\citenamefont {Valani}\ \emph {et~al.}(2018)\citenamefont {Valani},
  \citenamefont {Groszek},\ and\ \citenamefont {Simula}}]{Valani_2018}%
  \BibitemOpen
  \bibfield  {author} {\bibinfo {author} {\bibfnamefont {R.~N.}\ \bibnamefont
  {Valani}}, \bibinfo {author} {\bibfnamefont {A.~J.}\ \bibnamefont {Groszek}},
  \ and\ \bibinfo {author} {\bibfnamefont {T.~P.}\ \bibnamefont {Simula}},\
  }\bibfield  {title} {\enquote {\bibinfo {title} {Einstein{\textendash}Bose
  condensation of Onsager vortices},}\ }\href {\doibase
  10.1088/1367-2630/aac0bb} {\bibfield  {journal} {\bibinfo  {journal} {New
  Journal of Physics}\ }\textbf {\bibinfo {volume} {20}},\ \bibinfo {pages}
  {053038} (\bibinfo {year} {2018})}\BibitemShut {NoStop}%
\bibitem [{\citenamefont {Groszek}\ \emph {et~al.}(2020)\citenamefont
  {Groszek}, \citenamefont {Davis},\ and\ \citenamefont
  {Simula}}]{10.21468/SciPostPhys.8.3.039}%
  \BibitemOpen
  \bibfield  {author} {\bibinfo {author} {\bibfnamefont {A.~J.}\ \bibnamefont
  {Groszek}}, \bibinfo {author} {\bibfnamefont {M.~J.}\ \bibnamefont {Davis}},
  \ and\ \bibinfo {author} {\bibfnamefont {T.~P.}\ \bibnamefont {Simula}},\
  }\bibfield  {title} {\enquote {\bibinfo {title} {{Decaying quantum turbulence
  in a two-dimensional Bose-Einstein condensate at finite temperature}},}\
  }\href {\doibase 10.21468/SciPostPhys.8.3.039} {\bibfield  {journal}
  {\bibinfo  {journal} {SciPost Phys.}\ }\textbf {\bibinfo {volume} {8}},\
  \bibinfo {pages} {39} (\bibinfo {year} {2020})}\BibitemShut {NoStop}%
\bibitem [{\citenamefont {Myatt}\ \emph {et~al.}(1997)\citenamefont {Myatt},
  \citenamefont {Burt}, \citenamefont {Ghrist}, \citenamefont {Cornell},\ and\
  \citenamefont {Wieman}}]{Myatt_1997}%
  \BibitemOpen
  \bibfield  {author} {\bibinfo {author} {\bibfnamefont {C.~J.}\ \bibnamefont
  {Myatt}}, \bibinfo {author} {\bibfnamefont {E.~A.}\ \bibnamefont {Burt}},
  \bibinfo {author} {\bibfnamefont {R.~W.}\ \bibnamefont {Ghrist}}, \bibinfo
  {author} {\bibfnamefont {E.~A.}\ \bibnamefont {Cornell}}, \ and\ \bibinfo
  {author} {\bibfnamefont {C.~E.}\ \bibnamefont {Wieman}},\ }\bibfield  {title}
  {\enquote {\bibinfo {title} {Production of Two Overlapping Bose-Einstein
  Condensates by Sympathetic Cooling},}\ }\href {\doibase
  10.1103/PhysRevLett.78.586} {\bibfield  {journal} {\bibinfo  {journal} {Phys.
  Rev. Lett.}\ }\textbf {\bibinfo {volume} {78}},\ \bibinfo {pages} {586}
  (\bibinfo {year} {1997})}\BibitemShut {NoStop}%
\bibitem [{\citenamefont {Hall}\ \emph {et~al.}(1998)\citenamefont {Hall},
  \citenamefont {Matthews}, \citenamefont {Ensher}, \citenamefont {Wieman},\
  and\ \citenamefont {Cornell}}]{Hall_1998}%
  \BibitemOpen
  \bibfield  {author} {\bibinfo {author} {\bibfnamefont {D.~S.}\ \bibnamefont
  {Hall}}, \bibinfo {author} {\bibfnamefont {M.~R.}\ \bibnamefont {Matthews}},
  \bibinfo {author} {\bibfnamefont {J.~R.}\ \bibnamefont {Ensher}}, \bibinfo
  {author} {\bibfnamefont {C.~E.}\ \bibnamefont {Wieman}}, \ and\ \bibinfo
  {author} {\bibfnamefont {E.~A.}\ \bibnamefont {Cornell}},\ }\bibfield
  {title} {\enquote {\bibinfo {title} {Dynamics of Component Separation in a
  Binary Mixture of Bose-Einstein Condensates},}\ }\href {\doibase
  10.1103/PhysRevLett.81.1539} {\bibfield  {journal} {\bibinfo  {journal}
  {Phys. Rev. Lett.}\ }\textbf {\bibinfo {volume} {81}},\ \bibinfo {pages}
  {1539} (\bibinfo {year} {1998})}\BibitemShut {NoStop}%
\bibitem [{\citenamefont {Maddaloni}\ \emph {et~al.}(2000)\citenamefont
  {Maddaloni}, \citenamefont {Modugno}, \citenamefont {Fort}, \citenamefont
  {Minardi},\ and\ \citenamefont {Inguscio}}]{Maddaloni_2000}%
  \BibitemOpen
  \bibfield  {author} {\bibinfo {author} {\bibfnamefont {P.}~\bibnamefont
  {Maddaloni}}, \bibinfo {author} {\bibfnamefont {M.}~\bibnamefont {Modugno}},
  \bibinfo {author} {\bibfnamefont {C.}~\bibnamefont {Fort}}, \bibinfo {author}
  {\bibfnamefont {F.}~\bibnamefont {Minardi}}, \ and\ \bibinfo {author}
  {\bibfnamefont {M.}~\bibnamefont {Inguscio}},\ }\bibfield  {title} {\enquote
  {\bibinfo {title} {Collective Oscillations of Two Colliding Bose-Einstein
  Condensates},}\ }\href {\doibase 10.1103/PhysRevLett.85.2413} {\bibfield
  {journal} {\bibinfo  {journal} {Phys. Rev. Lett.}\ }\textbf {\bibinfo
  {volume} {85}},\ \bibinfo {pages} {2413} (\bibinfo {year}
  {2000})}\BibitemShut {NoStop}%
\bibitem [{\citenamefont {Tojo}\ \emph {et~al.}(2010)\citenamefont {Tojo},
  \citenamefont {Taguchi}, \citenamefont {Masuyama}, \citenamefont {Hayashi},
  \citenamefont {Saito},\ and\ \citenamefont {Hirano}}]{Satoshi_2010}%
  \BibitemOpen
  \bibfield  {author} {\bibinfo {author} {\bibfnamefont {S.}~\bibnamefont
  {Tojo}}, \bibinfo {author} {\bibfnamefont {Y.}~\bibnamefont {Taguchi}},
  \bibinfo {author} {\bibfnamefont {Y.}~\bibnamefont {Masuyama}}, \bibinfo
  {author} {\bibfnamefont {T.}~\bibnamefont {Hayashi}}, \bibinfo {author}
  {\bibfnamefont {H.}~\bibnamefont {Saito}}, \ and\ \bibinfo {author}
  {\bibfnamefont {T.}~\bibnamefont {Hirano}},\ }\bibfield  {title} {\enquote
  {\bibinfo {title} {Controlling phase separation of binary Bose-Einstein
  condensates via mixed-spin-channel Feshbach resonance},}\ }\href {\doibase
  10.1103/PhysRevA.82.033609} {\bibfield  {journal} {\bibinfo  {journal} {Phys.
  Rev. A}\ }\textbf {\bibinfo {volume} {82}},\ \bibinfo {pages} {033609}
  (\bibinfo {year} {2010})}\BibitemShut {NoStop}%
\bibitem [{\citenamefont {Egorov}\ \emph {et~al.}(2013)\citenamefont {Egorov},
  \citenamefont {Opanchuk}, \citenamefont {Drummond}, \citenamefont {Hall},
  \citenamefont {Hannaford},\ and\ \citenamefont {Sidorov}}]{Egorov_2013}%
  \BibitemOpen
  \bibfield  {author} {\bibinfo {author} {\bibfnamefont {M.}~\bibnamefont
  {Egorov}}, \bibinfo {author} {\bibfnamefont {B.}~\bibnamefont {Opanchuk}},
  \bibinfo {author} {\bibfnamefont {P.}~\bibnamefont {Drummond}}, \bibinfo
  {author} {\bibfnamefont {B.~V.}\ \bibnamefont {Hall}}, \bibinfo {author}
  {\bibfnamefont {P.}~\bibnamefont {Hannaford}}, \ and\ \bibinfo {author}
  {\bibfnamefont {A.~I.}\ \bibnamefont {Sidorov}},\ }\bibfield  {title}
  {\enquote {\bibinfo {title} {Measurement of $s$-wave scattering lengths in a
  two-component Bose-Einstein condensate},}\ }\href {\doibase
  10.1103/PhysRevA.87.053614} {\bibfield  {journal} {\bibinfo  {journal} {Phys.
  Rev. A}\ }\textbf {\bibinfo {volume} {87}},\ \bibinfo {pages} {053614}
  (\bibinfo {year} {2013})}\BibitemShut {NoStop}%
\bibitem [{\citenamefont {Modugno}\ \emph {et~al.}(2002)\citenamefont
  {Modugno}, \citenamefont {Modugno}, \citenamefont {Riboli}, \citenamefont
  {Roati},\ and\ \citenamefont {Inguscio}}]{Modugno_2002}%
  \BibitemOpen
  \bibfield  {author} {\bibinfo {author} {\bibfnamefont {G.}~\bibnamefont
  {Modugno}}, \bibinfo {author} {\bibfnamefont {M.}~\bibnamefont {Modugno}},
  \bibinfo {author} {\bibfnamefont {F.}~\bibnamefont {Riboli}}, \bibinfo
  {author} {\bibfnamefont {G.}~\bibnamefont {Roati}}, \ and\ \bibinfo {author}
  {\bibfnamefont {M.}~\bibnamefont {Inguscio}},\ }\bibfield  {title} {\enquote
  {\bibinfo {title} {Two Atomic Species Superfluid},}\ }\href {\doibase
  10.1103/PhysRevLett.89.190404} {\bibfield  {journal} {\bibinfo  {journal}
  {Phys. Rev. Lett.}\ }\textbf {\bibinfo {volume} {89}},\ \bibinfo {pages}
  {190404} (\bibinfo {year} {2002})}\BibitemShut {NoStop}%
\bibitem [{\citenamefont {Papp}\ \emph {et~al.}(2008)\citenamefont {Papp},
  \citenamefont {Pino},\ and\ \citenamefont {Wieman}}]{Papp_2008}%
  \BibitemOpen
  \bibfield  {author} {\bibinfo {author} {\bibfnamefont {S.~B.}\ \bibnamefont
  {Papp}}, \bibinfo {author} {\bibfnamefont {J.~M.}\ \bibnamefont {Pino}}, \
  and\ \bibinfo {author} {\bibfnamefont {C.~E.}\ \bibnamefont {Wieman}},\
  }\bibfield  {title} {\enquote {\bibinfo {title} {Tunable Miscibility in a
  Dual-Species Bose-Einstein Condensate},}\ }\href {\doibase
  10.1103/PhysRevLett.101.040402} {\bibfield  {journal} {\bibinfo  {journal}
  {Phys. Rev. Lett.}\ }\textbf {\bibinfo {volume} {101}},\ \bibinfo {pages}
  {040402} (\bibinfo {year} {2008})}\BibitemShut {NoStop}%
\bibitem [{\citenamefont {Thalhammer}\ \emph {et~al.}(2008)\citenamefont
  {Thalhammer}, \citenamefont {Barontini}, \citenamefont {De~Sarlo},
  \citenamefont {Catani}, \citenamefont {Minardi},\ and\ \citenamefont
  {Inguscio}}]{Thalhammer_2008}%
  \BibitemOpen
  \bibfield  {author} {\bibinfo {author} {\bibfnamefont {G.}~\bibnamefont
  {Thalhammer}}, \bibinfo {author} {\bibfnamefont {G.}~\bibnamefont
  {Barontini}}, \bibinfo {author} {\bibfnamefont {L.}~\bibnamefont {De~Sarlo}},
  \bibinfo {author} {\bibfnamefont {J.}~\bibnamefont {Catani}}, \bibinfo
  {author} {\bibfnamefont {F.}~\bibnamefont {Minardi}}, \ and\ \bibinfo
  {author} {\bibfnamefont {M.}~\bibnamefont {Inguscio}},\ }\bibfield  {title}
  {\enquote {\bibinfo {title} {Double Species Bose-Einstein Condensate with
  Tunable Interspecies Interactions},}\ }\href {\doibase
  10.1103/PhysRevLett.100.210402} {\bibfield  {journal} {\bibinfo  {journal}
  {Phys. Rev. Lett.}\ }\textbf {\bibinfo {volume} {100}},\ \bibinfo {pages}
  {210402} (\bibinfo {year} {2008})}\BibitemShut {NoStop}%
\bibitem [{\citenamefont {McCarron}\ \emph {et~al.}(2011)\citenamefont
  {McCarron}, \citenamefont {Cho}, \citenamefont {Jenkin}, \citenamefont
  {K\"oppinger},\ and\ \citenamefont {Cornish}}]{McCarron_2011}%
  \BibitemOpen
  \bibfield  {author} {\bibinfo {author} {\bibfnamefont {D.~J.}\ \bibnamefont
  {McCarron}}, \bibinfo {author} {\bibfnamefont {H.~W.}\ \bibnamefont {Cho}},
  \bibinfo {author} {\bibfnamefont {D.~L.}\ \bibnamefont {Jenkin}}, \bibinfo
  {author} {\bibfnamefont {M.~P.}\ \bibnamefont {K\"oppinger}}, \ and\ \bibinfo
  {author} {\bibfnamefont {S.~L.}\ \bibnamefont {Cornish}},\ }\bibfield
  {title} {\enquote {\bibinfo {title} {Dual-species Bose-Einstein condensate of
  $^{87}\mathrm{Rb}$ and $^{133}\mathrm{Cs}$},}\ }\href {\doibase
  10.1103/PhysRevA.84.011603} {\bibfield  {journal} {\bibinfo  {journal} {Phys.
  Rev. A}\ }\textbf {\bibinfo {volume} {84}},\ \bibinfo {pages} {011603}
  (\bibinfo {year} {2011})}\BibitemShut {NoStop}%
\bibitem [{\citenamefont {Kasamatsu}\ \emph {et~al.}(2005)\citenamefont
  {Kasamatsu}, \citenamefont {Tsubota},\ and\ \citenamefont
  {Ueda}}]{doi:10.1142/S0217979205029602}%
  \BibitemOpen
  \bibfield  {author} {\bibinfo {author} {\bibfnamefont {K.}~\bibnamefont
  {Kasamatsu}}, \bibinfo {author} {\bibfnamefont {M.}~\bibnamefont {Tsubota}},
  \ and\ \bibinfo {author} {\bibfnamefont {M.}~\bibnamefont {Ueda}},\
  }\bibfield  {title} {\enquote {\bibinfo {title} {Vortices in multicomponent
  Bose-Einstein condensates},}\ }\href {\doibase 10.1142/S0217979205029602}
  {\bibfield  {journal} {\bibinfo  {journal} {International Journal of Modern
  Physics B}\ }\textbf {\bibinfo {volume} {19}},\ \bibinfo {pages} {1835}
  (\bibinfo {year} {2005})}\BibitemShut {NoStop}%
\bibitem [{\citenamefont {Mueller}\ and\ \citenamefont
  {Ho}(2002)}]{Mueller_2002}%
  \BibitemOpen
  \bibfield  {author} {\bibinfo {author} {\bibfnamefont {E.~J.}\ \bibnamefont
  {Mueller}}\ and\ \bibinfo {author} {\bibfnamefont {T.-L.}\ \bibnamefont
  {Ho}},\ }\bibfield  {title} {\enquote {\bibinfo {title} {Two-Component
  Bose-Einstein Condensates with a Large Number of Vortices},}\ }\href
  {\doibase 10.1103/PhysRevLett.88.180403} {\bibfield  {journal} {\bibinfo
  {journal} {Phys. Rev. Lett.}\ }\textbf {\bibinfo {volume} {88}},\ \bibinfo
  {pages} {180403} (\bibinfo {year} {2002})}\BibitemShut {NoStop}%
\bibitem [{\citenamefont {Kasamatsu}\ \emph {et~al.}(2003)\citenamefont
  {Kasamatsu}, \citenamefont {Tsubota},\ and\ \citenamefont
  {Ueda}}]{Kasamatsu_2003}%
  \BibitemOpen
  \bibfield  {author} {\bibinfo {author} {\bibfnamefont {K.}~\bibnamefont
  {Kasamatsu}}, \bibinfo {author} {\bibfnamefont {M.}~\bibnamefont {Tsubota}},
  \ and\ \bibinfo {author} {\bibfnamefont {M.}~\bibnamefont {Ueda}},\
  }\bibfield  {title} {\enquote {\bibinfo {title} {Vortex Phase Diagram in
  Rotating Two-Component Bose-Einstein Condensates},}\ }\href {\doibase
  10.1103/PhysRevLett.91.150406} {\bibfield  {journal} {\bibinfo  {journal}
  {Phys. Rev. Lett.}\ }\textbf {\bibinfo {volume} {91}},\ \bibinfo {pages}
  {150406} (\bibinfo {year} {2003})}\BibitemShut {NoStop}%
\bibitem [{\citenamefont {Law}\ \emph {et~al.}(2010)\citenamefont {Law},
  \citenamefont {Kevrekidis},\ and\ \citenamefont {Tuckerman}}]{Law_2010}%
  \BibitemOpen
  \bibfield  {author} {\bibinfo {author} {\bibfnamefont {K.~J.~H.}\
  \bibnamefont {Law}}, \bibinfo {author} {\bibfnamefont {P.~G.}\ \bibnamefont
  {Kevrekidis}}, \ and\ \bibinfo {author} {\bibfnamefont {L.~S.}\ \bibnamefont
  {Tuckerman}},\ }\bibfield  {title} {\enquote {\bibinfo {title} {Stable
  Vortex--Bright-Soliton Structures in Two-Component Bose-Einstein
  Condensates},}\ }\href {\doibase 10.1103/PhysRevLett.105.160405} {\bibfield
  {journal} {\bibinfo  {journal} {Phys. Rev. Lett.}\ }\textbf {\bibinfo
  {volume} {105}},\ \bibinfo {pages} {160405} (\bibinfo {year}
  {2010})}\BibitemShut {NoStop}%
\bibitem [{\citenamefont {Danaila}\ \emph {et~al.}(2016)\citenamefont
  {Danaila}, \citenamefont {Khamehchi}, \citenamefont {Gokhroo}, \citenamefont
  {Engels},\ and\ \citenamefont {Kevrekidis}}]{ionut}%
  \BibitemOpen
  \bibfield  {author} {\bibinfo {author} {\bibfnamefont {I.}~\bibnamefont
  {Danaila}}, \bibinfo {author} {\bibfnamefont {M.~A.}\ \bibnamefont
  {Khamehchi}}, \bibinfo {author} {\bibfnamefont {V.}~\bibnamefont {Gokhroo}},
  \bibinfo {author} {\bibfnamefont {P.}~\bibnamefont {Engels}}, \ and\ \bibinfo
  {author} {\bibfnamefont {P.~G.}\ \bibnamefont {Kevrekidis}},\ }\bibfield
  {title} {\enquote {\bibinfo {title} {Vector dark-antidark solitary waves in
  multicomponent Bose-Einstein condensates},}\ }\href {\doibase
  10.1103/PhysRevA.94.053617} {\bibfield  {journal} {\bibinfo  {journal} {Phys.
  Rev. A}\ }\textbf {\bibinfo {volume} {94}},\ \bibinfo {pages} {053617}
  (\bibinfo {year} {2016})}\BibitemShut {NoStop}%
\bibitem [{\citenamefont {Eto}\ \emph {et~al.}(2011)\citenamefont {Eto},
  \citenamefont {Kasamatsu}, \citenamefont {Nitta}, \citenamefont {Takeuchi},\
  and\ \citenamefont {Tsubota}}]{Eto_2011}%
  \BibitemOpen
  \bibfield  {author} {\bibinfo {author} {\bibfnamefont {M.}~\bibnamefont
  {Eto}}, \bibinfo {author} {\bibfnamefont {K.}~\bibnamefont {Kasamatsu}},
  \bibinfo {author} {\bibfnamefont {M.}~\bibnamefont {Nitta}}, \bibinfo
  {author} {\bibfnamefont {H.}~\bibnamefont {Takeuchi}}, \ and\ \bibinfo
  {author} {\bibfnamefont {M.}~\bibnamefont {Tsubota}},\ }\bibfield  {title}
  {\enquote {\bibinfo {title} {Interaction of half-quantized vortices in
  two-component Bose-Einstein condensates},}\ }\href {\doibase
  10.1103/PhysRevA.83.063603} {\bibfield  {journal} {\bibinfo  {journal} {Phys.
  Rev. A}\ }\textbf {\bibinfo {volume} {83}},\ \bibinfo {pages} {063603}
  (\bibinfo {year} {2011})}\BibitemShut {NoStop}%
\bibitem [{\citenamefont {Kasamatsu}\ \emph {et~al.}(2016)\citenamefont
  {Kasamatsu}, \citenamefont {Eto},\ and\ \citenamefont
  {Nitta}}]{Kasamatsushortrange_2016}%
  \BibitemOpen
  \bibfield  {author} {\bibinfo {author} {\bibfnamefont {K.}~\bibnamefont
  {Kasamatsu}}, \bibinfo {author} {\bibfnamefont {M.}~\bibnamefont {Eto}}, \
  and\ \bibinfo {author} {\bibfnamefont {M.}~\bibnamefont {Nitta}},\ }\bibfield
   {title} {\enquote {\bibinfo {title} {Short-range intervortex interaction and
  interacting dynamics of half-quantized vortices in two-component
  Bose-Einstein condensates},}\ }\href {\doibase 10.1103/PhysRevA.93.013615}
  {\bibfield  {journal} {\bibinfo  {journal} {Phys. Rev. A}\ }\textbf {\bibinfo
  {volume} {93}},\ \bibinfo {pages} {013615} (\bibinfo {year}
  {2016})}\BibitemShut {NoStop}%
\bibitem [{\citenamefont {Cornell}\ \emph {et~al.}(1998)\citenamefont
  {Cornell}, \citenamefont {Hall}, \citenamefont {Matthews},\ and\
  \citenamefont {Wieman}}]{cornell1998having}%
  \BibitemOpen
  \bibfield  {author} {\bibinfo {author} {\bibfnamefont {E.~A.}\ \bibnamefont
  {Cornell}}, \bibinfo {author} {\bibfnamefont {D.}~\bibnamefont {Hall}},
  \bibinfo {author} {\bibfnamefont {M.}~\bibnamefont {Matthews}}, \ and\
  \bibinfo {author} {\bibfnamefont {C.}~\bibnamefont {Wieman}},\ }\bibfield
  {title} {\enquote {\bibinfo {title} {Having it both ways: Distinguishable yet
  phase-coherent mixtures of Bose-Einstein condensates},}\ }\href@noop {}
  {\bibfield  {journal} {\bibinfo  {journal} {Journal of low temperature
  physics}\ }\textbf {\bibinfo {volume} {113}},\ \bibinfo {pages} {151}
  (\bibinfo {year} {1998})}\BibitemShut {NoStop}%
\bibitem [{\citenamefont {Karl}\ \emph {et~al.}(2013)\citenamefont {Karl},
  \citenamefont {Nowak},\ and\ \citenamefont {Gasenzer}}]{MarkusUniversal2013}%
  \BibitemOpen
  \bibfield  {author} {\bibinfo {author} {\bibfnamefont {M.}~\bibnamefont
  {Karl}}, \bibinfo {author} {\bibfnamefont {B.}~\bibnamefont {Nowak}}, \ and\
  \bibinfo {author} {\bibfnamefont {T.}~\bibnamefont {Gasenzer}},\ }\bibfield
  {title} {\enquote {\bibinfo {title} {Universal scaling at nonthermal fixed
  points of a two-component Bose gas},}\ }\href {\doibase
  10.1103/PhysRevA.88.063615} {\bibfield  {journal} {\bibinfo  {journal} {Phys.
  Rev. A}\ }\textbf {\bibinfo {volume} {88}},\ \bibinfo {pages} {063615}
  (\bibinfo {year} {2013})}\BibitemShut {NoStop}%
\bibitem [{\citenamefont {Kobyakov}\ \emph {et~al.}(2014)\citenamefont
  {Kobyakov}, \citenamefont {Bezett}, \citenamefont {Lundh}, \citenamefont
  {Marklund},\ and\ \citenamefont {Bychkov}}]{kobyakov2014turbulence}%
  \BibitemOpen
  \bibfield  {author} {\bibinfo {author} {\bibfnamefont {D.}~\bibnamefont
  {Kobyakov}}, \bibinfo {author} {\bibfnamefont {A.}~\bibnamefont {Bezett}},
  \bibinfo {author} {\bibfnamefont {E.}~\bibnamefont {Lundh}}, \bibinfo
  {author} {\bibfnamefont {M.}~\bibnamefont {Marklund}}, \ and\ \bibinfo
  {author} {\bibfnamefont {V.}~\bibnamefont {Bychkov}},\ }\bibfield  {title}
  {\enquote {\bibinfo {title} {Turbulence in binary Bose-Einstein condensates
  generated by highly nonlinear Rayleigh-Taylor and Kelvin-Helmholtz
  instabilities},}\ }\href@noop {} {\bibfield  {journal} {\bibinfo  {journal}
  {Physical Review A}\ }\textbf {\bibinfo {volume} {89}},\ \bibinfo {pages}
  {013631} (\bibinfo {year} {2014})}\BibitemShut {NoStop}%
\bibitem [{\citenamefont {Han}\ and\ \citenamefont {Tsubota}(2018)}]{Han_2018}%
  \BibitemOpen
  \bibfield  {author} {\bibinfo {author} {\bibfnamefont {J.}~\bibnamefont
  {Han}}\ and\ \bibinfo {author} {\bibfnamefont {M.}~\bibnamefont {Tsubota}},\
  }\bibfield  {title} {\enquote {\bibinfo {title} {Onsager Vortex Formation in
  Two-component Bose-Einstein condensate},}\ }\href {\doibase
  10.7566/JPSJ.87.063601} {\bibfield  {journal} {\bibinfo  {journal} {Journal
  of the Physical Society of Japan}\ }\textbf {\bibinfo {volume} {87}},\
  \bibinfo {pages} {063601} (\bibinfo {year} {2018})},\ \Eprint
  {http://arxiv.org/abs/https://doi.org/10.7566/JPSJ.87.063601}
  {https://doi.org/10.7566/JPSJ.87.063601} \BibitemShut {NoStop}%
\bibitem [{\citenamefont {Han}\ and\ \citenamefont {Tsubota}(2019)}]{Han_2019}%
  \BibitemOpen
  \bibfield  {author} {\bibinfo {author} {\bibfnamefont {J.}~\bibnamefont
  {Han}}\ and\ \bibinfo {author} {\bibfnamefont {M.}~\bibnamefont {Tsubota}},\
  }\bibfield  {title} {\enquote {\bibinfo {title} {Phase separation of
  quantized vortices in two-component miscible Bose-Einstein condensates in a
  two-dimensional box potential},}\ }\href {\doibase
  10.1103/PhysRevA.99.033607} {\bibfield  {journal} {\bibinfo  {journal} {Phys.
  Rev. A}\ }\textbf {\bibinfo {volume} {99}},\ \bibinfo {pages} {033607}
  (\bibinfo {year} {2019})}\BibitemShut {NoStop}%
\bibitem [{\citenamefont {Bisset}\ \emph {et~al.}(2015)\citenamefont {Bisset},
  \citenamefont {Wilson},\ and\ \citenamefont {Ticknor}}]{tick}%
  \BibitemOpen
  \bibfield  {author} {\bibinfo {author} {\bibfnamefont {R.~N.}\ \bibnamefont
  {Bisset}}, \bibinfo {author} {\bibfnamefont {R.~M.}\ \bibnamefont {Wilson}},
  \ and\ \bibinfo {author} {\bibfnamefont {C.}~\bibnamefont {Ticknor}},\
  }\bibfield  {title} {\enquote {\bibinfo {title} {Scaling of fluctuations in a
  trapped binary condensate},}\ }\href {\doibase 10.1103/PhysRevA.91.053613}
  {\bibfield  {journal} {\bibinfo  {journal} {Phys. Rev. A}\ }\textbf {\bibinfo
  {volume} {91}},\ \bibinfo {pages} {053613} (\bibinfo {year}
  {2015})}\BibitemShut {NoStop}%
\bibitem [{Note1()}]{Note1}%
  \BibitemOpen
  \bibinfo {note} {It should be noted though that radial phase separation is
  possible as well, see e.g.~\cite {ionut2} for a recent discussion and
  azimuthal variations thereof.}\BibitemShut {Stop}%
\bibitem [{\citenamefont {Kwon}\ \emph {et~al.}(2014)\citenamefont {Kwon},
  \citenamefont {Moon}, \citenamefont {Choi}, \citenamefont {Seo},\ and\
  \citenamefont {Shin}}]{Kwon_2014}%
  \BibitemOpen
  \bibfield  {author} {\bibinfo {author} {\bibfnamefont {W.~J.}\ \bibnamefont
  {Kwon}}, \bibinfo {author} {\bibfnamefont {G.}~\bibnamefont {Moon}}, \bibinfo
  {author} {\bibfnamefont {J.-y.}\ \bibnamefont {Choi}}, \bibinfo {author}
  {\bibfnamefont {S.~W.}\ \bibnamefont {Seo}}, \ and\ \bibinfo {author}
  {\bibfnamefont {Y.-i.}\ \bibnamefont {Shin}},\ }\bibfield  {title} {\enquote
  {\bibinfo {title} {Relaxation of superfluid turbulence in highly oblate
  Bose-Einstein condensates},}\ }\href {\doibase 10.1103/PhysRevA.90.063627}
  {\bibfield  {journal} {\bibinfo  {journal} {Phys. Rev. A}\ }\textbf {\bibinfo
  {volume} {90}},\ \bibinfo {pages} {063627} (\bibinfo {year}
  {2014})}\BibitemShut {NoStop}%
\bibitem [{\citenamefont {Kwon}\ \emph {et~al.}(2015)\citenamefont {Kwon},
  \citenamefont {Seo},\ and\ \citenamefont {Shin}}]{Kwon_2015}%
  \BibitemOpen
  \bibfield  {author} {\bibinfo {author} {\bibfnamefont {W.~J.}\ \bibnamefont
  {Kwon}}, \bibinfo {author} {\bibfnamefont {S.~W.}\ \bibnamefont {Seo}}, \
  and\ \bibinfo {author} {\bibfnamefont {Y.-i.}\ \bibnamefont {Shin}},\
  }\bibfield  {title} {\enquote {\bibinfo {title} {Periodic shedding of vortex
  dipoles from a moving penetrable obstacle in a Bose-Einstein condensate},}\
  }\href {\doibase 10.1103/PhysRevA.92.033613} {\bibfield  {journal} {\bibinfo
  {journal} {Phys. Rev. A}\ }\textbf {\bibinfo {volume} {92}},\ \bibinfo
  {pages} {033613} (\bibinfo {year} {2015})}\BibitemShut {NoStop}%
\bibitem [{\citenamefont {Seo}\ \emph {et~al.}(2017)\citenamefont {Seo},
  \citenamefont {Ko}, \citenamefont {Kim},\ and\ \citenamefont
  {Shin}}]{seo2017observation}%
  \BibitemOpen
  \bibfield  {author} {\bibinfo {author} {\bibfnamefont {S.~W.}\ \bibnamefont
  {Seo}}, \bibinfo {author} {\bibfnamefont {B.}~\bibnamefont {Ko}}, \bibinfo
  {author} {\bibfnamefont {J.~H.}\ \bibnamefont {Kim}}, \ and\ \bibinfo
  {author} {\bibfnamefont {Y.-i.}\ \bibnamefont {Shin}},\ }\bibfield  {title}
  {\enquote {\bibinfo {title} {Observation of vortex-antivortex pairing in
  decaying 2D turbulence of a superfluid gas},}\ }\href@noop {} {\bibfield
  {journal} {\bibinfo  {journal} {Scientific reports}\ }\textbf {\bibinfo
  {volume} {7}},\ \bibinfo {pages} {4587} (\bibinfo {year} {2017})}\BibitemShut
  {NoStop}%
\bibitem [{\citenamefont {Thompson}\ \emph {et~al.}(2013)\citenamefont
  {Thompson}, \citenamefont {Bagnato}, \citenamefont {Telles}, \citenamefont
  {Caracanhas}, \citenamefont {dos Santos},\ and\ \citenamefont
  {Bagnato}}]{Thompson_2013}%
  \BibitemOpen
  \bibfield  {author} {\bibinfo {author} {\bibfnamefont {K.~J.}\ \bibnamefont
  {Thompson}}, \bibinfo {author} {\bibfnamefont {G.~G.}\ \bibnamefont
  {Bagnato}}, \bibinfo {author} {\bibfnamefont {G.~D.}\ \bibnamefont {Telles}},
  \bibinfo {author} {\bibfnamefont {M.~A.}\ \bibnamefont {Caracanhas}},
  \bibinfo {author} {\bibfnamefont {F.~E.~A.}\ \bibnamefont {dos Santos}}, \
  and\ \bibinfo {author} {\bibfnamefont {V.~S.}\ \bibnamefont {Bagnato}},\
  }\bibfield  {title} {\enquote {\bibinfo {title} {Evidence of power law
  behavior in the momentum distribution of a turbulent trapped
  Bose{\textendash}Einstein condensate},}\ }\href {\doibase
  10.1088/1612-2011/11/1/015501} {\ \textbf {\bibinfo {volume} {11}},\ \bibinfo
  {pages} {015501} (\bibinfo {year} {2013})}\BibitemShut {NoStop}%
\bibitem [{\citenamefont {Castilho}\ \emph {et~al.}(2019)\citenamefont
  {Castilho}, \citenamefont {Pedrozo-Pe{\~{n}}afiel}, \citenamefont
  {Gutierrez}, \citenamefont {Mazo}, \citenamefont {Roati}, \citenamefont
  {Farias},\ and\ \citenamefont {Bagnato}}]{Castilho_2019}%
  \BibitemOpen
  \bibfield  {author} {\bibinfo {author} {\bibfnamefont {P.~C.~M.}\
  \bibnamefont {Castilho}}, \bibinfo {author} {\bibfnamefont {E.}~\bibnamefont
  {Pedrozo-Pe{\~{n}}afiel}}, \bibinfo {author} {\bibfnamefont {E.~M.}\
  \bibnamefont {Gutierrez}}, \bibinfo {author} {\bibfnamefont {P.~L.}\
  \bibnamefont {Mazo}}, \bibinfo {author} {\bibfnamefont {G.}~\bibnamefont
  {Roati}}, \bibinfo {author} {\bibfnamefont {K.~M.}\ \bibnamefont {Farias}}, \
  and\ \bibinfo {author} {\bibfnamefont {V.~S.}\ \bibnamefont {Bagnato}},\
  }\bibfield  {title} {\enquote {\bibinfo {title} {{A compact experimental
  machine for studying tunable Bose{\textendash}Bose superfluid mixtures}},}\
  }\href {\doibase 10.1088/1612-202x/ab00fb} {\bibfield  {journal} {\bibinfo
  {journal} {Laser Physics Letters}\ }\textbf {\bibinfo {volume} {16}},\
  \bibinfo {pages} {035501} (\bibinfo {year} {2019})}\BibitemShut {NoStop}%
\bibitem [{\citenamefont {Stagg}\ \emph {et~al.}(2015)\citenamefont {Stagg},
  \citenamefont {Allen}, \citenamefont {Parker},\ and\ \citenamefont
  {Barenghi}}]{Stagg_2015}%
  \BibitemOpen
  \bibfield  {author} {\bibinfo {author} {\bibfnamefont {G.~W.}\ \bibnamefont
  {Stagg}}, \bibinfo {author} {\bibfnamefont {A.~J.}\ \bibnamefont {Allen}},
  \bibinfo {author} {\bibfnamefont {N.~G.}\ \bibnamefont {Parker}}, \ and\
  \bibinfo {author} {\bibfnamefont {C.~F.}\ \bibnamefont {Barenghi}},\
  }\bibfield  {title} {\enquote {\bibinfo {title} {Generation and decay of
  two-dimensional quantum turbulence in a trapped Bose-Einstein condensate},}\
  }\href {\doibase 10.1103/PhysRevA.91.013612} {\bibfield  {journal} {\bibinfo
  {journal} {Phys. Rev. A}\ }\textbf {\bibinfo {volume} {91}},\ \bibinfo
  {pages} {013612} (\bibinfo {year} {2015})}\BibitemShut {NoStop}%
\bibitem [{\citenamefont {Kevrekidis}\ \emph {et~al.}(2015)\citenamefont
  {Kevrekidis}, \citenamefont {Frantzeskakis},\ and\ \citenamefont
  {Carretero-Gonz{\'a}lez}}]{siambook}%
  \BibitemOpen
  \bibfield  {author} {\bibinfo {author} {\bibfnamefont {P.}~\bibnamefont
  {Kevrekidis}}, \bibinfo {author} {\bibfnamefont {D.}~\bibnamefont
  {Frantzeskakis}}, \ and\ \bibinfo {author} {\bibfnamefont {R.}~\bibnamefont
  {Carretero-Gonz{\'a}lez}},\ }\href@noop {} {\emph {\bibinfo {title} {The
  Defocusing Nonlinear Schrodinger Equation}}}\ (\bibinfo  {publisher} {Society
  for Industrial and Applied Mathematics},\ \bibinfo {address} {Philadelphia,
  PA},\ \bibinfo {year} {2015})\BibitemShut {NoStop}%
\bibitem [{\citenamefont {Raman}\ \emph {et~al.}(1999)\citenamefont {Raman},
  \citenamefont {K\"ohl}, \citenamefont {Onofrio}, \citenamefont {Durfee},
  \citenamefont {Kuklewicz}, \citenamefont {Hadzibabic},\ and\ \citenamefont
  {Ketterle}}]{PhysRevLett.83.2502}%
  \BibitemOpen
  \bibfield  {author} {\bibinfo {author} {\bibfnamefont {C.}~\bibnamefont
  {Raman}}, \bibinfo {author} {\bibfnamefont {M.}~\bibnamefont {K\"ohl}},
  \bibinfo {author} {\bibfnamefont {R.}~\bibnamefont {Onofrio}}, \bibinfo
  {author} {\bibfnamefont {D.~S.}\ \bibnamefont {Durfee}}, \bibinfo {author}
  {\bibfnamefont {C.~E.}\ \bibnamefont {Kuklewicz}}, \bibinfo {author}
  {\bibfnamefont {Z.}~\bibnamefont {Hadzibabic}}, \ and\ \bibinfo {author}
  {\bibfnamefont {W.}~\bibnamefont {Ketterle}},\ }\bibfield  {title} {\enquote
  {\bibinfo {title} {Evidence for a Critical Velocity in a Bose-Einstein
  Condensed Gas},}\ }\href {\doibase 10.1103/PhysRevLett.83.2502} {\bibfield
  {journal} {\bibinfo  {journal} {Phys. Rev. Lett.}\ }\textbf {\bibinfo
  {volume} {83}},\ \bibinfo {pages} {2502} (\bibinfo {year}
  {1999})}\BibitemShut {NoStop}%
\bibitem [{\citenamefont {Onofrio}\ \emph {et~al.}(2000)\citenamefont
  {Onofrio}, \citenamefont {Raman}, \citenamefont {Vogels}, \citenamefont
  {Abo-Shaeer}, \citenamefont {Chikkatur},\ and\ \citenamefont
  {Ketterle}}]{PhysRevLett.85.2228}%
  \BibitemOpen
  \bibfield  {author} {\bibinfo {author} {\bibfnamefont {R.}~\bibnamefont
  {Onofrio}}, \bibinfo {author} {\bibfnamefont {C.}~\bibnamefont {Raman}},
  \bibinfo {author} {\bibfnamefont {J.~M.}\ \bibnamefont {Vogels}}, \bibinfo
  {author} {\bibfnamefont {J.~R.}\ \bibnamefont {Abo-Shaeer}}, \bibinfo
  {author} {\bibfnamefont {A.~P.}\ \bibnamefont {Chikkatur}}, \ and\ \bibinfo
  {author} {\bibfnamefont {W.}~\bibnamefont {Ketterle}},\ }\bibfield  {title}
  {\enquote {\bibinfo {title} {Observation of Superfluid Flow in a
  Bose-Einstein Condensed Gas},}\ }\href {\doibase 10.1103/PhysRevLett.85.2228}
  {\bibfield  {journal} {\bibinfo  {journal} {Phys. Rev. Lett.}\ }\textbf
  {\bibinfo {volume} {85}},\ \bibinfo {pages} {2228} (\bibinfo {year}
  {2000})}\BibitemShut {NoStop}%
\bibitem [{\citenamefont {Weiler}\ \emph {et~al.}(2008)\citenamefont {Weiler},
  \citenamefont {Neely}, \citenamefont {Scherer}, \citenamefont {Bradley},
  \citenamefont {Davis},\ and\ \citenamefont {Anderson}}]{Weiler_2008}%
  \BibitemOpen
  \bibfield  {author} {\bibinfo {author} {\bibfnamefont {C.~N.}\ \bibnamefont
  {Weiler}}, \bibinfo {author} {\bibfnamefont {T.~W.}\ \bibnamefont {Neely}},
  \bibinfo {author} {\bibfnamefont {D.~R.}\ \bibnamefont {Scherer}}, \bibinfo
  {author} {\bibfnamefont {A.~S.}\ \bibnamefont {Bradley}}, \bibinfo {author}
  {\bibfnamefont {M.~J.}\ \bibnamefont {Davis}}, \ and\ \bibinfo {author}
  {\bibfnamefont {B.~P.}\ \bibnamefont {Anderson}},\ }\bibfield  {title}
  {\enquote {\bibinfo {title} {Spontaneous vortices in the formation of
  Bose--Einstein condensates},}\ }\href@noop {} {\bibfield  {journal} {\bibinfo
   {journal} {Nature}\ }\textbf {\bibinfo {volume} {455}},\ \bibinfo {pages}
  {948} (\bibinfo {year} {2008})}\BibitemShut {NoStop}%
\bibitem [{\citenamefont {Reeves}\ \emph {et~al.}(2012)\citenamefont {Reeves},
  \citenamefont {Anderson},\ and\ \citenamefont {Bradley}}]{Reeves_2012}%
  \BibitemOpen
  \bibfield  {author} {\bibinfo {author} {\bibfnamefont {M.~T.}\ \bibnamefont
  {Reeves}}, \bibinfo {author} {\bibfnamefont {B.~P.}\ \bibnamefont
  {Anderson}}, \ and\ \bibinfo {author} {\bibfnamefont {A.~S.}\ \bibnamefont
  {Bradley}},\ }\bibfield  {title} {\enquote {\bibinfo {title} {Classical and
  quantum regimes of two-dimensional turbulence in trapped Bose-Einstein
  condensates},}\ }\href {\doibase 10.1103/PhysRevA.86.053621} {\bibfield
  {journal} {\bibinfo  {journal} {Phys. Rev. A}\ }\textbf {\bibinfo {volume}
  {86}},\ \bibinfo {pages} {053621} (\bibinfo {year} {2012})}\BibitemShut
  {NoStop}%
\bibitem [{\citenamefont {Zhu}\ and\ \citenamefont
  {An}(2018)}]{zhu2018surface}%
  \BibitemOpen
  \bibfield  {author} {\bibinfo {author} {\bibfnamefont {Q.-L.}\ \bibnamefont
  {Zhu}}\ and\ \bibinfo {author} {\bibfnamefont {J.}~\bibnamefont {An}},\
  }\bibfield  {title} {\enquote {\bibinfo {title} {Surface Excitations, Shape
  Deformation, and the Long-Time Behavior in a Stirred Bose--Einstein
  Condensate},}\ }\href@noop {} {\bibfield  {journal} {\bibinfo  {journal}
  {Condensed Matter}\ }\textbf {\bibinfo {volume} {3}},\ \bibinfo {pages} {41}
  (\bibinfo {year} {2018})}\BibitemShut {NoStop}%
\bibitem [{\citenamefont {Groszek}\ \emph {et~al.}(2018)\citenamefont
  {Groszek}, \citenamefont {Davis}, \citenamefont {Paganin}, \citenamefont
  {Helmerson},\ and\ \citenamefont {Simula}}]{Groszek_2018}%
  \BibitemOpen
  \bibfield  {author} {\bibinfo {author} {\bibfnamefont {A.~J.}\ \bibnamefont
  {Groszek}}, \bibinfo {author} {\bibfnamefont {M.~J.}\ \bibnamefont {Davis}},
  \bibinfo {author} {\bibfnamefont {D.~M.}\ \bibnamefont {Paganin}}, \bibinfo
  {author} {\bibfnamefont {K.}~\bibnamefont {Helmerson}}, \ and\ \bibinfo
  {author} {\bibfnamefont {T.~P.}\ \bibnamefont {Simula}},\ }\bibfield  {title}
  {\enquote {\bibinfo {title} {Vortex Thermometry for Turbulent Two-Dimensional
  Fluids},}\ }\href {\doibase 10.1103/PhysRevLett.120.034504} {\bibfield
  {journal} {\bibinfo  {journal} {Phys. Rev. Lett.}\ }\textbf {\bibinfo
  {volume} {120}},\ \bibinfo {pages} {034504} (\bibinfo {year}
  {2018})}\BibitemShut {NoStop}%
\bibitem [{\citenamefont {Zhou}\ and\ \citenamefont {Zhai}(2004)}]{Zhou_2004}%
  \BibitemOpen
  \bibfield  {author} {\bibinfo {author} {\bibfnamefont {Q.}~\bibnamefont
  {Zhou}}\ and\ \bibinfo {author} {\bibfnamefont {H.}~\bibnamefont {Zhai}},\
  }\bibfield  {title} {\enquote {\bibinfo {title} {Vortex dipole in a trapped
  atomic Bose-Einstein condensate},}\ }\href {\doibase
  10.1103/PhysRevA.70.043619} {\bibfield  {journal} {\bibinfo  {journal} {Phys.
  Rev. A}\ }\textbf {\bibinfo {volume} {70}},\ \bibinfo {pages} {043619}
  (\bibinfo {year} {2004})}\BibitemShut {NoStop}%
\bibitem [{\citenamefont {Mithun}\ and\ \citenamefont
  {Kasamatsu}(2019)}]{Mithun_2019}%
  \BibitemOpen
  \bibfield  {author} {\bibinfo {author} {\bibfnamefont {T.}~\bibnamefont
  {Mithun}}\ and\ \bibinfo {author} {\bibfnamefont {K.}~\bibnamefont
  {Kasamatsu}},\ }\bibfield  {title} {\enquote {\bibinfo {title} {Modulation
  instability associated nonlinear dynamics of spin{\textendash}orbit coupled
  Bose{\textendash}Einstein condensates},}\ }\href {\doibase
  10.1088/1361-6455/aafbdd} {\bibfield  {journal} {\bibinfo  {journal} {Journal
  of Physics B: Atomic, Molecular and Optical Physics}\ }\textbf {\bibinfo
  {volume} {52}},\ \bibinfo {pages} {045301} (\bibinfo {year}
  {2019})}\BibitemShut {NoStop}%
\bibitem [{\citenamefont {Sasaki}\ \emph {et~al.}(2010)\citenamefont {Sasaki},
  \citenamefont {Suzuki},\ and\ \citenamefont {Saito}}]{Sasaki_2010}%
  \BibitemOpen
  \bibfield  {author} {\bibinfo {author} {\bibfnamefont {K.}~\bibnamefont
  {Sasaki}}, \bibinfo {author} {\bibfnamefont {N.}~\bibnamefont {Suzuki}}, \
  and\ \bibinfo {author} {\bibfnamefont {H.}~\bibnamefont {Saito}},\ }\bibfield
   {title} {\enquote {\bibinfo {title} {B\'enard--von K\'arm\'an Vortex Street
  in a Bose-Einstein Condensate},}\ }\href {\doibase
  10.1103/PhysRevLett.104.150404} {\bibfield  {journal} {\bibinfo  {journal}
  {Phys. Rev. Lett.}\ }\textbf {\bibinfo {volume} {104}},\ \bibinfo {pages}
  {150404} (\bibinfo {year} {2010})}\BibitemShut {NoStop}%
\bibitem [{\citenamefont {Reeves}\ \emph {et~al.}(2015)\citenamefont {Reeves},
  \citenamefont {Billam}, \citenamefont {Anderson},\ and\ \citenamefont
  {Bradley}}]{Reeves_2015}%
  \BibitemOpen
  \bibfield  {author} {\bibinfo {author} {\bibfnamefont {M.~T.}\ \bibnamefont
  {Reeves}}, \bibinfo {author} {\bibfnamefont {T.~P.}\ \bibnamefont {Billam}},
  \bibinfo {author} {\bibfnamefont {B.~P.}\ \bibnamefont {Anderson}}, \ and\
  \bibinfo {author} {\bibfnamefont {A.~S.}\ \bibnamefont {Bradley}},\
  }\bibfield  {title} {\enquote {\bibinfo {title} {Identifying a Superfluid
  Reynolds Number via Dynamical Similarity},}\ }\href {\doibase
  10.1103/PhysRevLett.114.155302} {\bibfield  {journal} {\bibinfo  {journal}
  {Phys. Rev. Lett.}\ }\textbf {\bibinfo {volume} {114}},\ \bibinfo {pages}
  {155302} (\bibinfo {year} {2015})}\BibitemShut {NoStop}%
\bibitem [{ani()}]{animation}%
  \BibitemOpen
  \href
  {https://www.dropbox.com/sh/bugpiobvfp5zqss/AAB09CzoTjPqQMhqBkr6zXAfa?dl=0}
  {\bibinfo  {journal} {Videos are available here}\ }\BibitemShut {NoStop}%
\bibitem [{\citenamefont {Anderson}\ \emph {et~al.}(2000)\citenamefont
  {Anderson}, \citenamefont {Haljan}, \citenamefont {Wieman},\ and\
  \citenamefont {Cornell}}]{anderson_VB}%
  \BibitemOpen
\bibfield  {journal} {  }\bibfield  {author} {\bibinfo {author} {\bibfnamefont
  {B.~P.}\ \bibnamefont {Anderson}}, \bibinfo {author} {\bibfnamefont {P.~C.}\
  \bibnamefont {Haljan}}, \bibinfo {author} {\bibfnamefont {C.~E.}\
  \bibnamefont {Wieman}}, \ and\ \bibinfo {author} {\bibfnamefont {E.~A.}\
  \bibnamefont {Cornell}},\ }\bibfield  {title} {\enquote {\bibinfo {title}
  {Vortex Precession in Bose-Einstein Condensates: Observations with Filled and
  Empty Cores},}\ }\href {\doibase 10.1103/PhysRevLett.85.2857} {\bibfield
  {journal} {\bibinfo  {journal} {Phys. Rev. Lett.}\ }\textbf {\bibinfo
  {volume} {85}},\ \bibinfo {pages} {2857} (\bibinfo {year}
  {2000})}\BibitemShut {NoStop}%
\bibitem [{\citenamefont {Schweikhard}\ \emph {et~al.}(2004)\citenamefont
  {Schweikhard}, \citenamefont {Coddington}, \citenamefont {Engels},
  \citenamefont {Tung},\ and\ \citenamefont {Cornell}}]{Schweikhard_2004}%
  \BibitemOpen
  \bibfield  {author} {\bibinfo {author} {\bibfnamefont {V.}~\bibnamefont
  {Schweikhard}}, \bibinfo {author} {\bibfnamefont {I.}~\bibnamefont
  {Coddington}}, \bibinfo {author} {\bibfnamefont {P.}~\bibnamefont {Engels}},
  \bibinfo {author} {\bibfnamefont {S.}~\bibnamefont {Tung}}, \ and\ \bibinfo
  {author} {\bibfnamefont {E.~A.}\ \bibnamefont {Cornell}},\ }\bibfield
  {title} {\enquote {\bibinfo {title} {Vortex-Lattice Dynamics in Rotating
  Spinor Bose-Einstein Condensates},}\ }\href {\doibase
  10.1103/PhysRevLett.93.210403} {\bibfield  {journal} {\bibinfo  {journal}
  {Phys. Rev. Lett.}\ }\textbf {\bibinfo {volume} {93}},\ \bibinfo {pages}
  {210403} (\bibinfo {year} {2004})}\BibitemShut {NoStop}%
\bibitem [{\citenamefont {Mukherjee}\ \emph {et~al.}(2020)\citenamefont
  {Mukherjee}, \citenamefont {Mukherjee}, \citenamefont {Mistakidis},
  \citenamefont {Kevrekidis},\ and\ \citenamefont
  {Schmelcher}}]{Mukherjee_2020}%
  \BibitemOpen
  \bibfield  {author} {\bibinfo {author} {\bibfnamefont {K.}~\bibnamefont
  {Mukherjee}}, \bibinfo {author} {\bibfnamefont {K.}~\bibnamefont
  {Mukherjee}}, \bibinfo {author} {\bibfnamefont {S.}~\bibnamefont
  {Mistakidis}}, \bibinfo {author} {\bibfnamefont {P.~G.}\ \bibnamefont
  {Kevrekidis}}, \ and\ \bibinfo {author} {\bibfnamefont {P.}~\bibnamefont
  {Schmelcher}},\ }\bibfield  {title} {\enquote {\bibinfo {title} {Quench
  induced vortex-bright-soliton formation in binary Bose-Einstein
  condensates},}\ }\href {\doibase 10.1088/1361-6455/ab678d} {\bibfield
  {journal} {\bibinfo  {journal} {Journal of Physics B: Atomic, Molecular and
  Optical Physics}\ } (\bibinfo {year} {2020}),\
  10.1088/1361-6455/ab678d}\BibitemShut {NoStop}%
\bibitem [{\citenamefont {Lobo}\ \emph {et~al.}(2004)\citenamefont {Lobo},
  \citenamefont {Sinatra},\ and\ \citenamefont {Castin}}]{Lobo_2004}%
  \BibitemOpen
  \bibfield  {author} {\bibinfo {author} {\bibfnamefont {C.}~\bibnamefont
  {Lobo}}, \bibinfo {author} {\bibfnamefont {A.}~\bibnamefont {Sinatra}}, \
  and\ \bibinfo {author} {\bibfnamefont {Y.}~\bibnamefont {Castin}},\
  }\bibfield  {title} {\enquote {\bibinfo {title} {Vortex Lattice Formation in
  Bose-Einstein Condensates},}\ }\href {\doibase 10.1103/PhysRevLett.92.020403}
  {\bibfield  {journal} {\bibinfo  {journal} {Phys. Rev. Lett.}\ }\textbf
  {\bibinfo {volume} {92}},\ \bibinfo {pages} {020403} (\bibinfo {year}
  {2004})}\BibitemShut {NoStop}%
\bibitem [{\citenamefont {Parker}\ and\ \citenamefont
  {Adams}(2005)}]{Parker_2005}%
  \BibitemOpen
  \bibfield  {author} {\bibinfo {author} {\bibfnamefont {N.~G.}\ \bibnamefont
  {Parker}}\ and\ \bibinfo {author} {\bibfnamefont {C.~S.}\ \bibnamefont
  {Adams}},\ }\bibfield  {title} {\enquote {\bibinfo {title} {Emergence and
  Decay of Turbulence in Stirred Atomic Bose-Einstein Condensates},}\ }\href
  {\doibase 10.1103/PhysRevLett.95.145301} {\bibfield  {journal} {\bibinfo
  {journal} {Phys. Rev. Lett.}\ }\textbf {\bibinfo {volume} {95}},\ \bibinfo
  {pages} {145301} (\bibinfo {year} {2005})}\BibitemShut {NoStop}%
\bibitem [{\citenamefont {Matthews}\ \emph {et~al.}(1999)\citenamefont
  {Matthews}, \citenamefont {Anderson}, \citenamefont {Haljan}, \citenamefont
  {Hall}, \citenamefont {Wieman},\ and\ \citenamefont
  {Cornell}}]{Matthews_1999}%
  \BibitemOpen
  \bibfield  {author} {\bibinfo {author} {\bibfnamefont {M.~R.}\ \bibnamefont
  {Matthews}}, \bibinfo {author} {\bibfnamefont {B.~P.}\ \bibnamefont
  {Anderson}}, \bibinfo {author} {\bibfnamefont {P.~C.}\ \bibnamefont
  {Haljan}}, \bibinfo {author} {\bibfnamefont {D.~S.}\ \bibnamefont {Hall}},
  \bibinfo {author} {\bibfnamefont {C.~E.}\ \bibnamefont {Wieman}}, \ and\
  \bibinfo {author} {\bibfnamefont {E.~A.}\ \bibnamefont {Cornell}},\
  }\bibfield  {title} {\enquote {\bibinfo {title} {Vortices in a Bose-Einstein
  Condensate},}\ }\href {\doibase 10.1103/PhysRevLett.83.2498} {\bibfield
  {journal} {\bibinfo  {journal} {Phys. Rev. Lett.}\ }\textbf {\bibinfo
  {volume} {83}},\ \bibinfo {pages} {2498} (\bibinfo {year}
  {1999})}\BibitemShut {NoStop}%
\bibitem [{\citenamefont {Garc{\'\i}a-Ripoll}\ and\ \citenamefont
  {P{\'e}rez-Garc{\'\i}a}(2000)}]{Garcia_2000}%
  \BibitemOpen
  \bibfield  {author} {\bibinfo {author} {\bibfnamefont {J.~J.}\ \bibnamefont
  {Garc{\'\i}a-Ripoll}}\ and\ \bibinfo {author} {\bibfnamefont {V.~M.}\
  \bibnamefont {P{\'e}rez-Garc{\'\i}a}},\ }\bibfield  {title} {\enquote
  {\bibinfo {title} {Stable and Unstable Vortices in Multicomponent
  Bose-Einstein Condensates},}\ }\href {\doibase 10.1103/PhysRevLett.84.4264}
  {\bibfield  {journal} {\bibinfo  {journal} {Phys. Rev. Lett.}\ }\textbf
  {\bibinfo {volume} {84}},\ \bibinfo {pages} {4264} (\bibinfo {year}
  {2000})}\BibitemShut {NoStop}%
\bibitem [{\citenamefont {P{\'e}rez-Garc{\'\i}a}\ and\ \citenamefont
  {Garcia-Ripoll}(2000)}]{Garcia_2000:a}%
  \BibitemOpen
  \bibfield  {author} {\bibinfo {author} {\bibfnamefont {V.~M.}\ \bibnamefont
  {P{\'e}rez-Garc{\'\i}a}}\ and\ \bibinfo {author} {\bibfnamefont {J.~J.}\
  \bibnamefont {Garcia-Ripoll}},\ }\bibfield  {title} {\enquote {\bibinfo
  {title} {Two-mode theory of vortex stability in multicomponent Bose-Einstein
  condensates},}\ }\href {\doibase 10.1103/PhysRevA.62.033601} {\bibfield
  {journal} {\bibinfo  {journal} {Phys. Rev. A}\ }\textbf {\bibinfo {volume}
  {62}},\ \bibinfo {pages} {033601} (\bibinfo {year} {2000})}\BibitemShut
  {NoStop}%
\bibitem [{\citenamefont {Numasato}\ \emph {et~al.}(2010)\citenamefont
  {Numasato}, \citenamefont {Tsubota},\ and\ \citenamefont
  {L'vov}}]{Numasato_2010}%
  \BibitemOpen
  \bibfield  {author} {\bibinfo {author} {\bibfnamefont {R.}~\bibnamefont
  {Numasato}}, \bibinfo {author} {\bibfnamefont {M.}~\bibnamefont {Tsubota}}, \
  and\ \bibinfo {author} {\bibfnamefont {V.~S.}\ \bibnamefont {L'vov}},\
  }\bibfield  {title} {\enquote {\bibinfo {title} {Direct energy cascade in
  two-dimensional compressible quantum turbulence},}\ }\href {\doibase
  10.1103/PhysRevA.81.063630} {\bibfield  {journal} {\bibinfo  {journal} {Phys.
  Rev. A}\ }\textbf {\bibinfo {volume} {81}},\ \bibinfo {pages} {063630}
  (\bibinfo {year} {2010})}\BibitemShut {NoStop}%
\bibitem [{\citenamefont {Horng}\ \emph {et~al.}(2009)\citenamefont {Horng},
  \citenamefont {Hsueh}, \citenamefont {Su}, \citenamefont {Kao},\ and\
  \citenamefont {Gou}}]{Horng_2009}%
  \BibitemOpen
  \bibfield  {author} {\bibinfo {author} {\bibfnamefont {T.-L.}\ \bibnamefont
  {Horng}}, \bibinfo {author} {\bibfnamefont {C.-H.}\ \bibnamefont {Hsueh}},
  \bibinfo {author} {\bibfnamefont {S.-W.}\ \bibnamefont {Su}}, \bibinfo
  {author} {\bibfnamefont {Y.-M.}\ \bibnamefont {Kao}}, \ and\ \bibinfo
  {author} {\bibfnamefont {S.-C.}\ \bibnamefont {Gou}},\ }\bibfield  {title}
  {\enquote {\bibinfo {title} {Two-dimensional quantum turbulence in a
  nonuniform Bose-Einstein condensate},}\ }\href {\doibase
  10.1103/PhysRevA.80.023618} {\bibfield  {journal} {\bibinfo  {journal} {Phys.
  Rev. A}\ }\textbf {\bibinfo {volume} {80}},\ \bibinfo {pages} {023618}
  (\bibinfo {year} {2009})}\BibitemShut {NoStop}%
\bibitem [{\citenamefont {Simula}\ \emph {et~al.}(2014)\citenamefont {Simula},
  \citenamefont {Davis},\ and\ \citenamefont {Helmerson}}]{Simula_2014}%
  \BibitemOpen
  \bibfield  {author} {\bibinfo {author} {\bibfnamefont {T.}~\bibnamefont
  {Simula}}, \bibinfo {author} {\bibfnamefont {M.~J.}\ \bibnamefont {Davis}}, \
  and\ \bibinfo {author} {\bibfnamefont {K.}~\bibnamefont {Helmerson}},\
  }\bibfield  {title} {\enquote {\bibinfo {title} {Emergence of Order from
  Turbulence in an Isolated Planar Superfluid},}\ }\href {\doibase
  10.1103/PhysRevLett.113.165302} {\bibfield  {journal} {\bibinfo  {journal}
  {Phys. Rev. Lett.}\ }\textbf {\bibinfo {volume} {113}},\ \bibinfo {pages}
  {165302} (\bibinfo {year} {2014})}\BibitemShut {NoStop}%
\bibitem [{\citenamefont {Billam}\ \emph {et~al.}(2015)\citenamefont {Billam},
  \citenamefont {Reeves},\ and\ \citenamefont {Bradley}}]{Billam_2015}%
  \BibitemOpen
  \bibfield  {author} {\bibinfo {author} {\bibfnamefont {T.~P.}\ \bibnamefont
  {Billam}}, \bibinfo {author} {\bibfnamefont {M.~T.}\ \bibnamefont {Reeves}},
  \ and\ \bibinfo {author} {\bibfnamefont {A.~S.}\ \bibnamefont {Bradley}},\
  }\bibfield  {title} {\enquote {\bibinfo {title} {Spectral energy transport in
  two-dimensional quantum vortex dynamics},}\ }\href {\doibase
  10.1103/PhysRevA.91.023615} {\bibfield  {journal} {\bibinfo  {journal} {Phys.
  Rev. A}\ }\textbf {\bibinfo {volume} {91}},\ \bibinfo {pages} {023615}
  (\bibinfo {year} {2015})}\BibitemShut {NoStop}%
\bibitem [{\citenamefont {Sasa}\ \emph {et~al.}(2011)\citenamefont {Sasa},
  \citenamefont {Kano}, \citenamefont {Machida}, \citenamefont {L’vov},
  \citenamefont {Rudenko},\ and\ \citenamefont {Tsubota}}]{sasa2011energy}%
  \BibitemOpen
  \bibfield  {author} {\bibinfo {author} {\bibfnamefont {N.}~\bibnamefont
  {Sasa}}, \bibinfo {author} {\bibfnamefont {T.}~\bibnamefont {Kano}}, \bibinfo
  {author} {\bibfnamefont {M.}~\bibnamefont {Machida}}, \bibinfo {author}
  {\bibfnamefont {V.~S.}\ \bibnamefont {L’vov}}, \bibinfo {author}
  {\bibfnamefont {O.}~\bibnamefont {Rudenko}}, \ and\ \bibinfo {author}
  {\bibfnamefont {M.}~\bibnamefont {Tsubota}},\ }\bibfield  {title} {\enquote
  {\bibinfo {title} {Energy spectra of quantum turbulence: Large-scale
  simulation and modeling},}\ }\href@noop {} {\bibfield  {journal} {\bibinfo
  {journal} {Physical Review B}\ }\textbf {\bibinfo {volume} {84}},\ \bibinfo
  {pages} {054525} (\bibinfo {year} {2011})}\BibitemShut {NoStop}%
\bibitem [{\citenamefont {Kobayashi}\ and\ \citenamefont
  {Tsubota}(2005)}]{Kobayashi_2005}%
  \BibitemOpen
  \bibfield  {author} {\bibinfo {author} {\bibfnamefont {M.}~\bibnamefont
  {Kobayashi}}\ and\ \bibinfo {author} {\bibfnamefont {M.}~\bibnamefont
  {Tsubota}},\ }\bibfield  {title} {\enquote {\bibinfo {title} {Kolmogorov
  Spectrum of Superfluid Turbulence: Numerical Analysis of the Gross-Pitaevskii
  Equation with a Small-Scale Dissipation},}\ }\href {\doibase
  10.1103/PhysRevLett.94.065302} {\bibfield  {journal} {\bibinfo  {journal}
  {Phys. Rev. Lett.}\ }\textbf {\bibinfo {volume} {94}},\ \bibinfo {pages}
  {065302} (\bibinfo {year} {2005})}\BibitemShut {NoStop}%
\bibitem [{\citenamefont {Mosk}(2005)}]{Mosk:2005}%
  \BibitemOpen
  \bibfield  {author} {\bibinfo {author} {\bibfnamefont {A.~P.}\ \bibnamefont
  {Mosk}},\ }\bibfield  {title} {\enquote {\bibinfo {title} {Atomic Gases at
  Negative Kinetic Temperature},}\ }\href {\doibase
  10.1103/PhysRevLett.95.040403} {\bibfield  {journal} {\bibinfo  {journal}
  {Phys. Rev. Lett.}\ }\textbf {\bibinfo {volume} {95}},\ \bibinfo {pages}
  {040403} (\bibinfo {year} {2005})}\BibitemShut {NoStop}%
\bibitem [{\citenamefont {White}\ \emph
  {et~al.}(2014{\natexlab{b}})\citenamefont {White}, \citenamefont
  {Proukakis},\ and\ \citenamefont {Barenghi}}]{White_2014}%
  \BibitemOpen
  \bibfield  {author} {\bibinfo {author} {\bibfnamefont {A.~C.}\ \bibnamefont
  {White}}, \bibinfo {author} {\bibfnamefont {N.~P.}\ \bibnamefont
  {Proukakis}}, \ and\ \bibinfo {author} {\bibfnamefont {C.~F.}\ \bibnamefont
  {Barenghi}},\ }\bibfield  {title} {\enquote {\bibinfo {title} {Topological
  stirring of two-dimensional atomic Bose-Einstein condensates},}\ }\href
  {http://stacks.iop.org/1742-6596/544/i=1/a=012021} {\bibfield  {journal}
  {\bibinfo  {journal} {Journal of Physics: Conference Series}\ }\textbf
  {\bibinfo {volume} {544}},\ \bibinfo {pages} {012021} (\bibinfo {year}
  {2014}{\natexlab{b}})}\BibitemShut {NoStop}%
\bibitem [{\citenamefont {Pakter}\ and\ \citenamefont
  {Levin}(2018)}]{Pakter_2018}%
  \BibitemOpen
  \bibfield  {author} {\bibinfo {author} {\bibfnamefont {R.}~\bibnamefont
  {Pakter}}\ and\ \bibinfo {author} {\bibfnamefont {Y.}~\bibnamefont {Levin}},\
  }\bibfield  {title} {\enquote {\bibinfo {title} {Nonequilibrium Statistical
  Mechanics of Two-Dimensional Vortices},}\ }\href {\doibase
  10.1103/PhysRevLett.121.020602} {\bibfield  {journal} {\bibinfo  {journal}
  {Phys. Rev. Lett.}\ }\textbf {\bibinfo {volume} {121}},\ \bibinfo {pages}
  {020602} (\bibinfo {year} {2018})}\BibitemShut {NoStop}%
\bibitem [{\citenamefont {Yatsuyanagi}\ \emph {et~al.}(2005)\citenamefont
  {Yatsuyanagi}, \citenamefont {Kiwamoto}, \citenamefont {Tomita},
  \citenamefont {Sano}, \citenamefont {Yoshida},\ and\ \citenamefont
  {Ebisuzaki}}]{Yatsuyanagi_2005}%
  \BibitemOpen
  \bibfield  {author} {\bibinfo {author} {\bibfnamefont {Y.}~\bibnamefont
  {Yatsuyanagi}}, \bibinfo {author} {\bibfnamefont {Y.}~\bibnamefont
  {Kiwamoto}}, \bibinfo {author} {\bibfnamefont {H.}~\bibnamefont {Tomita}},
  \bibinfo {author} {\bibfnamefont {M.~M.}\ \bibnamefont {Sano}}, \bibinfo
  {author} {\bibfnamefont {T.}~\bibnamefont {Yoshida}}, \ and\ \bibinfo
  {author} {\bibfnamefont {T.}~\bibnamefont {Ebisuzaki}},\ }\bibfield  {title}
  {\enquote {\bibinfo {title} {Dynamics of Two-Sign Point Vortices in Positive
  and Negative Temperature States},}\ }\href {\doibase
  10.1103/PhysRevLett.94.054502} {\bibfield  {journal} {\bibinfo  {journal}
  {Phys. Rev. Lett.}\ }\textbf {\bibinfo {volume} {94}},\ \bibinfo {pages}
  {054502} (\bibinfo {year} {2005})}\BibitemShut {NoStop}%
\bibitem [{\citenamefont {White}\ \emph {et~al.}(2012)\citenamefont {White},
  \citenamefont {Barenghi},\ and\ \citenamefont {Proukakis}}]{White_2012}%
  \BibitemOpen
  \bibfield  {author} {\bibinfo {author} {\bibfnamefont {A.~C.}\ \bibnamefont
  {White}}, \bibinfo {author} {\bibfnamefont {C.~F.}\ \bibnamefont {Barenghi}},
  \ and\ \bibinfo {author} {\bibfnamefont {N.~P.}\ \bibnamefont {Proukakis}},\
  }\bibfield  {title} {\enquote {\bibinfo {title} {Creation and
  characterization of vortex clusters in atomic Bose-Einstein condensates},}\
  }\href {\doibase 10.1103/PhysRevA.86.013635} {\bibfield  {journal} {\bibinfo
  {journal} {Phys. Rev. A}\ }\textbf {\bibinfo {volume} {86}},\ \bibinfo
  {pages} {013635} (\bibinfo {year} {2012})}\BibitemShut {NoStop}%
\bibitem [{\citenamefont {Reeves}\ \emph {et~al.}(2013)\citenamefont {Reeves},
  \citenamefont {Billam}, \citenamefont {Anderson},\ and\ \citenamefont
  {Bradley}}]{Reeves_2013}%
  \BibitemOpen
  \bibfield  {author} {\bibinfo {author} {\bibfnamefont {M.~T.}\ \bibnamefont
  {Reeves}}, \bibinfo {author} {\bibfnamefont {T.~P.}\ \bibnamefont {Billam}},
  \bibinfo {author} {\bibfnamefont {B.~P.}\ \bibnamefont {Anderson}}, \ and\
  \bibinfo {author} {\bibfnamefont {A.~S.}\ \bibnamefont {Bradley}},\
  }\bibfield  {title} {\enquote {\bibinfo {title} {Inverse Energy Cascade in
  Forced Two-Dimensional Quantum Turbulence},}\ }\href {\doibase
  10.1103/PhysRevLett.110.104501} {\bibfield  {journal} {\bibinfo  {journal}
  {Phys. Rev. Lett.}\ }\textbf {\bibinfo {volume} {110}},\ \bibinfo {pages}
  {104501} (\bibinfo {year} {2013})}\BibitemShut {NoStop}%
\bibitem [{\citenamefont {Skaugen}\ and\ \citenamefont
  {Angheluta}(2016)}]{Skaugen_2016}%
  \BibitemOpen
  \bibfield  {author} {\bibinfo {author} {\bibfnamefont {A.}~\bibnamefont
  {Skaugen}}\ and\ \bibinfo {author} {\bibfnamefont {L.}~\bibnamefont
  {Angheluta}},\ }\bibfield  {title} {\enquote {\bibinfo {title} {Vortex
  clustering and universal scaling laws in two-dimensional quantum
  turbulence},}\ }\href {\doibase 10.1103/PhysRevE.93.032106} {\bibfield
  {journal} {\bibinfo  {journal} {Phys. Rev. E}\ }\textbf {\bibinfo {volume}
  {93}},\ \bibinfo {pages} {032106} (\bibinfo {year} {2016})}\BibitemShut
  {NoStop}%
\bibitem [{\citenamefont {Mazenko}(2001)}]{Mazenko_2001}%
  \BibitemOpen
  \bibfield  {author} {\bibinfo {author} {\bibfnamefont {G.~F.}\ \bibnamefont
  {Mazenko}},\ }\bibfield  {title} {\enquote {\bibinfo {title} {Defect
  statistics in the two-dimensional complex Ginzburg-Landau model},}\ }\href
  {\doibase 10.1103/PhysRevE.64.016110} {\bibfield  {journal} {\bibinfo
  {journal} {Phys. Rev. E}\ }\textbf {\bibinfo {volume} {64}},\ \bibinfo
  {pages} {016110} (\bibinfo {year} {2001})}\BibitemShut {NoStop}%
\bibitem [{\citenamefont {Angheluta}\ \emph {et~al.}(2012)\citenamefont
  {Angheluta}, \citenamefont {Jeraldo},\ and\ \citenamefont
  {Goldenfeld}}]{Angheluta_2012}%
  \BibitemOpen
  \bibfield  {author} {\bibinfo {author} {\bibfnamefont {L.}~\bibnamefont
  {Angheluta}}, \bibinfo {author} {\bibfnamefont {P.}~\bibnamefont {Jeraldo}},
  \ and\ \bibinfo {author} {\bibfnamefont {N.}~\bibnamefont {Goldenfeld}},\
  }\bibfield  {title} {\enquote {\bibinfo {title} {Anisotropic velocity
  statistics of topological defects under shear flow},}\ }\href {\doibase
  10.1103/PhysRevE.85.011153} {\bibfield  {journal} {\bibinfo  {journal} {Phys.
  Rev. E}\ }\textbf {\bibinfo {volume} {85}},\ \bibinfo {pages} {011153}
  (\bibinfo {year} {2012})}\BibitemShut {NoStop}%
\bibitem [{\citenamefont {Henderson}\ \emph {et~al.}(2009)\citenamefont
  {Henderson}, \citenamefont {Ryu}, \citenamefont {MacCormick},\ and\
  \citenamefont {Boshier}}]{boshier}%
  \BibitemOpen
  \bibfield  {author} {\bibinfo {author} {\bibfnamefont {K.}~\bibnamefont
  {Henderson}}, \bibinfo {author} {\bibfnamefont {C.}~\bibnamefont {Ryu}},
  \bibinfo {author} {\bibfnamefont {C.}~\bibnamefont {MacCormick}}, \ and\
  \bibinfo {author} {\bibfnamefont {M.~G.}\ \bibnamefont {Boshier}},\
  }\bibfield  {title} {\enquote {\bibinfo {title} {Experimental demonstration
  of painting arbitrary and dynamic potentials for Bose{\textendash}Einstein
  condensates},}\ }\href {\doibase 10.1088/1367-2630/11/4/043030} {\bibfield
  {journal} {\bibinfo  {journal} {New Journal of Physics}\ }\textbf {\bibinfo
  {volume} {11}},\ \bibinfo {pages} {043030} (\bibinfo {year}
  {2009})}\BibitemShut {NoStop}%
\bibitem [{\citenamefont {Bersano}\ \emph {et~al.}(2018)\citenamefont
  {Bersano}, \citenamefont {Gokhroo}, \citenamefont {Khamehchi}, \citenamefont
  {D'Ambroise}, \citenamefont {Frantzeskakis}, \citenamefont {Engels},\ and\
  \citenamefont {Kevrekidis}}]{bersano}%
  \BibitemOpen
  \bibfield  {author} {\bibinfo {author} {\bibfnamefont {T.~M.}\ \bibnamefont
  {Bersano}}, \bibinfo {author} {\bibfnamefont {V.}~\bibnamefont {Gokhroo}},
  \bibinfo {author} {\bibfnamefont {M.~A.}\ \bibnamefont {Khamehchi}}, \bibinfo
  {author} {\bibfnamefont {J.}~\bibnamefont {D'Ambroise}}, \bibinfo {author}
  {\bibfnamefont {D.~J.}\ \bibnamefont {Frantzeskakis}}, \bibinfo {author}
  {\bibfnamefont {P.}~\bibnamefont {Engels}}, \ and\ \bibinfo {author}
  {\bibfnamefont {P.~G.}\ \bibnamefont {Kevrekidis}},\ }\bibfield  {title}
  {\enquote {\bibinfo {title} {Three-Component Soliton States in Spinor $F=1$
  Bose-Einstein Condensates},}\ }\href {\doibase
  10.1103/PhysRevLett.120.063202} {\bibfield  {journal} {\bibinfo  {journal}
  {Phys. Rev. Lett.}\ }\textbf {\bibinfo {volume} {120}},\ \bibinfo {pages}
  {063202} (\bibinfo {year} {2018})}\BibitemShut {NoStop}%
\bibitem [{\citenamefont {Chai}\ \emph {et~al.}(2020)\citenamefont {Chai},
  \citenamefont {Lao}, \citenamefont {Fujimoto}, \citenamefont {Hamazaki},
  \citenamefont {Ueda},\ and\ \citenamefont {Raman}}]{chai2019magnetic}%
  \BibitemOpen
  \bibfield  {author} {\bibinfo {author} {\bibfnamefont {X.}~\bibnamefont
  {Chai}}, \bibinfo {author} {\bibfnamefont {D.}~\bibnamefont {Lao}}, \bibinfo
  {author} {\bibfnamefont {K.}~\bibnamefont {Fujimoto}}, \bibinfo {author}
  {\bibfnamefont {R.}~\bibnamefont {Hamazaki}}, \bibinfo {author}
  {\bibfnamefont {M.}~\bibnamefont {Ueda}}, \ and\ \bibinfo {author}
  {\bibfnamefont {C.}~\bibnamefont {Raman}},\ }\bibfield  {title} {\enquote
  {\bibinfo {title} {Magnetic Solitons in a Spin-1 Bose-Einstein Condensate},}\
  }\href {\doibase 10.1103/PhysRevLett.125.030402} {\bibfield  {journal}
  {\bibinfo  {journal} {Phys. Rev. Lett.}\ }\textbf {\bibinfo {volume} {125}},\
  \bibinfo {pages} {030402} (\bibinfo {year} {2020})}\BibitemShut {NoStop}%
\bibitem [{\citenamefont {Kim}\ \emph {et~al.}(2020)\citenamefont {Kim},
  \citenamefont {Hong},\ and\ \citenamefont {Shin}}]{kim2020observation}%
  \BibitemOpen
  \bibfield  {author} {\bibinfo {author} {\bibfnamefont {J.~H.}\ \bibnamefont
  {Kim}}, \bibinfo {author} {\bibfnamefont {D.}~\bibnamefont {Hong}}, \ and\
  \bibinfo {author} {\bibfnamefont {Y.-i.}\ \bibnamefont {Shin}},\ }\bibfield
  {title} {\enquote {\bibinfo {title} {Observation of two sound modes in a
  binary superfluid gas},}\ }\href@noop {} {\bibfield  {journal} {\bibinfo
  {journal} {Physical Review A}\ }\textbf {\bibinfo {volume} {101}},\ \bibinfo
  {pages} {061601} (\bibinfo {year} {2020})}\BibitemShut {NoStop}%
\bibitem [{\citenamefont {Norrie}\ \emph {et~al.}(2006)\citenamefont {Norrie},
  \citenamefont {Ballagh},\ and\ \citenamefont {Gardiner}}]{Norrie_2006}%
  \BibitemOpen
  \bibfield  {author} {\bibinfo {author} {\bibfnamefont {A.~A.}\ \bibnamefont
  {Norrie}}, \bibinfo {author} {\bibfnamefont {R.~J.}\ \bibnamefont {Ballagh}},
  \ and\ \bibinfo {author} {\bibfnamefont {C.~W.}\ \bibnamefont {Gardiner}},\
  }\bibfield  {title} {\enquote {\bibinfo {title} {Quantum turbulence and
  correlations in Bose-Einstein condensate collisions},}\ }\href {\doibase
  10.1103/PhysRevA.73.043617} {\bibfield  {journal} {\bibinfo  {journal} {Phys.
  Rev. A}\ }\textbf {\bibinfo {volume} {73}},\ \bibinfo {pages} {043617}
  (\bibinfo {year} {2006})}\BibitemShut {NoStop}%
\bibitem [{\citenamefont {Tsatsos}\ and\ \citenamefont
  {Lode}(2015)}]{tsatsos2014vortex}%
  \BibitemOpen
  \bibfield  {author} {\bibinfo {author} {\bibfnamefont {M.~C.}\ \bibnamefont
  {Tsatsos}}\ and\ \bibinfo {author} {\bibfnamefont {A.~U.~J.}\ \bibnamefont
  {Lode}},\ }\bibfield  {title} {\enquote {\bibinfo {title} {Resonances and
  Dynamical Fragmentation in a Stirred Bose–Einstein Condensate},}\ }\href
  {\doibase doi.org/10.1007/s10909-015-1335-5} {\bibfield  {journal} {\bibinfo
  {journal} {J. Low Temp. Phys.}\ }\textbf {\bibinfo {volume} {181}},\ \bibinfo
  {pages} {171} (\bibinfo {year} {2015})}\BibitemShut {NoStop}%
\bibitem [{\citenamefont {Weiner}\ \emph {et~al.}(2017)\citenamefont {Weiner},
  \citenamefont {Tsatsos}, \citenamefont {Cederbaum},\ and\ \citenamefont
  {Lode}}]{weiner2017phantom}%
  \BibitemOpen
  \bibfield  {author} {\bibinfo {author} {\bibfnamefont {S.~E.}\ \bibnamefont
  {Weiner}}, \bibinfo {author} {\bibfnamefont {M.~C.}\ \bibnamefont {Tsatsos}},
  \bibinfo {author} {\bibfnamefont {L.~S.}\ \bibnamefont {Cederbaum}}, \ and\
  \bibinfo {author} {\bibfnamefont {A.~U.}\ \bibnamefont {Lode}},\ }\bibfield
  {title} {\enquote {\bibinfo {title} {Phantom vortices: hidden angular
  momentum in ultracold dilute Bose-Einstein condensates},}\ }\href@noop {}
  {\bibfield  {journal} {\bibinfo  {journal} {Scientific reports}\ }\textbf
  {\bibinfo {volume} {7}},\ \bibinfo {pages} {1} (\bibinfo {year}
  {2017})}\BibitemShut {NoStop}%
\bibitem [{\citenamefont {Katsimiga}\ \emph {et~al.}(2017)\citenamefont
  {Katsimiga}, \citenamefont {Koutentakis}, \citenamefont {Mistakidis},
  \citenamefont {Kevrekidis},\ and\ \citenamefont
  {Schmelcher}}]{katsimiga2017dark}%
  \BibitemOpen
  \bibfield  {author} {\bibinfo {author} {\bibfnamefont {G.}~\bibnamefont
  {Katsimiga}}, \bibinfo {author} {\bibfnamefont {G.}~\bibnamefont
  {Koutentakis}}, \bibinfo {author} {\bibfnamefont {S.}~\bibnamefont
  {Mistakidis}}, \bibinfo {author} {\bibfnamefont {P.}~\bibnamefont
  {Kevrekidis}}, \ and\ \bibinfo {author} {\bibfnamefont {P.}~\bibnamefont
  {Schmelcher}},\ }\bibfield  {title} {\enquote {\bibinfo {title} {Dark--bright
  soliton dynamics beyond the mean-field approximation},}\ }\href@noop {}
  {\bibfield  {journal} {\bibinfo  {journal} {New Journal of Physics}\ }\textbf
  {\bibinfo {volume} {19}},\ \bibinfo {pages} {073004} (\bibinfo {year}
  {2017})}\BibitemShut {NoStop}%
\bibitem [{\citenamefont {Sreenivasan}(1999)}]{RevModPhys.71.S383}%
  \BibitemOpen
  \bibfield  {author} {\bibinfo {author} {\bibfnamefont {K.~R.}\ \bibnamefont
  {Sreenivasan}},\ }\bibfield  {title} {\enquote {\bibinfo {title} {Fluid
  turbulence},}\ }\href {\doibase 10.1103/RevModPhys.71.S383} {\bibfield
  {journal} {\bibinfo  {journal} {Rev. Mod. Phys.}\ }\textbf {\bibinfo {volume}
  {71}},\ \bibinfo {pages} {S383} (\bibinfo {year} {1999})}\BibitemShut
  {NoStop}%
\bibitem [{\citenamefont {Katsimiga}\ \emph {et~al.}(2020)\citenamefont
  {Katsimiga}, \citenamefont {Mistakidis}, \citenamefont {Bersano},
  \citenamefont {Ome}, \citenamefont {Mossman}, \citenamefont {Mukherjee},
  \citenamefont {Schmelcher}, \citenamefont {Engels},\ and\ \citenamefont
  {Kevrekidis}}]{ionut2}%
  \BibitemOpen
  \bibfield  {author} {\bibinfo {author} {\bibfnamefont {G.~C.}\ \bibnamefont
  {Katsimiga}}, \bibinfo {author} {\bibfnamefont {S.~I.}\ \bibnamefont
  {Mistakidis}}, \bibinfo {author} {\bibfnamefont {T.~M.}\ \bibnamefont
  {Bersano}}, \bibinfo {author} {\bibfnamefont {M.~K.~H.}\ \bibnamefont {Ome}},
  \bibinfo {author} {\bibfnamefont {S.~M.}\ \bibnamefont {Mossman}}, \bibinfo
  {author} {\bibfnamefont {K.}~\bibnamefont {Mukherjee}}, \bibinfo {author}
  {\bibfnamefont {P.}~\bibnamefont {Schmelcher}}, \bibinfo {author}
  {\bibfnamefont {P.}~\bibnamefont {Engels}}, \ and\ \bibinfo {author}
  {\bibfnamefont {P.~G.}\ \bibnamefont {Kevrekidis}},\ }\bibfield  {title}
  {\enquote {\bibinfo {title} {Observation and analysis of multiple
  dark-antidark solitons in two-component Bose-Einstein condensates},}\ }\href
  {\doibase 10.1103/PhysRevA.102.023301} {\bibfield  {journal} {\bibinfo
  {journal} {Phys. Rev. A}\ }\textbf {\bibinfo {volume} {102}},\ \bibinfo
  {pages} {023301} (\bibinfo {year} {2020})}\BibitemShut {NoStop}%
\end{thebibliography}%
\end{document}